\documentclass[final,authoryear,5p,times,twocolumn]{elsarticle}
\usepackage{amsmath,amssymb,natbib,latexsym,times}
\usepackage[usenames,dvipsnames]{xcolor}

\usepackage[OT2,T1]{fontenc}
\usepackage[draft]{todonotes}
\usepackage{listings}
\usepackage{footmisc}
\usepackage[hidelinks]{hyperref}
\usepackage{url}
\usepackage{breakurl}

\newcommand\YAMLcommentstyle{\footnotesize \color{blue}\mdseries}
\newcommand\YAMLcolonstyle{\footnotesize \color{purple}\bfseries}
\newcommand\YAMLkeystyle{\footnotesize \color{black}\bfseries}
\newcommand\YAMLvaluestyle{\footnotesize \color{red}\bfseries}
\newcommand\ProcessThreeDashes{\llap{\color{black}\mdseries-{-}-}}
\makeatletter
\newcommand\language@yaml{yaml}
\expandafter\expandafter\expandafter\lstdefinelanguage
\expandafter{\language@yaml}
{
  keywords={true,false,null,y,n},
  keywordstyle=\color{darkgray}\bfseries,
  basicstyle=\YAMLkeystyle,                                 % assuming a key comes first
  sensitive=false,
  comment=[l]{\#},
  morecomment=[s]{/*}{*/},
  commentstyle=\YAMLcommentstyle,
  stringstyle=\YAMLvaluestyle,
  moredelim=[l][\color{orange}]{\&},
  moredelim=**[il][\YAMLcolonstyle{:}\YAMLvaluestyle]{:},   % switch to value style at :
  moredelim=**[il][\ \YAMLcolonstyle{\{}\YAMLkeystyle]{\{\ },   % switch to key style at {, [, etc.
  moredelim=**[il][\YAMLcolonstyle{\}}\YAMLkeystyle]{\}},
  moredelim=**[il][\ \YAMLcolonstyle{[}\YAMLkeystyle]{[\ },
  moredelim=**[il][\YAMLcolonstyle{]}\YAMLkeystyle]{]},
  moredelim=**[il][\YAMLcolonstyle{,}\YAMLkeystyle]{,},
  morestring=[b]',
  morestring=[b]",
  literate =    {---}{{\ProcessThreeDashes}}3
                {>}{{\textcolor{red}\textgreater}}1     
                {|}{{\textcolor{red}\textbar}}1 
                {\ -\ }{{\bfseries\ -\ }}3,
}
\lst@AddToHook{EveryLine}{\ifx\lst@language\language@yaml\YAMLkeystyle\fi}
\makeatother

\newcommand{\kmax}{\ensuremath{k_\mathrm{max}}}
\newcommand{\kmin}{\ensuremath{k_\mathrm{min}}}
\newcommand{\sinc}{\ensuremath{\mathrm{sinc}}}
\newcommand{\rmd}{\ensuremath{\mathrm{d}}}
\newcommand{\vecx}{\ensuremath{\mathbf{x}}}
\newcommand{\veck}{\ensuremath{\mathbf{k}}}
\newcommand{\matA}{\ensuremath{\bf A}}
\newcommand{\Si}{\ensuremath{\mathrm{Si}}}
\newcommand{\Ci}{\ensuremath{\mathrm{Ci}}}

\newcommand\conj[1]{\ensuremath{\overline{#1}}}
\newcommand\fourier[1]{\ensuremath{\widetilde{#1}}}

\newcommand{\mi}{{\rm i}}
\newcommand{\me}{{\rm e}}

\newcommand{\hst}{{\em HST}}
\newcommand\shera{\textsc{shera}}
\newcommand\galsim{\textsc{GalSim}}
\newcommand\version{1.2}
\newcommand\numpy{\textsc{NumPy}}
\newcommand\fftw{\textsc{fftw}}  % They use italics on their website for the title, but usually it's just plain FFTW.
\newcommand\sersic{S\'{e}rsic}
\newcommand\disk{disk}  % Consistent choice of spelling for disk/disc.

\newcommand\eqn[1]{equation~\eqref{#1}}
\newcommand\eqnb[2]{equations~\eqref{#1}~\& \eqref{#2}}

\newcommand\app[1]{\ref{#1}}

\newcommand\degree{\ensuremath{\,^\circ}}

\newcommand*\justify{%
  \fontdimen2\font=0.4em% interword space
  \fontdimen3\font=0.2em% interword stretch
  \fontdimen4\font=0.1em% interword shrink
  \fontdimen7\font=0.1em% extra space
  \hyphenchar\font=`\-% allowing hyphenation
}

\newcommand\code[1]{\texttt{\justify #1}}

\DeclareSymbolFont{cyrletters}{OT2}{wncyr}{m}{n}
\DeclareMathSymbol{\Sha}{\mathalpha}{cyrletters}{"58}

\allowdisplaybreaks

\newcommand\conv{\circ}

\newcommand\edit{}%\color{blue}}

\begin{document}

\begin{frontmatter}

\title{\galsim: The modular galaxy image simulation toolkit}

\author[ucl,jpl,caltech]{Barnaby~Rowe\corref{cor}}
\ead{browe@star.ucl.ac.uk}

\author[penn]{Mike~Jarvis\corref{cor}}
\ead{michael@jarvis.net}

\author[cmu]{Rachel~Mandelbaum\corref{cor}}
\ead{rmandelb@andrew.cmu.edu}

\author[penn]{Gary~M.~Bernstein}
\author[princeton]{James~Bosch}
\author[cmu]{Melanie~Simet}
\author[kipac1]{Joshua~E.~Meyers}
\author[ucl,manc]{Tomasz~Kacprzak}
\author[aifa]{Reiko~Nakajima}
\author[manc]{Joe~Zuntz}
\author[princeton,ipmu]{Hironao~Miyatake}
\author[usm,ecu]{J\"{o}rg~P.~Dietrich}
\author[princeton]{Robert~Armstrong}
\author[ccapp]{Peter~Melchior}
\author[kipac2]{Mandeep~S.~S.~Gill}

\cortext[cor]{Corresponding author}

\address[ucl]{Department of Physics \& Astronomy, University
  College London, Gower Street, London, WC1E~6BT, United Kingdom}

\address[jpl]{Jet Propulsion Laboratory, California Institute of
  Technology, 4800 Oak Grove Drive, Pasadena, CA 91109, United States
  of America}

\address[caltech]{California Institute of Technology, 1200 East
  California Boulevard, Pasadena, CA 91106, United States of America}

\address[penn]{Department of Physics \& Astronomy, University of
  Pennsylvania, Philadelphia, PA 19104, United States of America}

\address[cmu]{McWilliams Center for Cosmology, Department of
  Physics, Carnegie Mellon University, 5000 Forbes Ave., Pittsburgh,
  PA 15213, United States of America}

\address[princeton]{Department of Astrophysical
  Sciences, Princeton University, Peyton Hall, Princeton, NJ 08544,
  United States of America}

\address[kipac1]{Kavli Institute for Particle Astrophysics and Cosmology, Department of Physics, Stanford
University, Stanford, CA 94305, United States of America}

\address[manc]{Jodrell Bank Centre for Astrophysics, University of
  Manchester, Manchester, M13~9PL, United Kingdom}

\address[aifa]{Argelander-Institut f\"{u}r Astronomie, Universit\"{a}t
  Bonn, Auf dem H\"{u}gel 71, D-53121 Bonn, Germany}

\address[ipmu]{Kavli Institute for the Physics and Mathematics of the
  Universe (Kavli IPMU, WPI), The University of Tokyo, Kashiwa, Chiba
  277-8582, Japan}

\address[usm]{Universit\"ats-Sternwarte M\"unchen, Scheinerstr.\ 1, 81679 M\"unchen,
Germany}

\address[ecu]{Excellence Cluster Universe, 85748 Garching b.\ M\"unchen, Germany}

\address[ccapp]{Center for Cosmology and Astro-Particle Physics and
  Department of Physics, The Ohio State University, Columbus, OH
  43210, United States of America}

\address[kipac2]{Kavli Institute for Particle Astrophysics and Cosmology,
SLAC National Accelerator Laboratory, Menlo Park, CA 94025-7015, United States of America}

%\address[help]{\bf Please send your current
%  affiliation(s), any name corrections and any required text for the
%  acknowledgements to browe@star.ucl.ac.uk}

% Authorship tier for people who did coding, per e-mails from a long
% time ago:
%   Joe Zuntz, Reiko Nakajima, Melanie Simet, Joerg Dietrich, Peter
%   Melchior, Bob Armstrong, Tomek Kacprzak, Hironao Miyatake,
%   Josh Meyers
% All except for Peter expressed interest in working on the paper.
% Peter does not seem to have replied to the e-mail, or if he did, his
% response did not go to me [Rachel].
%
% Authorship tier for people who did not do coding but discussed the
% project:
%   Michael Hirsch (did not respond to query about paper), Mandeep
%   Gill (confirmed), Lisa Voigt (did not respond to query about
%   paper), David Kirkby (declined), Paul Price (declined, but offered
%   to read, could be very helpful...).

\begin{abstract}
  \galsim\ is a collaborative, open-source project aimed at providing
  an image simulation tool of enduring benefit to the astronomical
  community.  It provides a software library for generating images of
  astronomical objects such as stars and galaxies in a variety of
  ways, efficiently handling image transformations and operations such
  as convolution and rendering at high precision.  We describe the
  \galsim\ software and its capabilities, including necessary
  theoretical background.  We demonstrate that the performance of
  \galsim\ meets the stringent requirements of high precision image
  analysis applications such as weak gravitational lensing, for
  current datasets and for the Stage IV dark energy surveys of the
  Large Synoptic Survey Telescope, ESA's \emph{Euclid} mission, and
  NASA's \emph{WFIRST-AFTA} mission.  The \galsim\ project repository
  is public and includes the full code history, all open and closed
  issues, installation instructions, documentation, and wiki pages
  (including a Frequently Asked Questions section).  The \galsim\
  repository can be found at
  \url{https://github.com/GalSim-developers/GalSim}.
\end{abstract}


\end{frontmatter}

%
% Section 1
% introduction
%

\section{Introduction}
Rapid advances in instrumentation and telescope technology are forcing
changes in the techniques used to analyse astronomical data.  As data
volumes increase, and statistical uncertainties decrease
correspondingly, systematic biases resulting from imperfect or
approximate inference must be reduced to ensure {\edit full return on}
investments made into increasingly large projects.

An area of research where technology and data volumes are placing
increasingly stringent requirements on data analysis methodology is
observational cosmology.  Recent years have seen a number of wide
area, long exposure imaging surveys of the extragalactic sky from both
ground-based telescopes (e.g.\
SDSS\footnote{\url{http://www.sdss.org/}},
CFHTLS\footnote{\url{http://www.cfht.hawaii.edu/Science/CFHTLS/}}\textsuperscript{,}\footnote{\url{http://www.cfhtlens.org/}}:
see \citealp[respectively]{sdssdr709,heymansetal12cfhtlens}) and space
(COSMOS\footnote{\url{http://cosmos.astro.caltech.edu/}}: see
\citealp{scovilleetal07}).  Data from these projects continue to be
exploited for their rich scientific content.

More ambitious ground-based projects are already underway,
including the Dark Energy
Survey\footnote{\url{http://www.darkenergysurvey.org/}} (DES:
\citealp[see e.g.][]{sanchezetal10}), Hyper
Suprime-Cam\footnote{\url{http://www.naoj.org/Projects/HSC/}} (HSC:
\citealp[see][]{miyazakietal12}) and the Kilo Degree
Survey\footnote{\url{http://kids.strw.leidenuniv.nl/}} (KiDS:
\citealp[see][]{dejongetal13}).  By most measures these upcoming
surveys of the deep extragalactic sky will bring an order of magnitude
more imaging data than their recent predecessors.  In the 2020s, the
ground-based Large Synoptic Survey Telescope
(LSST\footnote{\url{http://www.lsst.org/lsst/}}), and the ESA
\emph{Euclid}\footnote{\url{http://sci.esa.int/euclid/}, \url{http://www.euclid-ec.org}} and NASA
\emph{WFIRST-AFTA}\footnote{\url{http://wfirst.gsfc.nasa.gov/}} space
missions, will be taking extragalactic imaging data in vast
quantities.  These successive generations of projects are examples of
``Stage III'' and ``Stage IV'' dark energy surveys
\citep[][]{albrechtetal06}.

The increasing data volumes in these planned surveys are due to
increases in survey area and, to a somewhat lesser extent, depth, as
motivated by their science goals
\citep[][]{albrechtetal06,peacocketal06}.  Increasing area and depth
brings a greater number of galaxy objects, and thus a decrease in the
statistical uncertainties on final measurements.  However, this means
that the tolerable \emph{systematic} error in the inferred properties
of each galaxy needs to be reduced commensurately.  As surveys become
larger, the accurate estimation of galaxy properties from noisy images
becomes increasingly important.  {\edit In this context the
\galsim\ project was conceived, aiming to provide a
common image simulation tool for use across multiple surveys and to
aid comparison between measurement methods.}

Weak gravitational lensing \citep[for reviews see,
e.g.,][]{schneider06,bartelmann10,huterer10} is a prime example of a
scientific application that relies on reliable inference regarding the
properties of astrophysical objects from imperfect images.  Here the
\emph{shape} of the galaxy (typically some property related to its
ellipticity) is used to construct a noisy estimate of gravitational
\emph{shear}, which can be related to second derivatives of the
projected gravitational potential and is thus sensitive to all matter
{\edit (including dark matter)}.
Weak lensing can be used to constrain both cosmic expansion and the
growth of matter structure over time, and is thus valuable as a test
of dark energy and modified gravity models
\citep[e.g.][]{albrechtetal06,peacocketal06}.

To achieve this potential as a probe of cosmology, however, the weak
lensing community must control
additive and multiplicative systematic biases in shear estimates to
$\sim 2\times 10^{-4}$ and $2\times 10^{-3}$, respectively 
\citep[e.g.][]{hutereretal06,amararefregier08,masseyetal13,handbook}.
This accuracy must be achieved in the presence of detector
imperfections, telescope blurring and distortion, atmospheric effects
(for ground-based surveys), and noise
(galaxies used for lensing typically have a signal-to-noise ratio
of the flux as low as 10).
Weak lensing inference for cosmology presents a
significant technical challenge.

Simulations of astronomical imaging data will play an important part
in meeting this challenge.  For example, weak lensing estimators
typically invoke non-linear combinations of image pixels, and can be
shown to suffer from generic systematic biases when applied to noisy
galaxy images
\citep[e.g.][]{2000ApJ...537..555K,BJ02,hirataetal04,refregieretal12,melchiorviola12}
unless constructed extremely carefully
\citep{2000ApJ...537..555K,bernsteinarmstrong14}.  
{\edit Simulations of weak
lensing survey data in the GREAT08 (GRavitational lEnsing Accuracy
Testing 2008) challenge \citep{bridleetal10}
clearly showed the impact of noise in a blind comparison of
shear estimation methods. This helped reinvigorate activity aimed at better
characterizing these ``noise biases''
\citep{refregieretal12,melchiorviola12,kacprzaketal12} or eliminating
them \citep{bernsteinarmstrong14}.}

Another type of bias arises when true galaxy surface brightness
profiles do not match the models being fit, often referred to as model
bias or underfitting bias
\citep[e.g.][]{voigtbridle10,bernstein10,melchioretal10,kacprzaketal14}.  One of the
goals of GREAT3 \citep{handbook}, the third challenge in the GREAT
series, is to explore the impact of model bias across a range of shear
measurement methods.  For one of its ``experiments'', it takes real
galaxy images from the \emph{Hubble Space Telescope} (\hst) as the
underlying surface brightness profiles of the galaxies, and draws
sheared versions of these profiles using a method derived from that of
\citealp{2012MNRAS.420.1518M}: see \S\ref{sect:shera} in this article.
GREAT3 also {\edit involved} controlled tests of the impact of
multiepoch (i.e.\ multiple exposure) imaging and realistic
uncertainty about the point spread function (PSF) in survey data
\citep{handbook}. {\edit The initial imperative for developing \galsim\ was
specifically to enable the creation of the required images for
the GREAT3 challenge.}

But there are other issues in both space and ground-based data that
are still to be fully addressed for precision photometry and shape
estimation.  These include object confusion (or deblending),
PSF-object colour mismatching \citep{cyprianoetal10}, differential
atmospheric chromatic refraction \citep{meyersburchat14}, galaxy
colour gradients \citep{2012MNRAS.421.1385V,2013MNRAS.432.2385S}, and
non-linear detector effects
\citep[e.g.][]{rhodesetal10,seshadrietal13,antilogusetal14}.
Simulation investigations into the many different aspects of the shear
measurement problem will continue to improve our understanding of how
to meet these challenges for upcoming surveys.

Systematic biases are likely to be present, to some degree, in all
\emph{practical} estimators of shear, and thus simulations of weak
lensing observations will be \emph{required} for precision cosmology:
either to estimate (and thus calibrate) biases for a given survey, or
(ideally) to demonstrate that they have been controlled to tolerable
levels.  {\edit The same is also true of the many applications that rely on
highly accurate photometry 
and astrometry. These measurements often require steps such as
the classification and removal of
outliers, the application of calibration corrections,
and the fitting of models to data; simulated tests of these techniques 
are typically necessary.} Those who study galaxy properties by fitting
parametric models {\edit in order} to
learn something about galaxy evolution also
benefit from controlled simulations with which to test their analysis
methods.  Motivated by these requirements, the \galsim\ project has
drawn contributions from members of multiple Stage III and Stage IV
survey collaborations, aimed at producing an open-source,
community-vetted toolkit for building these indispensable image
simulations.

The structure of this paper is as follows.  In \S\ref{sect:software}
we provide an overview of the \galsim\ software toolkit, and describe
the motivation and structure of the different component parts of
\galsim.  In \S\ref{sect:gsstructure} we provide references to where
these component parts are described, in greater detail, in the rest of
the paper.

In \S\ref{sect:objects} we describe the types of astronomical objects
that \galsim\ can currently represent.  
In \S\ref{sect:lensingengine} we describe the \galsim\ ``lensing
engine'', which generates cosmologically motivated shear fields.
In \S\ref{sect:wcs} we
describe how transformations between coordinate systems are
represented, including between image pixel coordinates and ``world
coordinate systems'' (which may use either celestial coordinates or a
Euclidean tangent plane approximation).  In \S\ref{sect:rendering} we
describe how \galsim\ objects are rendered to form images, and
\S\ref{sect:noise} describes the noise models that can be invoked to
add noise to images.  In \S\ref{sect:hsm} we describe some of the
shear estimation routines that come as part of \galsim.

In \S\ref{sect:validation} we describe the numerical validation of
\galsim\ via series of tests and comparisons, and
\S\ref{sect:performance} discusses performance.  In
\S\ref{sect:notingalsim} we highlight some important effects in real
data that \galsim\ does \emph{not} currently include, and we end with
a summary and conclusions in \S\ref{sect:conclusions}.

%
% Section 2
% software
%

\section{Software overview}\label{sect:software}

The design and characteristics of \galsim\ arose through the need to
meet goals and requirements that were set down early in the life of
the project.  We describe these,
{\edit along with some aspects of the collaborative development process,}
and show how they led to the
current state of the \galsim\ software.
{\edit All of the code described in
this paper can be found in version 1.2 of \galsim.}

\subsection{Science requirements}\label{sect:scirequirements}
The basic capabilities of \galsim\ were driven by the need to meet the
following requirements for simulating astronomical images:
\begin{enumerate}
\item The ability to flexibly represent and render a broad range of
  models for astronomical objects in imaging data, including
  observationally-motivated PSF and galaxy
  profiles.\label{item:flexibility}
\item Handling of coordinate transformations such as shear, dilation
  and rotation for all objects being rendered, motivated by the need
  to simulate weak lensing effects at high precision.
\item Representation of the convolution of two or more objects, to 
  describe the convolution of galaxy surface brightness
  profiles with PSFs.
\item Accuracy in object transformation, convolution and rendering  at
  the level required for Stage IV surveys.
\item Flexibility in specifying image pixel noise models and data
  configurations (e.g. overlapping objects), so as to allow the
  importance of these data properties to be evaluated in controlled
  tests.
\item The ability to do all the above for galaxy or PSF models
  generated from an input image, e.g.\ from the \hst,
  so as to be able to compare results against those obtained
  from simpler parametric prescriptions for galaxies and PSF profiles.
\end{enumerate}
As can be seen, many of these requirements are driven by the need to
simulate weak lensing observations.  Precision simulation of
astrometry and photometry places similar
requirements.\footnote{{\edit For
  example, see \url{http://www.lsst.org/files/docs/SRD.pdf}.}}  Software which
meets the above requirements is therefore of great value for
simulations that require accurate photometry (e.g.\ for photometric
redshift estimation).

\subsection{Software design requirements}
In addition to the scientific goals for the \galsim\ project, an
emphasis was also placed on software design and implementation
considerations.  \galsim\ was conceived to be freely available under
an open-source (BSD-style) license, and written in non-proprietary programming
languages, making it available to all.  This aim is satisfied by the
choice of Python in combination with C++.

\galsim\ was also designed to be \emph{modular}, i.e.\ separated into
various functional components, and thus easily extensible.  Enforcing
modularity allows parts of the code to be written and maintained
independently by multiple individuals.  This design feature was
important in allowing \galsim\ to be developed collaboratively.
Within a modular program design there is scope for different modules
(with a common interface) to be used interchangeably, and this
property was crucial in providing the flexibility demanded by
\galsim's science goals (\S\ref{sect:scirequirements}: item
\ref{item:flexibility}).  Future extensions to the project also occur
naturally within this framework as additional modules.

To ensure that the learning curve for \galsim\ users is
as easy as possible, we require clear documentation on all new features
before they are deployed, and major features are demonstrated in
{\edit
heavily-commented example scripts that amount to a kind of tutorial for
how to use GalSim.  Writing these example scripts has been very useful
as a} design tool to help determine what user interface is most natural for 
realistic applications.

{\edit
Similarly, all new code also requires an accompanying test suite before
being merged into the main code base.  While we have not tried to impose
the precepts of test-driven development within the collaboration, we 
do take seriously the requirement that all code be tested as comprehensively
as possible.  In addition to checking the general validity of new code,}
the extensive unit tests have proved extremely valuable for dealing
with issues of cross-platform portability: we regularly run the tests
on a wide variety of systems with various idiosyncrasies.

{\edit
Finally, we require all new code to undergo
extensive code review by the rest of the development team, including
review of the code, the documentation, and the unit tests. 
This review process is facilitated by the ``Pull Request'' feature
of \textsc{GitHub}\footnote{\url{https://github.com}}, where the
code is hosted.
The other members of the collaboration team can easily see the set of 
changes being proposed and even comment on individual lines of code.}
While bugs are inevitable at some level
no matter how much care is taken to avoid them, these steps have been
quite effective at catching bugs and documentation errors before
code is deployed.

\subsection{The structure of \galsim\ }\label{sect:gsstructure}

Given the emphasis on modularity, and the desire to make the software
easily useable for as many scientific projects and applications as
possible, \galsim\ was conceived from the outset to be a toolkit of
library routines rather than as a ``monolithic'' package.  From the
user's perspective \galsim\ is fundamentally a Python \emph{class
  library}, providing a number of objects that can be employed for
astronomical image simulation in almost any combination specified by
the calling code.

{\edit
In addition, for users who may be more comfortable using configuration files
than writing Python code, we also provide a stand-alone executable that
reads relatively simple configuration files, which can carry out most
of the important \galsim\ functionality.  In particular, each
of the tutorial example scripts have a corresponding configuration file that
produces the same output files\footnote{
Actually, as of version \version, the chromatic functionality has yet
to be ported over to the configuration interface, so the example script for
that (demo12) does not yet have a corresponding configuration file.  We plan 
to add this functionality in the near future.}.  Information about the configuration
file interface can be found on the \galsim\ wiki\footnote{
\url{https://github.com/GalSim-developers/GalSim/wiki/Config-Documentation}}.
}

The classes and functions in the \galsim\ toolkit can be separated
into the following broad categories by functionality.  These are (with
references to relevant Sections of this article):
\begin{itemize}
\item Representing astronomical objects, including any
  transformations, distortions,
and convolutions (see \S\ref{sect:objects}).
\item Generating gravitational lensing distortions to be
  applied to astronomical objects (see \S\ref{sect:lensingengine}).
\item Characterizing the connection between image coordinates and
  world coordinates (see \S\ref{sect:wcs}).
\item Rendering the profiles into images (see \S\ref{sect:rendering}).
\item Generating random numbers, and using these to apply
  noise to images according to physically-motivated models (see
  \S\ref{sect:noise}).
\item Estimating the shapes of objects once these have been rendered
  into images (useful for testing, see \S\ref{sect:hsm}).
\end{itemize}
For information about the practical use of these tools, we refer the
reader to the online documentation available at the \galsim\ project
page\footnote{See \url{https://github.com/GalSim-developers/GalSim}.  The
  example scripts and Quick Reference Guide provide a good starting
  point for documentation, in addition to the Python docstrings for
  individual functions and class methods.}. {\edit Further information
  and a forum for questions regarding the structure or usage of the
  \galsim\ software can be found in the repository Issues pages, or
  via the \texttt{galsim} tag on the
StackOverflow web site\footnote{\url{http://stackoverflow.com/tags/galsim/info}}.}

{\edit While Python serves as the principal user interface to \galsim, many
of the numerical calculations are actually implemented in C++. 
However, this should be considered a mere implementation detail; 
the structure of \galsim\ is intended to prevent
 the user from needing to interface with the C++ layer directly.}

In this paper, we will
focus on a scientific, theoretical description of the output generated
by {\edit \galsim's} classes and functions, and the validation of those outputs to
ensure they meet our accuracy requirements.

%
% Section 3
% objects
%

{\edit
\section{Surface brightness profiles}\label{sect:objects}
}

In this section we describe the kinds of surface brightness
profiles that are available in \galsim.  We include their analytic
formulae where appropriate, and the physical and observational
motivations for the models.

\subsection{Galaxy models}\label{sect:galaxies}

\subsubsection{The exponential \disk\ profile}
First identified in M33 by \citet{patterson40}, and observed
systematically by \citet{devaucouleurs59} as a characteristic
component of the light profile in a sample of the brightest nearby
galaxies, the exponential \disk\ profile provides a good
description of the outer, star-forming regions of spiral galaxies
\citep[e.g.][]{lacknergunn12}.  

The surface brightness of an exponential
\disk\ profile varies as
\begin{align}
I(r) &= \frac{F}{2\pi r_0^2} {\rm e}^{-r/r_0} \\
& = \frac{F}{2.23057 r_{\rm e}^2}{\rm e}^{-1.67835 r/r_{\rm e}},
\end{align}
where $F$ is the total flux, $r_0$ is the scale radius, and
  $r_{\rm e} =
1.67835 r_0$ is the half-light radius, the radius that encloses half
of the total flux.

It is represented in \galsim\ by the \code{Exponential} class, and
the size can be specified using either $r_0$ or the half-light radius.

\subsubsection{The de Vaucouleurs profile}
This profile \citep[first used by][]{devaucouleurs48} is found to give
a good fit to the light profiles of both elliptical galaxies and the
central bulge regions of galaxies that are still actively star-forming
\citep[e.g.][]{lacknergunn12}.

The surface brightness of a de Vaucouleurs profile
varies as 
\begin{align}
I(r) &= \frac{F}{7! \cdot 8\pi r_0^2} {\rm e}^{-(r/r_0)^{1/4}} \\
&= \frac{F}{0.010584 r_{\rm e}^2} {\rm e}^{-7.66925(r/r_{\rm e})^{1/4}},
\end{align}
where $F$ is the total flux, $r_0$ is the scale radius, and $r_{\rm e}$ is 
the half-light radius.

De Vaucouleurs profiles are notorious for having both very cuspy
cores and also very broad wings.  The cusp occurs around the size
of the scale radius $r_0$, but because of the broad wings, the
half-light radius is several orders of magnitude larger\footnote{ 
More precisely, $r_{\rm e} \simeq 3459.485 r_0$}.  As such, the second
formula is more often used in practice.

It is represented in \galsim\ by the \code{DeVaucouleurs} class, and
the size can be specified using either $r_0$ or the half-light radius.

Because de Vaucouleurs profiles have such broad wings, it is sometimes
desirable to truncate the profile at some radius, rather than
allow it extend to infinity.  Thus, \galsim\ provides the option of 
specifying a truncation radius beyond which to have the surface brightness
drop to zero.

\subsubsection{The \sersic\ profile}\label{sect:sersic}
This profile, developed by \citet{sersic63}, is a generalization 
of both the exponential and de Vaucouleurs profiles.  
The surface brightness of a \sersic\ profile
profile varies as 
\begin{align}
I(r) &= \frac{F}{2n\pi \Gamma(2n) r_0^2} {\rm e}^{-(r/r_0)^{1/n}} \\
&= \frac{F}{a(n) r_{\rm e}^2} {\rm e}^{-b(n)(r/r_{\rm e})^{1/n}},
\end{align}
where $F$ is the total flux, $r_0$ is the scale radius, $r_{\rm e}$ is the
half-light radius, and $a(n)$ and $b(n)$ are known functions with
numerical solutions.  Note that the index $n$ need not be an integer.
{\edit Flux normalization determines $a(n)$.}
To calculate $b(n)$ in \galsim\ we start with the approximation given
by {\edit \citet{ciottibertin99}}, then perform a small number of iterations
of a numerical, non-linear root solver to increase accuracy.

As with the de Vaucouleurs profile, it is standard practice to use the
second formula using the half-light radius, since that size more
closely matches the apparent size of the galaxy, particularly for 
larger $n$ values ($n \gtrsim 2.5$).

The \sersic\ profile is represented in \galsim\ by the \code{Sersic}
class, and the size can be specified using either $r_0$ or the
half-light radius.  The $n$ parameter is allowed to range from $n =
0.3$, below which there are serious numerical problems, to $n=6.2$,
above which rendering inaccuracy may exceed the requirements set for
\galsim\ (see \S\ref{sect:validate-dft}).

As with \code{DeVaucouleurs}, it is also possible to specify a truncation
radius for the profile, rather than allow it to extend indefinitely.

\subsubsection{Galaxy profiles from direct observations}\label{sect:realgalaxy}
Observations made using the \hst\ provide high resolution images of
individual galaxies that can be used as direct models of light
profiles for simulations
\citep{2000ApJ...537..555K,2012MNRAS.420.1518M}.  These profiles
naturally include realistic morphological variation and irregular
galaxies, a contribution of increasing importance to the galaxy
population at high redshift.

The details of the implementation of such galaxy profiles are covered
in two later Sections: \S\ref{sect:interpolatedimage}, which describes
how objects are represented via two-dimensional lookup tables, and
\S\ref{sect:shera}, which discusses the specifics of how \hst\ galaxy
images and their PSFs are handled by \galsim.  However, it is worth
stating here that such a model, based on direct observations of a
large sample of individual galaxies, is available in \galsim.  It is
represented in \galsim\ by the \code{RealGalaxy} class.

\subsection{Stellar \& PSF models}\label{sect:psfs}

\subsubsection{The Airy \& obscured Airy profile}

The finite circular aperture of a telescope gives rise to a
diffraction pattern for the light passing through the aperture, an
effect first explained by \citet{airy}.  The resulting image has a
central peak surrounded by a series of rings.

The theoretical intensity distribution for a perfectly circular aperture is given by
\begin{align}
I(r) &= \left[ \frac{J_1(\nu)}{\nu}\right]^2, \\
\nu &\equiv \pi r D/\lambda \, ,
\end{align}
where $D$ is the diameter of the telescope, $\lambda$ is the
wavelength of the light \citep[see, e.g.,][]{bornwolf99}, {\edit and $J_1$ is
the Bessel function of the first kind of order one.}

In many telescopes, the light is additionally diffracted by a central
circular obscuration -- either a secondary mirror or a prime focus camera.  The
diffraction pattern in this case is also analytic:
\begin{equation}
I(r) = \left[ \frac{J_1(\nu) - \epsilon J_1(\epsilon \nu)}{\nu}\right]^2,
\end{equation}
where $\epsilon$ is the fraction of the pupil radius (as a linear
dimension, not by area) that is obscured.

It is represented in \galsim\ by the \code{Airy} class, and the size
is specified in terms of $\lambda/D$, which dimensionally gives a size
in radians, but which is typically converted to arcsec in practice.

\subsubsection{The Zernike aberration model}\label{sect:opticalpsf}

Complex, aberrated wavefronts incident on a circular pupil can often
be well approximated by a sum of Zernike polynomials
\citep{zernike34}.  These functions form an orthogonal set over a
circle of unit radius.  

In \galsim\ the convention of \citet{noll76} is adopted, which labels
these polynomials by an integer $j$.  It is also useful to define the
polynomials in terms of $n$ and $m$, which are functions of $j$
satisfying $|m| \le n$ with $n-|m|$ being even\footnote{ Specifically,
  the rule is that the Noll indices $j$ are in order of increasing
  $n$, and then increasing $|m|$, and negative $m$ values have odd $j$
  \citep{noll76}.}.  The integers $n$ and $m$ specify the Zernike
polynomials in polar coordinates as
\begin{align}
Z_{{\rm even}~j} &=  \sqrt{2(n + 1)} R^m_n(r) \cos{\left(m\theta \right)}\\
Z_{{\rm odd}~j} &= \sqrt{2(n + 1)} R^m_n(r) \sin{\left(m\theta\right)}
\end{align}
for integer $m \ne 0$ and integer $n$, and
\begin{equation}
Z_j = \sqrt{n + 1} R^0_n(r),
\end{equation}
for $m=0$, where
\begin{equation}
R^m_n(r)  =  
\sum_{s=0}^{(n {-} m) / 2} r^{(n {-} 2s)}  \frac{(-\mbox{\small 1})^s (n {-} s)!}{s!
\left[\frac{n{+}m}{2} {-} s \right]! \left[ \frac{n {-} m}{2} {-} s \right]!} .
\end{equation}

The low order
Zernike polynomials map to the low order aberrations commonly found in
telescopes, such as defocus ($j=4$), astigmatism ($j=5,6$), coma
($j=7, 8$), trefoil ($j=9, 10$) etc.  In practice, Zernike polynomials
have been found to provide a convenient, and compact, approximate
model for telescope aberrations.

Given such a model for the wavefront incident at the pupil, the PSF is
also determined {\edit at a given image plane location}. 
In the Fraunhofer diffraction limit, the PSF is
calculated as the Fourier transform of the autocorrelation of the
pupil wavefront.

The resulting profile is represented in \galsim\ by the
\code{OpticalPSF} class.  The size is specified in terms of
$\lambda/D$, just as for \code{Airy}.  There are also options for
specifying a circular central obscuration as well as rectangular
support struts.  In \galsim\ versions through 1.1, aberrations up to
Noll index $j=11$ (spherical aberration) can be included, but there
are plans to allow for Zernike terms to arbitrary order.

Because the profile is not analytic in either real or Fourier space,
the implementation of this class in \galsim\ involves creating an
image of the PSF in real space by Discrete Fourier Transform, and then
using the \code{InterpolatedImage} class to handle the interpolation
across this image.  Therefore, the techniques that we will describe in
\S\ref{sect:rendering} for interpolated images also apply to
\code{OpticalPSF}.

\subsubsection{The Moffat profile}

\citet{moffat69} investigated the hypothesis that stellar PSFs could
be modelled as a Gaussian seeing kernel convolved by an obstructed Airy
diffraction pattern.  He discovered that the Gaussian was not a viable
model for the seeing component.  Instead, he developed an analytic
model for stellar PSFs with broader wings that provides a better fit
to observations, a model which now bears his name.

The surface brightness of a Moffat profile varies as 
\begin{equation}
I(r) = \frac{F (\beta-1)}{\pi r_0^2} \left[1+\left( r/r_0 \right)^2 \right]^{-\beta}
\end{equation}
where $F$ is the total flux, $r_0$ is the scale radius, and $\beta$
typically ranges from 2 to 5.  In the limit $\beta \rightarrow \infty$, the
Moffat profile tends towards a Gaussian.

It is represented in \galsim\ by the \code{Moffat} class, and the size
can be specified using $r_0$, the full-width at half-maximum (FWHM), or
the half-light radius.  It is also possible to specify a truncation
radius for the Moffat profile, rather than allow it to extend
indefinitely.

\subsubsection{The Kolmogorov profile}

In a long ground-based exposure, the PSF is predicted to follow a particular
functional form due to the Kolmogorov spectrum of turbulence in
the atmosphere.
\citet{racine96} showed that this prediction
was indeed a better fit to the observations of stars in long exposures than
a Moffat profile.

The surface brightness of a Kolmogorov profile is defined in Fourier space as
\begin{equation}
\fourier{I}(k) = F {\rm e}^{-(k/k_0)^{5/3}},
\end{equation}
where $F$ is the total flux, $k_0 = A \lambda/r_0$, $A \simeq 2.99$
is a constant given by \citet{Fried66}, and {\edit here $r_0$ is the Fried
parameter (to be distinguished from the use of $r_0$ to denote the scale radius
in other profiles)}. 
Typical values for the Fried parameter are on the order of
10~cm for most observatories and up to 20~cm for excellent sites. The
values are usually quoted at $\lambda$ = 500~nm, and {\edit the Fried
  parameter}
$r_0$ depends on wavelength as $r_0 \propto \lambda^{-6/5}$.

This profile is represented in \galsim\ by the \code{Kolmogorov} class, and
the size can be specified using $\lambda/r_0$, the FWHM, or the half-light radius.

\subsection{Generic models}

\subsubsection{The Gaussian profile}
The Gaussian profile has convenient properties: it is relatively
compact and analytic in both real space and Fourier space, and
convolutions of Gaussian profiles can themselves be written as a new
Gaussian profile.  Observational properties of Gaussian profiles, such
as their second moments, are also typically analytic.

For this reason the Gaussian profile is often used to represent
galaxies and PSFs in simple
simulations where speed is valued above realism. 
And thanks to its analytic properties, it has great value for
testing purposes.  {\edit Furthermore, while a single Gaussian profile
is generally a poor approximation to both real PSFs and real
galaxies, linear superpositions of Gaussian profiles
can be used to approximate realistic profiles with much
greater accuracy \citep[e.g.][]{hogglang13,sheldon14}.}

The surface brightness of a Gaussian profile varies as 
\begin{equation}
I(r) = \frac{F}{2\pi \sigma^2} {\rm e}^{-r^2/2\sigma^2},
\end{equation}
where $F$ is the total flux and $\sigma$ is the usual Gaussian 
scale parameter.

It is represented in \galsim\ by the \code{Gaussian} class, and
the size can be specified using $\sigma$, the FWHM, or the half-light radius.

\subsubsection{Shapelet profiles}\label{sect:shapelets}

Shapelets were developed independently by \citet{BJ02} and
\citet{Refregier03} as an effective means of characterizing compact
surface brightness profiles such as galaxies or stars.  The shapelets
basis set consists of two-dimensional Gaussian profiles multiplied by
polynomials, and they have a number of useful properties.  For
example, they constitute a complete basis set, so in theory any image
can be decomposed into a shapelet vector; however, profiles that are
not well matched to the size of the Gaussian will require very high
order shapelet terms, such that the decomposition is unfeasible in
practice \citep[e.g.][]{melchioretal10}.

For images that are relatively well approximated by a Gaussian, they
provide a compact representation of the object, since most of the
information in the image is described by low order corrections to the
Gaussian, which is what the shapelet decomposition provides.

The \code{Shapelet} class in \galsim\ follows the notation of
\citet{BJ02}, although it is also similar to what has been called
``polar shapelets'' by \citet{MasseyRefregier05}.  The shapelet
expansion is indexed by two numbers, $p$ and $q$\footnote{ The
  shapelet functions happen to be eigenfunctions of the 2D quantum
  harmonic oscillator.  In that framework, $p$ and $q$ count the
  number of quanta with positive and negative angular momentum,
  respectively.  $N=p+q$ is the total energy and $m=p-q$ is the net
  angular momentum.}:
\begin{align}
I(r,\theta) &= \sum_{p,q\ge0} b_{pq}\psi_{pq}^\sigma(r,\theta) \\
\psi_{pq}^\sigma(r,\theta) &= 
    \frac{ (-1)^q}{\sqrt\pi \sigma^2}
    \sqrt{ \frac{q!}{p!} }
    \left( \frac{r}{\sigma} \right)^m e^{ {\rm i}m\theta}
    e^{-r^2/2\sigma^2} \nonumber \\*
& \qquad \times
    L_q^{(m)}(r^2/\sigma^2) \qquad (p\ge q) 
\label{eq:realshapelet} \\
\psi_{qp}^\sigma(r,\theta) &= \conj{\psi_{pq}^\sigma}(r,\theta) \\
m &\equiv p-q.
\end{align}
$L_q^{(m)}(x)$ are the Laguerre polynomials, which 
satisfy the recurrence relation:
\begin{align}
L_0^{(m)}(x) &= 1 \\
L_1^{(m)}(x) &= (m+1)-x \\
(q+1) L_{q+1}^{(m)}(x) &= [(2q+m+1)-x]L_{q}^{(m)}(x) \nonumber \\*
&\qquad    - (q+m)L_{q-1}^{(m)}(x).
\end{align}
The functions may also be indexed by $N=p+q$ and $m=p-q$, which is
sometimes more convenient.  Both indexing conventions are implemented
in \galsim.

One of the handy features of shapelets is that their Fourier transforms are also shapelets:
\begin{align}
\fourier\psi_{pq}^\sigma(k,\phi)  &=  
    \frac{ (-i)^m}{\sqrt\pi }
    \sqrt{ \frac{q!}{p!} }
    \left( k\sigma \right)^m e^{{\rm i}m\phi}
    e^{-k^2\sigma^2/2} \nonumber \\
& \qquad \times
    L_q^{(m)}(k^2\sigma^2) \qquad (p\ge q).
\label{eq:fouriershapelet}
\end{align}
This means convolving shapelets in Fourier space is very efficient.

\subsubsection{Interpolated images}\label{sect:interpolatedimage}

In some cases, it is useful to be able to take a given image and treat
it as a surface brightness profile.  This requires defining how to
interpolate between the pixel values at which the surface brightness
is sampled.

One application for this is the use of direct observations of
individual galaxies (e.g., from \hst) as the models for further
simulations, already mentioned in \S\ref{sect:realgalaxy}.  See
\S\ref{sect:shera} for more about this possibility.

Another application was also mentioned above in
\S\ref{sect:opticalpsf}.  The general aberrated optical PSF is too
complicated to model analytically, so the \code{OpticalPSF} class is
internally evaluated by interpolation over a finite grid of
samples of the surface brightness profile.

The \code{InterpolatedImage} class converts an arbitrary $n\times n$
image array with elements $I_{ij}$ on pixels of size $s$ into a
continuous surface brightness distribution
\begin{equation}
\label{eq:interpolatedimage}
I(x,y)  = \sum_{i,j} I_{ij} \kappa(x/s{-}i, y/s{-}j) , 
\end{equation}
where $\kappa$ is a real-space kernel chosen from those listed in
Table~\ref{table:interpolants}.  Each interpolant uses $K\times K$
input pixel values around the given $(x,y)$ to render a sample at that
location.  Rendering an $M\times M$ output image hence requires $K^2
M^2$ operations, and the footprint of the \code{InterpolatedImage} is
extended by $K s/2$ beyond the original input image.

The delta-function kernel yields infinite (or zero) values in direct
rendering and is hence a highly ill-advised choice for an
\code{InterpolatedImage} that is to be rendered without further
convolutions. The sinc kernel has infinite extent and hence uses all
$N\times N$ input samples to reconstruct each output sample, leading
to $N^2 M^2$ operations in a direct-space rendering.  In a high-volume
simulation this will also be ill-advised.  Approximations to the sinc
kernel that provide finite support are often found to give a good
compromise.  Examples include the kernel that represents cubic and
quintic (i.e.\ 3rd and 5th order) polynomial interpolation, and the
Lanczos kernel (see Table~\ref{table:interpolants}, also
\citealp{BernsteinGruen}).  As will be demonstrated in
  \S\ref{sect:validate-ii} and \S\ref{sect:validate-reconv}, these
  higher-order interpolation kernels meet stringent performance
  requirements for the representation of galaxy images.

\begin{table*}
\begin{center}
\caption{Properties of interpolants}\label{table:interpolants}
\begin{tabular}{lclc}
\hline
%  \tablewidth{0pt} \tabletypesize{\small}
Name & Formula & Notes$^{a}$ & No.\ of points $K$ \\
\hline
%  } \startdata
\code{Delta} & $\delta(x)$ & Cannot be rendered
  in $x$ domain;
  extreme aliasing $\propto 1$. & 0,1  \\
  \code{Nearest} &  $1, \quad |x|<1/2$ & Aliasing $\propto k^{-1}$  & 1 \\
  \code{Linear} & $1-|x|, \quad |x|<1$ & Aliasing $\propto k^{-2}$ & 2 \\
  \code{Cubic} & piecewise cubic$^{\rm b}$ & Common choice
  for
  $K$ & 4\\
  \code{Quintic} & piecewise quintic$^{\rm b}$ & Optimized
  choice for $K_k$ (cf.\ \S\ref{sect:fourierinterpolatedimage})  & 6 \\
  \code{Lanczos} & $\sinc\, x \; \sinc(x/n), \quad |x|<n$ &
  Common choice for
  $K$  with $n=3$--5 & 2n \\
  \code{SincInterpolant} & $\sinc\, x$ & Perfect
  band-limited interpolation;
  slowest  & $\infty$\\
\hline
\end{tabular}
{\footnotesize $^{a}${Coordinate distance
    from the origin in Fourier space is denoted $k$.}}\\
{\footnotesize $^{b}${Formulae for \code{Cubic} and \code{Quintic} interpolants
are given in the Appendix to \citet{BernsteinGruen}.}}
\end{center}
\end{table*}

\subsection{Transformations}
\label{sect:transformations}
Many of the profiles we have described so far are circularly symmetric
and centred on the coordinate origin.  \galsim\ can, however, use
these profiles to represent (for example) elliptical, off-centred
surface brightness profiles.  Various transformations can be applied
to a \galsim\ object representing a surface brightness distribution
$I(\vecx)$, to return a new object representing $I'(\vecx)$.

The overall amplitude of the profile can be rescaled by some factor
simply by using the \code{*} operator.
It is also possible to get a rescaled profile
with a specified value for the new total flux.
%MJ: Peter was confused by this footnote. 
%    And it's a pretty unusual edge case, so I think better to just omit this.
%\footnote{
%If the input flux $f_I=0$, it is impossible to rescale to a new flux value,
%although the whole profile may still be scaled by an arbitrary factor.}.
Either way, the new profile is $I'(\vecx) = c I(\vecx)$
for some constant $c$.

Any one-to-one transformation $T$ of the plane defines an image
transformation via
\begin{equation}
I'(\vecx)  = I\left[ T^{-1}(\vecx)\right].
\end{equation}
The \galsim\ operation \code{transform} implements local linear
transformations defined by a transformation matrix $\matA$ as
\begin{equation}
T(\vecx) = \matA \vecx.
\end{equation}
This can account for any arbitrary distortion, rotation, parity flip 
or expansion of the coordinate system.  The user can specify $\matA$
directly; alternatively the \galsim\ operations \code{rotate,
  expand, magnify, shear,} and \code{lens} implement restricted
versions of the transformation that are often convenient in practice.

There is also a command called \code{dilate} that combines an
expansion of the linear size (i.e. $\matA$ being a multiple of the
identity matrix) with an amplitude scale factor to keep the overall
flux of the resulting profile unchanged.  This is merely a convenience
function, but it is handy, since this operation is fairly common.

Finally, the \code{shift} function can translate the entire profile by
some amount $\vecx_0$.  This corresponds to the transformation
\begin{equation}
T(\vecx) = \vecx + \vecx_0.
\end{equation}
Together, \code{shift} and \code{transform} enable any arbitrary affine
transformation.

All of these operations are implemented in \galsim\ via a wrapper
object that exists independently of, and interfaces with, the original
profile.  This means the code for implementing these transformations
(e.g.\ the effect in Fourier space, or the deflections to apply for
photon shooting, see \S\ref{sect:photon}) exists in a single place, which helps minimize
unnecessary duplication and ensure the reliability of the
transformation routines.

For reasons discussed in \S\ref{sect:rendering}, \galsim\ does not
currently handle non-affine transformations acting on a single
profile (as required by simulations of flexion such as
\citealp{velanderetal11,roweetal13}), but see \S\ref{sect:wcs} for how
to handle such coordinate transformation across a full image,
using the appropriate locally linear approximation at the location of
each object.

\subsection{Compositions}\label{sect:compositions}

Two or more individual surface brightness profiles may be added
together using the \code{+} operator to obtain a profile with
$I'(\vecx) = I_1(\vecx)+I_2(\vecx)+\ldots$.

The convolution of two or more objects $I_1, I_2$ is defined as usual via
\begin{equation}
I'(\vecx) = (I_1 \conv I_2)(\vecx) = \int
I_1(\vecx^\prime) I_2(\vecx - \vecx^\prime) \,  \rmd^2 \vecx^\prime. 
\label{eq:convintegral}
\end{equation}
This is implemented in \galsim\ with the \code{Convolve} function,
which takes a list of two or more objects to be convolved together.

There are also two special functions that can afford some modest
efficiency gains if they apply.  \code{AutoConvolve} performs a
convolution of an object with itself, i.e.\ with $I_2(\vecx) = I_1(\vecx)$ in
\eqn{eq:convintegral}.  \code{AutoCorrelate} performs a convolution of
an object with a 180\degree\ rotation of itself, i.e.\ with $I_2(\vecx) =
I_1(-\vecx)$.

Finally, \galsim\ can implement a deconvolution as well.  This lets
users solve for the solution $I'$ to the equation $I_1 = I' \conv
I_2$, which in Fourier space is simply
\begin{equation}
\fourier{I}'(\veck) = \fourier{I_1}(\veck) / \fourier{I_2}(\veck)
\end{equation}
The function \code{Deconvolve} implements the inverse of a profile
(e.g. $I_2$ above) in Fourier space, which is a deconvolution in real
space.  The returned object is not something that can be rendered
directly.  It must be convolved by some other profile to produce
something {\edit renderable} ($I_1$ in this example).

One common use case for this is to deconvolve an
observed image (represented by an \code{InterpolatedImage} object) by
its original PSF and pixel response,
producing an object that would then be reconvolved
by some other PSF corresponding to a different telescope and rendered
on an image with a new pixel scale.  This is the
functionality at the heart of \galsim's implementation of the
reconvolution algorithm described in \S\ref{sect:shera} and used by
the \code{RealGalaxy} class (\S\ref{sect:realgalaxy}).  As will be
discussed in \S\ref{sect:shera}, the latter PSF must be band-limited at
a lower spatial frequency than the original (deconvolved) PSF to avoid
serious artifacts in the final image.

\subsection{Chromatic objects}
\label{sect:chromatic}

Real astronomical stars and galaxies have wavelength-dependent
intensity distributions.  To model this, it is possible in \galsim\ to
define objects whose surface brightness is a function not only of
position, but also of wavelength.  These objects can then be rendered
as seen through a particular bandpass (the throughput as a function of
wavelength).

The simplest of the wavelength-dependent classes is the
\code{Chromatic} class, which is constructed from an achromatic
profile (i.e. any of the profiles described above) and a Spectral
Energy Distribution (SED).  The SED defines a wavelength-dependent
flux for the object.

Chromatic objects can be transformed, added, and convolved using
precisely the same syntax as for achromatic objects. 
The addition of multiple chromatic objects provides a simple way to
creating galaxies with non-trivial wavelength-dependent morphologies.
A bulge and a \disk\ can have different SEDs; star forming regions
or HII regions can also be added in with their own SEDs.

In addition, the various
%\code{shift}, \code{expand}, \code{dilate} and flux rescaling
% MJ: Why do we limit to just these?  We should allow any transformation
% to be wavelength-dependent.  I'll implement this.
transformations can take functions of wavelength for their arguments.
For example, differential chromatic refraction is implemented by a
wavelength-dependent shift, and the effects of chromatic variation in Kolmogorov
turbulence can be approximately modelled with a wavelength-dependent
dilation.  \galsim\ provides the helper function
\code{ChromaticAtmosphere}, which encapsulates both of these effects
for an atmospheric PSF.

A more detailed discussion of this recent addition to the \galsim\
toolkit is reserved for future work, but it provides a valuable
resource for simulating a number of important wavelength-dependent
effects in cosmology
(\citealp[e.g.][]{cyprianoetal10,2012MNRAS.421.1385V,plazasbernstein12,2013MNRAS.432.2385S,meyersburchat14}).

%
% Section 4
% lensingengine
%

{\edit
\section{Lensing shear and magnification}\label{sect:lensingengine}
}

A primary purpose of \galsim\ is to make simulated images that can be
used to test weak gravitational lensing data analysis algorithms,
which means that a framework for simulating lensing shear and
convergence \citep[see, e.g.,][for a review]{schneider06} is critical.  In
  the weak gravitational lensing limit, the transformation between
  unlensed coordinates ($x_u$, $y_u$; with the origin taken to be at
  the center of a distant galaxy light source) and the lensed
  coordinates in which we observe galaxies ($x_l$, $y_l$; with the
  origin at the center of the observed image) is linear: 
\begin{equation}\label{eq:lensingshear}
  \left( \begin{array}{c} x_u \\ y_u \end{array} \right)
  = \left( \begin{array}{cc} 1 - \gamma_1 - \kappa & - \gamma_2 \\
-\gamma_2 & 1+\gamma_1 - \kappa \end{array} \right)
  \left( \begin{array}{c} x_l \\ y_l \end{array} \right).
\end{equation}
Here we have introduced the two components of lensing shear $\gamma_1$
and $\gamma_2$, and the lensing convergence $\kappa$.  The shear
describes the anisotropic stretching of galaxy images in weak lensing.
The convergence $\kappa$ describes an isotropic change in apparent
object size: areas of the sky for which $\kappa$ is {\edit non-zero} have
apparent changes in area (at fixed surface brightness).  Lensing
magnification is produced from a combination of the shear and
convergence.

It is worth noting that even when $\kappa =0$ the
  transformation matrix in equation \eqref{eq:lensingshear} will not
  have a unit determinant in general.  Object area (and thus flux when
  conserving surface brightness) is therefore not
  conserved by weak lensing shear, an effect which is not always
  desired when simulating images. For convenience, \galsim\ implements
  a unit determinant
  shear transformation that conserves object area, while also
  implementing the non-area conserving shear defined by the weak
  lensing transformation of equation \eqref{eq:lensingshear}.

In the simplest use case, \galsim\ will take single values for the
lensing shear and convergence and apply them to objects.  In addition,
however, \galsim\ is able to generate fields of coherent shear and
convergence values corresponding to two physical scenarios described
below: cosmological weak lensing fields with some power spectrum, and
the weak lensing that arises from a spherical NFW halo
\citep*{1996ApJ...462..563N}.

\subsection{Cosmological lensing fields}

{\edit \galsim\ provides routines to simulate a Gaussian random
field approximation
to a cosmological lensing signal, characterized wholly by its power
spectrum.  In reality, shear and convergence fields show significant
non-Gaussianity.
The intention, therefore, is not
highly cosmologically accurate shear and convergence fields, such as
would be suitable
for an end-to-end test of cosmological parameter determination.  The
intention is rather to provide basic functionality for making semi-realistic,
spatially varying lensing fields so as to be able to test measurement
algorithms that find such regimes challenging in general. 

There is
nothing to prevent any user from using outputs from a more realistic
cosmological ray-tracing simulation \citep[e.g.][]{becker13} to create an observed
galaxy shape catalog that includes a realistic shear and convergence field.
If realism is required, this procedure should be followed.}

\subsubsection{Basic capabilities}
Any shear field can be projected into orthogonal $E$- and $B$-mode
components, named for their similarity to the electromagnetic vector
fields with zero curl and zero divergence, respectively \citep[see,
e.g.][]{schneider06,handbook}.  \galsim\ is capable of taking input
$E$- and $B$-mode shear power spectra as user supplied functions, and
generating random realizations of shear and convergence fields 
 drawn from those power spectra.  The basic functionality is
carried out by the \code{PowerSpectrum} class. 
Shears are generated on a grid of positions, but \galsim\ can 
interpolate to arbitrary positions within the bounds of the grid using
the interpolants discussed in \S\ref{sect:dft}.

Observable quantities such as the shear correlation functions are
defined using the vectors connecting pairs of galaxies at separation
$\theta$,
\begin{align}
  \xi_{\pm}(\theta) &=  \xi_{++}(\theta)\pm \xi_{\times \times}(\theta) \\
  &= \int_0^{\infty} \frac{1} {2 \pi} [P_E(k)\pm P_B(k)]
  J_{0/4}(k\theta) \, k \, \rmd k \label{eq:xi}
\end{align}
where $J_{0/4}$ denotes the 0th and 4th Bessel function of the first
kind, as is appropriate for $\xi_+$ and $\xi_-$, respectively.  See
\citet{Schneider02} for more details about relating the correlation
functions to the $E$- and $B$-modes of the flat-sky power
spectra\footnote{Many standard references regarding lensing power
  spectra work in terms of the spherical harmonics $\ell$, with the
  power spectrum denoted $C_\ell$.  In the flat-sky limit we can
  simply swap $\ell$ with wavenumber $k$ and $C_\ell$ with $P(k)$.}.

For cosmological shear correlations {\edit $P_B\simeq 0$ to a very good
approximation, although higher order effects (e.g.\ source redshift clustering)
can introduce physical B-modes \citep[see, e.g.,][]{schneider06}.
It is useful} to include $P_B$ when generating shears for other purposes,
such as when drawing atmospheric PSF anisotropies according to some
power spectrum (which typically requires similar amounts of $E$ and
$B$ power).  Here $P_E(k)$
and $P_B(k)$ have dimensions of angle$^2$; \galsim\ can accept the
power spectra in various formats and sets of units.

These definitions of observables rely on continuous Fourier
transforms, but in practice we implement the calculations using
discrete Fourier Transforms (DFT), and we must be careful about the
DFT conventions.  We assume we have a grid with length $L$ along one
dimension and spacing $d$ between grid points.  Given a Fourier-space
grid of size consistent with that of the input real-space grid, the
minimum and maximum one-dimensional wavenumbers 
$|k|$  on the grid are $k_\text{min}=2\pi/L$ and
$k_\text{max}=\pi/d$.  The lensing
engine finds the power $P(k)$ according to the value of
$k=\sqrt{k_1^2+k_2^2}$ on the grid, {\edit draws random amplitudes
  from a Rayleigh distribution based on $\sqrt{P(k)}$ and
  random phases from a uniform distribution}, and transforms back to our real-space
  grid to get the real-space shear field, with periodic boundary
conditions.  The standard definition of the correlation function
(Eq.~\ref{eq:xi}) uses the angular frequency, non-unitary definition
of the 1D Fourier transform:
\begin{align}
\fourier{f}(k) & =  \int_{-\infty}^{\infty} f(x) \me^{-\mi k x} \,
\rmd x ~ \equiv ~ \mathcal{F} \{ f(x) \} \label{eq:fwdft} ; \\
f(x) & =  \frac{1}{2 \pi} \int_{-\infty}^{\infty} \fourier{f}(k)
\me^{\mi k x} \,
\rmd k ~ \equiv ~ \mathcal{F}^{-1}\{ \fourier{f}(k) \} \label{eq:invft}.
\end{align}
If we trace through the impact of this convention on the discrete,
finitely-sampled power spectrum, then our use of the
\numpy\footnote{\url{http://www.numpy.org/}} package for Fourier transforms
(with the more common unitary definition of the DFT) necessitates
various normalization factors related to the input grid configuration.
For further details, see \app{sect:appendix}.
%\textbf{May be able to reference appendix of this document, depending
%on what we include there.}

\subsubsection{Implementation details}

Any DFT-based algorithm to generate shears according to a power
spectrum is subject to limitations due to the implicit
choice of a finite range in wavenumber and the
difference between a discrete and continuous Fourier transform.
\galsim\ includes options to ameliorate these limitations:

\begin{itemize}
\item \galsim\ can optionally decrease
  (increase) the minimum (maximum) value of $k$ by internally
  expanding {\edit (contracting)} 
  the real-space grid before generating shears.  This helps
  avoid problems with missing shear power on various scales; in the
  case of cosmological power spectra it is particularly recommended
  for those who wish to properly reproduce the large-scale shear
  correlation function (on scales comparable to $\sim 1/4$ the total
  grid extent).  There is an important tradeoff implicit in the use of
  these options: the power spectra that result from using
  internally enlarged grids is not strictly the same as the input
  one, and there can be $E$ and $B$ mode leakage as a result of using
  these options.
  %Thus, users who care about reproduction of the shear
  %correlation function may wish to use these options; those who care
  %about reproduction of the shear power spectrum should not.
\item \galsim\ has a utility to predict the shear correlation function
  resulting from the use of a limited $k$ range when generating
  shears.  While the output is not exactly what will be generated in
  reality since the algorithm does not account for the use of a DFT,
  it permits users to assess in advance whether their grid choices
  permit them to roughly reproduce shear correlations on the scales
  they want.
\item A natural consequence of the limited $k$ range is that aliasing
  can occur if users supply a power spectrum with power below the
  grid Nyquist scale.  \galsim\ shear generation routines can
  optionally band-limit the
  input power spectrum by applying a hard or soft cutoff at the
  Nyquist scale to avoid aliasing.
\end{itemize}

\subsubsection{Interpolation to non-gridded positions}

After building a grid of shears and convergences, users can
interpolate them to arbitrary positions within the grid.  However,
when using this functionality, some care must be employed to prevent
the interpolation procedure from spuriously modifying the shear power
spectrum or correlation function at scales of interest.

The key effect of interpolation is to multiply the shear power
spectrum by a quantity proportional to the square of the Fourier
transform of the interpolant.  As a result, we found that all
interpolants modify the shear 2-point functions in a significant way
($>10$\% but sometimes much more) for scales below $3$ times the
original grid spacing.   Thus, when interpolating shears to random
points, the grid spacing should be chosen with care, keeping in mind
the minimum scale above which the shear two-point functions should be
preserved.  Edge effects can also be important
depending on the intended use for the resulting interpolated shears and
the choice of interpolant (see Table~\ref{table:interpolants};
\S\ref{sect:interpolatedimage}).

\subsection{Lensing by individual dark matter halos}

The \code{NFWHalo} class can produce shears and convergences
corresponding to a dark matter halo following a spherical NFW
\citep{1996ApJ...462..563N} profile, using analytic formulae from
\citet{bartelmann96} and
\cite{2000ApJ...534...34W}.  The formulae depend on the 
cosmology being assumed, which in \galsim\ is allowed to be
an arbitrary $\Lambda$CDM cosmology.  It would not be difficult to 
extend this to more sophisticated cosmological models.

The NFW lensing halo is defined in terms of its mass, concentration,
redshift, and position in the image.  Then for any given position
  and redshift of the source galaxy to be lensed, the appropriate
  lensing quantities can be computed analytically.  It is, however,
  important to note that the implementation in \galsim\ adopts the
  coordinate frame of the observed galaxies, i.e.\ no deflection
  angles are calculated or applied to galaxy positions. As a
  consequence the density of galaxies behind NFW halos is not altered
  as would happen in nature. If needed, e.g.\ for studies of
  magnification from galaxy clusters, this effect should be taken into
  account. 
  %Adding this effect is straightforward, and
  %work is in progress to allow users to specify \emph{source} plane
  %galaxy positions in \galsim.

%
% Section 5
% wcs
%

\section{World coordinate systems}\label{sect:wcs}

The galaxy models described above are generally defined in terms of
angular units on the sky.  For example,
a \sersic\ profile might be constructed to have a half-light radius of 1.6
arcsec.  Before rendering this object onto an image, one must 
specify how the image pixel coordinates relate to sky coordinates
(also known as world coordinates).

The simplest such relationship is to assign a pixel scale to the
image, typically in units of arcsec/pixel.  However, this is not the
only possible such relationship; in general, the ``world coordinates''
on the sky can be rather complicated functions of the image
coordinates.  \galsim\ allows for a variety of ``world coordinate
systems'' (WCS) ranging from a simple pixel scale to the kinds of WCS
functions typically found in the FITS (Flexible Image Transport
System, \citealp[see e.g.][]{penceetal10}) headers of real data
images.

\subsection{{\edit Types} of WCS functions}

The WCS classes used by \galsim\ can be broken into two basic types:
celestial and euclidean coordinate systems.

Celestial coordinate systems are defined in terms of right ascension
(RA) and declination (Dec).  In \galsim\ this kind of WCS is
represented by subclasses of \code{CelestialWCS}.  This includes
\code{RaDecFunction}, which can represent any arbitrary function the
user supplies for RA$(x,y)$ and Dec$(x,y)$, and \code{FitsWCS}, which
reads in the WCS information from a FITS header.

Euclidean coordinate systems are defined relative to a tangent plane
projection of the sky.  Taking the sky coordinates to be on an
actual sphere with a particular radius, the tangent plane would be
tangent to that sphere.  We use the labels $(u,v)$ for the coordinates
in this system, where $+v$ points north and $+u$ points
west\footnote{This can be counterintuitive to some, but it matches
  the view from Earth.  When looking up into the sky, if north is up,
  then west is to the right.}.
  
In \galsim\ this kind of WCS is represented by subclasses of
\code{EuclideanWCS}.  The most general is \code{UVFunction}, which can
represent any arbitrary function the user supplies for $u(x,y)$ and
$v(x,y)$.  \code{AffineTransform} is the most general \emph{uniform}
coordinate transformation where all pixels have the same size and
shape, but that shape is an arbitrary parallelogram.  

Other classes representing restricted specializations of
\code{AffineTransform} are available.  The simplest one,
\code{PixelScale}, describes uniform square pixels.  To date, this is
what has been used in simulations testing shear measurement methods,
including the most recent GREAT3 challenge \citep{handbook}, as well
as all previous shear testing projects
\citep{heymansetal06,masseyetal07,bridleetal10,kitchingetal12,kitchingetal13}.
\galsim\ therefore makes it easy to ignore all of the WCS options and
just use a pixel scale instead.  Any function that can take a WCS
parameter, can also take a pixel scale parameter, which is interpreted
as a \code{PixelScale} WCS.

\subsection{Converting profiles}\label{sect:convertingprofiles}

It is relatively straightforward to convert a surface brightness
profile from the sky coordinates in which it is defined to 
image coordinates.  We only need to assume
that the first derivatives of the WCS functions are approximately
constant over the size of the object being modelled, which is generally
a safe assumption.  That is, we use the local linear approximation of
the WCS transformation at the location of the object.  Currently, \galsim\ cannot
accurately handle transformations that vary significantly over the
size of the profile being rendered; in fact this operation would 
be very similar to what is required for implementing flexion, and
would most simply be incorporated in the context of photon shooting
{\edit (see \S\ref{sect:photon})}.

The first step if we are dealing with a celestial coordinate system is
to do a local tangent projection at the position of the object.
Specifically, we use the TAN projection described by
\citet{CalabrettaGreisen02}, but it is likely that any of the tangent
plane projections discussed in that work would provide an equivalent
conversion to a local Euclidean coordinate system.

The next step is to calculate the Jacobian of the WCS at the location
of the object:
\begin{equation}
  \mathbf{J} = \left( \begin{array}{cc}  \rmd u/ \rmd x & \rmd u/ \rmd
      y \\  \rmd v/ \rmd x & \rmd v/ \rmd y \end{array} \right)
\end{equation}
This Jacobian defines the local affine approximation of the WCS at the
location of the object, which we assume we can take as uniform over
the extent of the profile we need to convert.

Applying a general Jacobian transformation in \galsim\ was described
above in \S\ref{sect:transformations}.  Here, we are transforming from
$(u,v)$ to $(x,y)$, which means the transformation to be applied is
really $\mathbf{J}^{-1}$.  To preserve the total flux of the profile, we
also need to multiply the resulting profile by $|{\rm det} \,\, \mathbf{J}|$.

The \galsim\ WCS classes have functions \code{toImage} and
\code{toWorld} that effect the conversion in either direction, using
whatever subset of the above steps are required for the given WCS.  If
the WCS transformation is not uniform (so it matters where the object
is in the image), then the position of the object (in either
coordinate system) must be specified.

\subsection{Integrating over pixels}\label{sect:integratingpixels}

While the galaxy profiles are naturally defined in world coordinates,
the pixel response is most naturally defined in image coordinates,
where it can typically be modelled as a unit square top hat profile.
The PSF profile may be more naturally
considered in either coordinate system depending on where the model is
coming from.  Thus, careful attention is required to handle all of these
correctly when using a complicated WCS.

We start by ignoring the PSF
%for the moment
to show how to properly
handle a galaxy drawn onto uniform square pixels of dimension $s$ in
an image (corresponding to a \code{PixelScale} WCS).  Let the galaxy
surface brightness be given in world coordinates as $I^w(u,v)$.
According to the methods in the previous section, the corresponding
profile in image coordinates would be
\begin{equation}
I^i(x,y) = s^2 I^w(xs,ys)
\end{equation}
where $s^2$ is the $|{\rm det}\,\,\mathbf{J}|$ factor mentioned above,
which ensures that the integral of each version of the
profile gives the same total flux.

The pixel response in the two coordinate systems is given by a
2-dimensional top hat function with unit flux:
\begin{align}
P^w(u,v) &= T_s(u,v) \equiv \begin{cases} \frac{1}{s^2} & |u|{<}s/2, |v|{<}s/2 \\ 
   0 & \textrm{otherwise} \end{cases}\\
P^i(x,y) &= T_1(x,y)
\end{align}
in world and image coordinates respectively, 
where in the latter the pixel size is, by definition, equal to 1.

When observed on an image, the galaxy's surface brightness profile is
integrated over the area of the pixel.  In a particular pixel $(i,j)$,
the integrated flux $I_{ij}$ is
\begin{align}
\!\!
I_{ij} &= \int_{is-s/2}^{is+s/2}\!\! \int_{js-s/2}^{js+s/2} I^w(u,v)
\, \rmd u \, \rmd v \\
&= \int_{-\infty}^{+\infty}\!\! \int_{-\infty}^{+\infty} \!\!\!s^2
I^w(u,v) P^w(is{-}u,js{-}v) \, \rmd u \, \rmd v \\
&= s^2 (I^w \conv P^w) (is,js) \\
&= \int_{-\infty}^{+\infty}\!\! \int_{-\infty}^{+\infty}
\!\!\!I^i(x,y) P^i(i{-}x,j{-}y)\,  \rmd x\,  \rmd y \\
&= (I^i \conv P^i)(i,j)
\end{align}
We thus find that $I_{ij}$ is the convolution of the galaxy profile
with the pixel response evaluated at the centre of the pixel, and then
multiplied by the pixel area.  Furthermore, this calculation can be
done in either coordinate system.  This well-known result
\citep[e.g.][]{lauer99} is the basis of how \galsim\ implements the
integration of a surface brightness profile over the pixel area.

For more complicated WCS functions, we can still apply the same
procedure.  We either convert the galaxy profile to image coordinates
and convolve by a unit square pixel, or convert the unit square pixel
to world coordinates and do the convolution there.  In \galsim\, the
\code{drawImage} method that performs the rendering has access to the
WCS of the image, and this convolution is handled automatically.  The
pattern adopted in the code is to transform the galaxy profile into
image coordinates and then convolve by a unit square pixel, but the
converse would have been equally valid.

For the PSF, when using a non-trivial WCS, some care is required
regarding the coordinate system in which the PSF profile is defined.
The processes that cause the image coordinates to become distorted
from a square pixel are related to the causes of the PSF, namely the
atmosphere and the optics of the telescope.  Therefore, when choosing
to apply a distorted WCS, care is needed about whether the PSF is
defined in world or image coordinates.  If it is defined in image
coordinates, then it should be converted to world coordinates (as
described in \S\ref{sect:convertingprofiles}) prior to convolution by
the galaxy profile.

Finally, if a PSF is measured from an image, such as a real data image
of a star, then it already includes a convolution by a pixel
response\footnote{ It is possible to remove the pixel response from
  the PSF image with the \code{Deconvolve} operator in \galsim\, which
  deconvolves by the original pixel.  This will only be numerically
  viable if the resulting object is drawn onto an image with pixels
  that are either the same size or (preferably) larger.  }.  To apply
this consistently to simulated data using the same WCS and pixel
scale, an additional convolution by a pixel should not be applied,
since that would effectively convolve by the pixel response twice.
This kind of rendering without the extra pixel convolution is possible
in \galsim\ using the \code{no\_pixel} option of \code{drawImage}.

%
% Section 6
% rendering
% 

\section{Image rendering}\label{sect:rendering}

A given object can be rendered onto an image (``drawn'') through three
methods in \galsim: direct drawing in real space; drawing via discrete
Fourier transform (DFT);
{\edit
or drawing via ``photon shooting,'' whereby the 
surface brightness profile is treated as a probability distribution, and a
finite number of ``photons'' sampled from this distribution are ``shot'' onto the
image.
}

In principle all three rendering methods should 
{\edit
be equivalent,
}
apart from the shot noise that is inherent to the
photon-shooting method.  However, there is one additional difference
that is worth noting.
The photon-shooting method bins the photons according to the pixels
they fall into
and hence automatically integrates the profile over the pixels.
{\edit
Thus, the \code{no\_pixel} option of \code{drawImage} mentioned above
in \S\ref{sect:integratingpixels} will always use either direct rendering
or DFT as appropriate.
}

There are in practice also further minor differences between the
rendered output of the different methods that are due to
approximations inherent to each technique.  These we highlight below.

Not all objects can be rendered with all three methods.
{\edit
In particular, convolution in real space is only implemented for convolution of
two profiles, one of which is typically the pixel response (cf.\ \S\ref{sect:realconv}).
Deconvolution (see \S\ref{sect:compositions}; \S\ref{sect:shera}) is
only possible with the DFT method.  
Non-affine transformations such as flexion
\citep[][]{goldbergbacon05,irwinshmakova05,baconetal06} cannot be 
done in Fourier space;
flexion is not currently implemented in \galsim\ for this reason.
}
Table~\ref{table:rendering} summarizes the advantages and
disadvantages of each rendering method.

\begin{table*}
\begin{center}
\caption{Characteristics of different rendering methods}\label{table:rendering}
\begin{tabular}{lccc}
\hline
Characteristic &
Direct real-space &
Fourier &
Photon shooting \\
\hline
Operations for setup$^{\rm a}$ & 0 & $O(N^2 \log N)$ & $O(N^2 \log N)$ \\
Operations for rendering$^{\rm a}$ & $O(N^2)$ & $O(N^2 \log N)$ & $O(N_{\gamma} \log N)$ \\
Operations for addition$^{\rm a}$ & $O(N^2)$ & $O(N^2)$ & $O(N_{\gamma})$ \\
Operations for convolution$^{\rm a}$ & $O(N^4)$$^{\rm b}$ & $O(N^2)$ & $O(N_{\gamma})$ \\
Operations for deconvolution$^{\rm a}$ & impossible & $O(N^2)$  & impossible \\
Easy to apply affine transformations? & yes & yes & yes \\
Easy to apply non-affine transformations? & yes & no & yes \\
Inaccuracies & - & band-limiting, folding  & shot noise, truncation \\
Fastest for & analytic objects & high $S/N$ & low $S/N$\\
\hline
\end{tabular}
\newline
{\footnotesize$^{\rm a}${For operation counts, $N$ is the typical linear size of an
  input or output image, and $N_\gamma$ is the number of photons
  traced during photon shooting.}}\\
{\footnotesize $^{\rm b}${Direct rendering can only handle convolution
    of two profiles.}}
\end{center}
\end{table*}

\subsection{Direct rendering}\label{sect:direct}

The default way to draw an object in \galsim\ is to integrate its surface
brightness over the pixel response
by convolving the object's profile with a square pixel
(cf.\ \S\ref{sect:integratingpixels}).  This means the object that is
actually being drawn will really be a convolution, which is normally
rendered using the DFT method (\S\ref{sect:dft}).

However, it is possible to tell \galsim\ not to convolve by the pixel
and instead draw the profile directly by sampling the surface
brightness at the centre of each pixel.  This is the easiest rendering
method to understand, so we describe it first.  But of course, it does
not directly correspond to a real image, so the 
sum of the pixel values may not match the input flux, for instance.

\subsubsection{Analytic objects}

If an analytic surface brightness profile is drawn without convolving
by the pixel response, \galsim\ will use direct rendering, which just
involves calculating $I(x,y)$ at the location of each output pixel.  The
formulae for $I(x,y)$ for our various analytic objects were given in
\S\ref{sect:objects}, and they are summarized in
Table~\ref{tab:analytic-objects}.

Many objects have an analytic formula in real space, so the direct
rendering is straightforward. However, the \code{Kolmogorov} profile is only
analytic in Fourier space, so the implementation of $I(r)$ in real
space uses a cubic spline lookup table for the Hankel transform of
$\fourier{I}(k)$.  This calculation is done the first time a
\code{Kolmogorov} object is instantiated and saved for any further
instantiations, so the setup cost is only required once.

\begin{table*}
\begin{center}
\caption{Analytic \galsim\ objects}\label{tab:analytic-objects}
\begin{tabular}{lccc}
%\tabletypesize{\small}
%\tablewidth{0pt}
%\tablehead{
\hline
Class name &
Real-space formula$^{\rm a}$ &
Fourier-space formula$^{\rm a}$ &
Photon-shooting method$^{\rm b}$ \\
\hline
\code{Exponential} & $e^{-r/r_{\rm s}}$ & $\left(1+k^2r_{\rm s}^2\right)^{-3/2}$ & Interval \\
\code{DeVaucouleurs} & $e^{-7.669(r/r_{\rm e})^{1/4}}$ & LUT & Interval \\
\code{Sersic} & $e^{-b(n)(r/r_{\rm e})^{1/n}}$ & LUT & Interval \\
\code{Airy} & $\left[ \frac{J_1(\nu)-\epsilon J_1(\epsilon
    \nu)}{\nu}\right]^2, \quad \nu = \pi r D / \lambda$ &
Analytic$^{\rm c}$ & Interval \\
\code{Moffat} & $\left(1+\frac{r^2}{r_0^2}\right)^{-\beta}$ &
LUT$^{\rm d}$ &
Analytic \\
\code{Kolmogorov} & LUT & $e^{-(k/k_0)^{5/3}}$ & Interval \\
\code{Gaussian} & $e^{-r^2/2\sigma^2}$
  & $e^{-k^2\sigma^2/2}$ & Analytic \\
\code{Shapelet} & See \S\ref{sect:shapelets} & See
\S\ref{sect:shapelets} & Not implemented \\
\code{Pixel} & $1, |x|<s/2, |y|<s/2$ & $\sinc (s
k_x)\; \sinc(s k_y)$ & Analytic \\
\hline
\end{tabular}
\newline
{\footnotesize $^{\rm a}${Normalization factors have been omitted from the given
  formulae for brevity; the integral of the real-space profile and the $k=0$ value
  in Fourier space should both equal the flux $F$ of the object.
  ``LUT'' means that the function requires integrations that are
  precomputed and stored in a cubic-spline look-up table.}}\\
{\footnotesize $^{\rm b}${The photon-shooting methods are the following:
  ``Analytic'' means that photon locations are
  derived from a remapping of one or more uniform deviates into \vecx;
  ``Interval'' involves a combination of weighting and rejection
  sampling, as described in Section~\ref{sect:photon};
  ``Not implemented'' means that \galsim\ cannot currently render this
  with photon shooting.}}\\
{\footnotesize $^{\rm c}${Fourier representation of the Airy function with
  obscuration is analytic but too complex for the
  table.}}\\
{\footnotesize $^{\rm d}${For generic $\beta$ values, a lookup table is used.
 However, some particular values of $\beta$ have analytic Fourier-space
 formulae, which \galsim\ uses in these cases.}}
\end{center}
\end{table*}

For the radially symmetric profiles, we can
take advantage of the known symmetry to speed up the calculation.
There are also usually vectorization techniques that lead to further
improvements in efficiency.

\subsubsection{Shapelet profiles}

The shapelet functions are not radially symmetric, but the code to
directly render \code{Shapelet} objects uses fast recurrence relations
for the shapelet functions given in \citet{BJ02}, which vectorize
conveniently when applied to the pixels in the image.  They are thus
also relatively efficient to draw.

\subsubsection{Interpolated images}

\code{InterpolatedImage} objects are usually the slowest to directly
render, since \eqn{eq:interpolatedimage}
involves a sum over a number of the original pixels
for each output pixel value, the exact number depending on the interpolant being
used (see Table~\ref{table:interpolants}).

\subsubsection{Transformations}

All of the transformations described in \S\ref{sect:transformations}
are very simple to apply when directly rendering.  The value of the
transformed profile, including all three kinds of effects, is simply
\begin{equation}
I'(\vecx)  = c \,I\left[ \matA^{-1}(\vecx - \vecx0)\right].
\end{equation}
where $I(\vecx)$ is the original profile.

\subsubsection{Compositions}\label{sect:realconv}

Directly rendering a sum of profiles is trivial, since each profile
can be rendered individually, adding the flux to whatever has already
been rendered.

The rendering of a convolution is almost always done via DFT or
photon-shooting methods.  However, when convolving two objects that
are both known to have high-frequency content, such that an unaliased
DFT is impractical, \galsim\ may elect to render it directly through a
Gauss-Kronrod-Patterson evaluation of the convolution integral,
\eqn{eq:convintegral}.

Generally such circumstances do not represent any physically
realizable image, since real telescopes and detectors lead to
band-limited images.  However, in simulated images, it can happen if
two objects being convolved both have hard edges.  Here, a ``hard
edge'' means a place where the surface brightness abruptly drops from
some finite value to zero.  For example a \code{Pixel} convolved with
a truncated \code{Moffat} profile would by default be convolved using
direct integration.  In this and similar cases, the direct integration
is generally both faster and more accurate.

\subsection{Discrete Fourier transform rendering}\label{sect:dft}
When rendering images that include convolution by a pixel
response (and often other profiles such as a PSF),
\galsim\ uses DFTs by default.

\subsubsection{Band limiting and folding}\label{sect:bandlimitingfolding}
To sample a surface brightness profile $I(\vecx)$ onto an $M\times M$
grid with sample pitch $\Delta x$ via DFT, we will need to know its
Fourier transform $\fourier I(\veck)$ at all points with $k_x, k_y$
being multiples of $\Delta k = 2\pi/(M\Delta x)$.  If any frequency
modes are present in $\fourier I(\veck)$ beyond the Nyquist frequency
$\pi / \Delta x$ of the output grid these will be aliased in the
output image, as required for the correct rendering of undersampled
data.

The input to the DFT will
be the $M\times M$ Hermitian array of Fourier coefficients summed over
all aliases:
\begin{align}
\fourier a_{ij} &= \sum_{p,q} \fourier I\left[ (i+pM)\Delta k, (j+qM)
  \Delta k)\right], \nonumber \\
  & \qquad \qquad -M/2\le i,j < M/2.
\label{aliassum}
\end{align}
These sums must be truncated at some finite $(p,q)$ range.  We must
select a spatial frequency $k_{\rm max}$ such that we approximate
$\fourier I(k_x,k_y)=0$ for $|k_x|>k_{\rm max}$ or $|k_y|>k_{\rm max}.$  Any
real telescope {\edit produces bandlimited images, transmitting} zero power beyond a 
wave vector of $k_{\rm max}=2\pi D / \lambda$,
where $D$ is the maximum entrance aperture dimension and $\lambda$ is the
wavelength of observation.
For space-based observations we are typically creating images sampled
near the Nyquist scale of $\lambda/2D$ for this frequency, and we can
run the sum (\ref{aliassum}) over all physically realizable frequencies
without approximation.

For ground-based observations, the PSF is typically much larger than
the diffraction limit, and the pixel scale has Nyquist frequency well
below the physical $k_{\rm max}$.  To render images with acceptable
speed it is often necessary to approximate the Fourier transform
$\fourier I(\veck)$ as being identically zero beyond a chosen
$k_{\rm max}$. This is called band-limiting the image.  Common
idealizations of PSFs, such as the Gaussian, Moffat, and Kolmogorov
functions, are unbounded in $\veck$, and therefore we must select a
$k_{\rm max}$ which yields an acceptable approximation.

Every kind of surface brightness profile in \galsim\ {\edit is able to
calculate and return} its $k_{\rm max}$, which we take to be the value
of $k$ where the Fourier-domain profile drops  below a threshold of
$10^{-3}$ of the peak.  See \S\ref{sect:tolerances} for 
{\edit more detail about the parameter (\code{maxk\_threshold}) that
 controls this value.  For some profiles $k_{\rm max}$ can be
 calculated analytically; where this is not possible $k_{\rm max}$
 is determined numerically using an appropriate method for each
 profile.}

Another characteristic of the DFT is that each output value
$I_{ij} = I(i \Delta x, j\Delta x)$ is
actually the true function folded at the period $P= M \Delta x = 2\pi
/ \Delta k$:
\begin{equation}
I_{ij} = \sum_{p,q} I( i\Delta x + pP, j\Delta x + qP).
\label{folding}
\end{equation}
Since it is impossible for $I(\vecx)$ to be compact after being
band-limited, this leads to some degree of inaccuracy
due to folding of the image.  We can limit
the aliasing errors by ensuring that every DFT is done with a
sufficiently large period.  Every surface brightness profile in \galsim\
also knows what $k_{\rm step}$ value should be used for the DFT (cf.
\code{folding\_threshold} in \S\ref{sect:tolerances}).
The DFT is executed with period $P \ge 2 \pi /
k_{\rm step}$, implying $\Delta k \le k_{\rm step}$ and $M\ge 1/
k_{\rm step}\Delta x$.

If the user has requested rendering of an image that does not meet
this criterion, \galsim\ will execute the DFT on a grid of larger $M$
that does, and then extract the output image from the result of this
larger DFT.  If the user does not specify an image size, \galsim\ will
select a value that satisfies this bound.

The DFT image must have dimension $M\ge 2k_{\rm max} / k_{\rm
  step}$.  Some of the \galsim\ profiles contain sharp features
that cause this minimum $M$ to exceed the maximum allowed DFT size, in
particular the \code{Pixel} class and the \code{Sersic} class with
higher indices, including the \code{DeVaucouleurs} class.
\galsim\ will raise an error if one attempts to render such an object via
DFT without first convolving by another object that attenuates the
high-frequency information.  This should not be an issue for rendering
of realistic images, in which convolution with the PSF and pixel response will
typically limit the bandwidth of the image to a manageable value.

\subsubsection{Analytic objects}

Table~\ref{tab:analytic-objects} lists the
profiles for which \galsim\ calculates Fourier-domain values
analytically.

For those radial profiles without fast analytic formulae in Fourier
domain, we tabulate numerical transforms into a lookup table and use
cubic spline interpolation to return $\fourier I(k)$.  The initial
setup of these tables can take some time, but they
are cached so that later
Fourier-domain evaluations of the same type of profile are
accomplished in constant time.

At small values of $k=|\veck|$ we generally replace the spline interpolation
with an analytic quadratic (or higher) order Taylor expansion of
$\fourier I(k)$, because the behaviour of $\fourier I(\veck)$ near the
origin has a strong effect on the appearance of poorly-resolved
objects.  In particular the second derivative of $\fourier I (\veck)$
determines the second central moments that are critical for
measurement of weak gravitational lensing, and the spline
interpolation can produce incorrect derivatives at the origin.

\subsubsection{Shapelet profiles}

Since shapelets are their own Fourier transforms (cf.\ \eqn{eq:fouriershapelet}),
the code to draw a \code{Shapelet} object in Fourier space is essentially identical 
to the code to directly render it.  It uses fast recurrence 
relations, which are efficient when applied to the pixels in the Fourier image.

\subsubsection{Interpolated images}
\label{sect:fourierinterpolatedimage}

Given an $N\times N$ array of samples at pitch $\Delta x$, we can
perform a DFT to yield samples of the Fourier transform at a grid of
values $\veck_{ij}=(i/N\Delta x, j/N\Delta x)$.  To render this
function back onto a different $x$-domain grid, or to execute
transformations, we need to evaluate $\fourier I(\veck)$ at
arbitrary \veck\ between the grid of DFT values.  This requires
some $k$-space interpolation scheme.

Furthermore we need to account for the facts that (a) the interpolated
image is finite in extent, while the DFT will yield the transform of a
periodic function, and (b) it represents a continuous, interpolated
version of the input sample grid.  Thus two distinct interpolants are
required: the $x$-domain interpolant $K_x$ is chosen by the user and
is part of the definition of $I(\vecx)$; the $k$-domain interpolant
$K_k$ is used to generate $\fourier I(\veck)$ by interpolating over
the DFT of the input image and the Fourier transform of $K_x$.

Correct evaluation of the interpolated function in $k$-space is not
trivial.  \citet{BernsteinGruen} present a detailed description of the
steps required, and we summarize their results here.  First, they find
that the rigorously correct form of $K_k$ is a sinc function wrapped
at the extent of the $x\rightarrow k$ DFT.  Since this interpolant
spans the entire plane, it means that the interpolation of $\fourier
I(\veck)$ to all the $M\times M$ points needed for the DFT requires
$N^2M^2$ operations, which can be prohibitively slow.  As a
consequence we normally elect to approximate this using a compact
interpolant for $K_k$.

\citet{BernsteinGruen} demonstrate that the consequences of this
interpolation are a slight multiplicative scaling of the image, plus
the generation of 4 ``ghost images'' displaced by $N\Delta x$ along
each axis.  They find that these artifacts can be reduced below 0.001
of the input flux by a combination of zero-padding the input image by
$4\times$ in each direction before conducting the $x\rightarrow k$
DFT, producing a denser set of samples in $k$ space, and then using a
custom-designed, 6-point piecewise quintic polynomial interpolant for
$K_k$.  This recipe is the default for Fourier-domain rendering of the
\code{InterpolatedImage} in \galsim.  Users can override the default
4-fold zero-padding factor and/or the choice of the \code{Quintic}
interpolant when 
intializing an \code{InterpolatedImage} object.

The value of $k_{\rm step}$ for an interpolated image
of input size $N$ and pitch $\Delta x$ would
naively be set at $2\pi / (N\Delta x)$ to reflect the bounded size of
the input image.  We must recall, however, that the interpolation of
the input image by the kernel $K_x$ will extend the support of the
object and slightly lower the desired $k_{\rm step}$ (with the sinc
interpolant extending the footprint of the image enormously).  

The value of $k_{\rm max}$ for a raw sampled image is infinite because it
is composed of $\delta$ functions.  However the application of the
interpolant $K_x$ to the samples will cause $\fourier I(\veck)$ to roll off
yielding $k_{\rm max} = c \pi / \Delta x$, where $c$ is a constant
characteristic to the chosen $K_x$.  In this respect
the band-limited sinc interpolant, with $c=1$, is ideal, but the
Lanczos interpolants with $n=3$--5 offer much more compact support
with $2<c<3$ sufficing to contain all aliases with amplitude
$>0.001$.  See Table~1 in \citet{BernsteinGruen} for the relevant
characteristics of the interpolants.

\subsubsection{Transformations}

Transformations of functions have well-known
correspondents in Fourier domain, which we quickly review here.

A shift $I'(\vecx) = I(\vecx - \vecx_0)$ has
Fourier-domain equivalent $\fourier I'(\veck) = e^{ {\rm i} \veck\cdot\vecx_0}
\fourier I(\veck)$.   A shift leaves $k_{\rm max}$ and $k_{\rm step}$ unchanged.

Linear transformations of the plane, such that $I'(\vecx) =
I(\matA^{-1}\vecx),$ are equivalent to $\fourier I'(\veck) = \fourier
I(\matA^T \veck).$  We identify the maximum and minimum eigenvalues
$\lambda_+$ and $\lambda_-$ of \matA, and set
\begin{align}
k_{\rm max} & \rightarrow \lambda_-^{-1} k_{\rm max} \\
k_{\rm step} & \rightarrow \lambda_+^{-1} k_{\rm step} 
\end{align}

\subsubsection{Compositions}

The Fourier transform is linear, so that if $I({\vecx}) = a I_A(\vecx) + b
I_B(\vecx)$, we also have $\fourier I(\veck) = a \fourier I_A(\veck) + b
\fourier I_B(\veck)$.  When we are dealing with a sum of multiple
components $I_i$, we set
\begin{align}
k_{\rm max} & = {\rm max}_i (k_{{\rm max},i}) \\
k_{\rm step} & = {\rm min}_i (k_{{\rm step},i})
\end{align}

The convolution $I(\vecx) = (I_A \conv I_B)(\vecx)$ is equivalent
to $\fourier I(\veck) = \fourier I_A(\veck) \fourier I_B(\veck)$.  Under a
convolution of multiple elements, we set
\begin{align}
k_{\rm max} & = {\rm min}_i(k_{{\rm max},i}) \\
\label{eq:convolutionkstep}
k_{\rm step} & = \left(\sum_i k_{{\rm step},i}^{-2}\right)^{-1/2}
\end{align}
While the propagation of the band limit $k_{\rm max}$ is rigorously
correct for convolution, the propagation of $k_{\rm step}$ in
\eqn{eq:convolutionkstep}
is merely heuristic.  It is based on the exact statement
that the central second moments of objects sum in quadrature under
convolution.  But there is no general rule for the propagation of the
radius enclosing some chosen fraction of the total object flux, so 
our heuristic is known to be correct only for Gaussian
objects.  For strictly compact functions such as a \code{Pixel}, the
maximum nonzero element $x_{\rm max}$ actually adds linearly, not in
quadrature, under convolution.  \galsim\ provides the option to
manually increase the size of DFTs if one is worried that the $k_{\rm
  step}$ heuristic will lead to excessive aliasing.

Finally, a \code{Deconvolve}
operation applied to some image $I_A(\vecx)$ yields $1/\fourier I_A(\veck)$
in the Fourier domain.
We leave $k_{\rm max}$ and $k_{\rm step}$ 
unaltered, because the deconvolved object is usually ill-defined
unless it is later re-convolved with an object that has a smaller $k_{\rm
  max}$ value.

\subsection{Photon shooting}\label{sect:photon}

``Photon shooting'' was used successfully by
\citet{bridleetal09,bridleetal10} and
\citet{kitchingetal12,kitchingetal13} to generate the simulated images
for the GREAT08 and GREAT10 weak lensing challenges. The objects were
convolutions of elliptical \sersic-profile galaxies with
Moffat-profile PSFs.  \galsim\ extends this technique to enable photon
shooting for nearly all of its possible objects, except for
deconvolutions.

When we ``shoot'' a \galsim\ object, $N_\gamma$ photons are created
with weights $w_i$ and positions $\vecx_i$.  The total weight within
any region has an expectation value of the flux in that region, and
the total weights in any two regions are uncorrelated\footnote{
This will not be true if you turn off the Poisson variation in the
total flux, which is optional but turned on by default, since then
the fixed total flux will lead to some correlation in the pixel values.}.

We allow for non-uniform weights primarily so that we can represent
negative values of surface brightness.  This is necessary to realize
interpolation with kernels that have negative regions (as will any
interpolant that approximates band-limited behaviour), and to correctly
render interpolated images that have negative pixel values, such
as might arise from using empirical, noisy galaxy images.

For photon shooting to be possible on a \galsim\ object, it must be
able to report $f_{\rm pos}$ and $f_{\rm neg},$ the absolute values of
the flux in regions with $I(\vecx)>0$ and $I(\vecx)<0$, respectively.
When $N_\gamma$ photons are requested, we draw their positions from
the distribution defined by $|I(\vecx)|$ and then assign weights
\begin{equation}
w_i = \frac{f_{\rm abs}}{N_\gamma} {\rm sign}\left[ I(\vecx_i)\right]
= \frac{f_{\rm pos}+f_{\rm neg}}{N_\gamma}{\rm sign}\left[
  I(\vecx_i)\right] .
\label{nominalweight}
\end{equation}

Shot noise in the fraction of photons that end up in the
negative-brightness regions will lead to variations in the total flux
of the photons.  \galsim\ mostly accounts for this effect automatically
by selecting an appropriate number of photons to get the noise correct
(cf.\ \S\ref{sect:nphotons}).
However, the partial cancellations between positive- and negative-flux
photons means that the resulting noise will not actually be
distributed precisely according to a Poisson distribution for the
given flux; only the variance of the noise will be approximately
accurate.  For objects constructed purely from positive-flux profiles,
this effect is absent, and the noise is indeed Poisson.

We will see below that in some cases, $|w_i|$ is allowed to
depart slightly from the constant $f_{\rm abs}/N_\gamma$ value, which
also alters the noise properties slightly.

\subsubsection{Analytic objects}\label{sect:analytic-objects-photon}

Photon shooting a surface-brightness function $I(\vecx)$ is simplest
when there is a known transformation from the unit square or circle to
the plane whose Jacobian determinant is $\propto I$.
{\edit
\code{Pixel} is the simplest such case, since it is just a scaling of the unit square.
}

Many other profiles are circularly symmetric functions with
a cumulative radial distribution function
\begin{equation}
C(r) = \frac{ \int_0^r I(r^\prime) r^\prime\,  \rmd r^\prime}{
  \int_0^\infty I(r^\prime) r^\prime\, \rmd r^\prime}.
\end{equation}
If $u$ is a uniform deviate between 0 and 1, $C^{-1}(u)$ will be distributed as
$rI(r)$, the distribution of flux with radius.  A second uniform
deviate can determine the azimuthal angle of each photon.  

\galsim\ uses fast analytic $C^{-1}(u)$ functions
for the \code{Gaussian} \citep[e.g.][]{NumericalRecipes} and \code{Moffat}
classes.

For the other circularly symmetric profiles we use the following
algorithm to convert a uniform deviate into the radial flux
distribution of $|I|$.  The algorithm is provided with the function
$I(r)$ and with a finite set of points $\{R_0, R_1, R_2, \ldots, R_M\}$
with the guarantee that each interval $(R_i, R_{i+1})$ contains no
sign changes and at most one extremum of $I(r)$. The
absolute flux $|f_i|$ in each annulus is obtained with standard
numerical integration techniques.

Note that this algorithm
requires a maximum radius, so the photon shooting rendering requires
truncation of unbounded profiles.  The fraction of excluded flux due
to this truncation is given by the \code{shoot\_accuracy} parameter
described in \S\ref{sect:tolerances}.

The algorithm proceeds as follows:
\begin{enumerate}
\item The code first inserts additional nodes into the $R_i$ series at
  the locations of any extrema.
\item The radial intervals are recursively bisected until either the
  absolute flux in the interval is small, or the ratio $\max|I|
  / \min|I| $ over the interval is below a chosen threshold.
\item The integral $f_i$ of the absolute flux is calculated for each
annular interval along with the probability $p_i = f_i / \sum_i f_i$ of a
photon being shot into the interval.  We can also calculate the
cumulative probability $P_i = \sum_{j<i} p_j$ of the photon being
interior to the annulus.
\item For each photon we draw a uniform deviate $u$.  It is mapped
  to a radius $r$ by finding the interval $i$ such that $P_i \le u <
  P_{i+1}$ and then assigning
\begin{equation}
r^2 = R_i^2 + \frac{u - P_i}{p_i} \left( R_{i+1}^2 - R_i^2 \right).
\label{eq:annulardeviate}
\end{equation}
In other words we draw $r$ from a flux distribution that approximates
$|I(r)|$ as a ``wedding cake'' of constant-valued annuli.
\item (a) If $r$ is in an interval that has a narrow range of brightness variation,
  we re-weight the photon by a factor equal to the brightness at $r$
  relative to the mean brightness in the annulus:
\begin{equation}
w = \frac{f_{\rm abs}}{N_\gamma}\frac{ I(r) \pi \left( R_{i+1}^2 - R_i^2 \right)}{f_i}.
\label{eq:reweight}
\end{equation}

(b) If the range of variation of $I$ in the annulus is large, which
  can happen when the annulus is near a zero crossing, we
  implement rejection sampling by drawing another uniform deviate
  $u^\prime$, and keeping the photon $r$ if $|I(r)| / {\rm max} |I|
  > u^\prime.$  If the photon is rejected, we use another uniform
  deviate $u$ to select a new trial radius within the interval:
\begin{equation}
r^2 = R_i^2 + u \left( R_{i+1}^2 - R_i^2 \right).
\end{equation}
The rejection test is repeated, and the process is iterated until a
photon radius is accepted.  This photon is given the nominal weight 
(\ref{nominalweight}).
\item The azimuthal angle of the photon is selected with a uniform deviate.
\end{enumerate}

This procedure yields $N_\gamma$ photon locations and weights,
employing $2N_\gamma(1+\epsilon)$ uniform deviates and $N_\gamma(1+\epsilon)$
evaluations of $I(r)$, where $\epsilon$ is the fraction of photons
rejected in step 5b.  {\edit We note that the finite widths of the
  annuli must be small for the approximation to be good: the close
  correspondence of photon shooting and DFT rendering results in
  \S\ref{sect:validate-dft} demonstrates that this is achieved in
  \galsim.}

In practice we
re-order the intervals into a binary tree {\edit
 \citep[see, e.g.,][]{makinson2012sets}} to optimize the selection
of an interval with the initial uniform deviate.  With this tree
structure the mean number of operations to select an interval
approaches the optimal $-\sum p_i \log p_i < \log M$, where $M$ is the
final number of radius intervals.

The \code{Sersic}, \code{Exponential}, \code{DeVaucouleurs},
\code{Airy}, and \code{Kolmogorov} classes use this interval-based
photon shooting algorithm.  The construction of the interval structure
and flux integrals are done only once for the \code{Exponential}
and \code{DeVaucouleurs} profiles and once for each distinct \sersic\ index
$n$ for \code{Sersic} profiles or distinct central
obscuration for \code{Airy}. These results are 
cached for use by future instances of the classes in each case.

\subsubsection{Shapelets} 

Because shapelet profiles are not radially symmetric, implementing
photon shooting for such objects represents a significant challenge
(although it is far from impossible).  As of \galsim\ version
1.1, we have not attempted to support photon shooting for 
\code{Shapelet} objects.

\subsubsection{Interpolated images}\label{sect:photon-interp}

Photon shooting an \code{InterpolatedImage} object requires two steps:
first we distribute the photons among the grid points $(i\Delta x, j
\Delta x)$ with probabilities $p_{ij} = |a_{ij}|/\sum |a_{ij}|$. We
select an ordering for these probabilities and create the cumulative
probabilities $P_{ij}$.  For each photon, we draw a uniform deviate
$u$, and assign an $(i,j)$ value
as the last one in the ordering with $P_{ij}<u$.  The weight is given
the sign of $a_{ij}$. As we did with the radial 
function intervals above, we create a tree structure which reduces the
selection of each $(i,j)$ to $O(\log N)$ for an $N\times N$ input
image.

The second step in photon shooting the \code{InterpolatedImage} is to
convolve the gridded samples with the interpolation kernel $K_x$ by
adding to each photon's location a displacement drawn from the kernel.
Since our 2D interpolants are all separable into 1D
functions, we can draw an $x$ and $y$ displacement from the 1D
functions.  The $\delta$-function interpolant can of course skip this
step, and the nearest-neighbour and linear interpolants are trivially
generated from uniform deviates.  Photon shooting is implemented for
the other interpolants using the same algorithm described for the
radial analytic functions, with the replacement of $r^2\rightarrow x$
in \eqnb{eq:annulardeviate}{eq:reweight} because of
the linear instead of circular geometry.

Attempts to shoot photons through an image with a sinc interpolant
will throw an exception.  The long oscillating tail of the sinc
function means that $f_{\rm abs}$ is divergent.

Photon shooting an \code{InterpolatedImage} thus requires
$3N_\gamma(1+\epsilon)$ random deviates plus, for most interpolants,
$2N_\gamma(1+\epsilon)$ evaluations of the 1D kernel function.

\subsubsection{Transformations}

All the transformations described in \S\ref{sect:transformations}
are very simply implemented in photon shooting.
Flux rescaling is simply a rescaling of the $w_i$.
{\edit Any} transformation of
the plane can be executed by $\vecx_i\rightarrow T(\vecx_i)$.

\subsubsection{Compositions}

To shoot $N_\gamma$ photons from a sum of $M$ profiles, we draw
the number of photons per summand $\{n_1, n_2, \ldots, n_M\}$ from a
multinomial distribution with expectation value $\langle n_j \rangle =
N_\gamma f_{{\rm abs},j} / \sum_j f_{{\rm abs},j}.$ Then we
concatenate the lists of $n_i$ photons generated by shooting
the individual summands.

Convolution is similarly straightforward: to shoot $N_\gamma$ photons
through $I_A \conv I_B$, we draw $N_\gamma$ photons each from $I_A$
and $I_B$, set the output weights to $w_i = w_{i,A} w_{i,B}$, and
sum the positions to get $\vecx_i = \vecx_{i,A} + \vecx_{i,B}.$ 
To ensure that there is no correlation between the photon positions
of $A$ and $B$ in the sequence, \galsim\ randomizes the order of the
$B$ photons.  This procedure generalizes straightforwardly
to the case of convolving more than two profiles together.

We are not aware of a practical means of implementing deconvolution
via photon shooting, so \galsim\ cannot use photon shooting
for any profile that involves a deconvolution.

\subsubsection{Selecting an appropriate number of photons}
\label{sect:nphotons}

One of the main advantages of using the photon-shooting method for
rendering images is that for objects with low signal-to-noise ($S/N$),
it can be much faster to shoot a relatively small number of photons
compared to performing one or more Fourier transforms.  However, for
this to be effective, one must know how many photons to shoot
so as to achieve the desired final $S/N$ value.

The simplest case is when there are no components that require
negative-flux photons.  Then the flux $f$ of the object given in
photons\footnote{ Technically, in \galsim\ the flux is defined in
  so-called analog-to-digital units, ADU.  These are the units of the
  numbers on astronomical images.  We allow a \code{gain} parameter,
  given in photons/ADU, to convert between these units and the actual
  number of photons, combining both the normal gain in e-/ADU and the
  quantum efficiency in e-/photon.  The default behaviour is to ignore
  all these issues and use a gain of 1.}
is exactly equal to the
number of photons that would be detected by the CCD.  In such a case,
the flux itself provides the number of photons to shoot.

This simple procedure is complicated by a number of considerations.
The first is that the flux itself is a random variable.  So
by default, \galsim\ will vary the total number of photons according
to Poisson statistics, which is the natural behaviour for real photons.
This means that the actual realized flux of the object varies
according to a Poisson distribution with the given mean value.  This
feature can also be turned off if desired, since for simulations,
users may want a specific flux value, rather than just something
with the right expectation value.

A more serious complication happens when there are
components that require negative-flux photons to be shot, such as
those involving interpolants (cf.\ \S\ref{sect:photon-interp}).
We need more photons to get the right total flux, since some of them
have negative flux, but then the increased count leads to more noise than
we want.  The solution we adopt to address this is to have each photon
carry somewhat less flux and use even more photons.

Let us define $\eta$ to be the fraction of photons being shot that
have negative flux, and let each photon carry a flux of $\pm g$.
We want both the total flux shot and its variance to equal the
desired flux, $f$:
\begin{align}
f &= N_\mathrm{photons} (1-2\eta) g \\
Var(f) &= N_\mathrm{photons} g^2
\end{align}
Setting these equal, we obtain
\begin{align}
g &= 1-2\eta \\
N_\mathrm{photons} &= f / (1-2\eta)^2
\end{align}

\galsim's function \code{drawImage} automatically does the above
calculation by default when photon shooting,
although users can override it and set a
different number of photons to be shot if they prefer.

Note that the above calculation assumes that $\eta$ is a constant
over the extent of the profile being shot.  In reality, it is not
constant, which leads to correlations between different regions of 
a photon-shot image using photons of mixed sign.  The resulting
noise pattern will roughly resemble the noise of a real photon image,
but one should not rely on the noise being either stationary or
uncorrelated in this case.  Nor will the noise in any pixel follow Poisson
statistics.  These inaccuracies gets more pronounced as $\eta$ nears 0.5.

There is one additional feature, which can be useful for objects with
moderately high flux, but which are still {\edit dominated by sky background light}
(a relatively common use case).
It may take many photons to get the noise level correct,
a computational cost which is wasted if a larger amount of noise is then
added from the sky background.
The same final $S/N$ can be obtained (to a very good approximation) by
shooting fewer photons, each with flux larger than the normal
choice of $g = 1-2\eta$.

To address this, \galsim\ has an option to allow the photon shooting
process to add an additional noise per pixel over the Poisson noise
expected for each pixel.  Each pixel may be allowed to have $\nu$ excess
variance.  The brightest pixel in the image, with flux $I_\mathrm{max}$,
can then have a variance of up to $I_\mathrm{max} + \nu$. 
Thus the number of photons can be reduced by as much as 
$I_\mathrm{max} / (I_\mathrm{max} + \nu)$.  The flux carried by each
photon is increased to keep the total flux equal to the target value, $f$:

\begin{align}
N_\mathrm{photons} &\ge \frac{I_\mathrm{max} f}{(I_\mathrm{max} + \nu) (1-2\eta)^2}
\label{eq:nphotwithnu}
\\
g &= \frac{f}{N_\mathrm{photons} (1-2\eta)}
\label{eq:gwithnu}
\end{align}
Clearly, this is only helpful if $\nu \gg I_\mathrm{max}$, 
which is indeed the case when the image is sky noise dominated.
This is a common use case in realistic simulations, particularly
ones that are meant to simulate ground-based observations.

Users of this behaviour in \galsim\ must explicitly indicate how
much extra noise per pixel, $\nu$, is tolerable.
Typically this will be something like 1\% of the sky
noise per pixel.  This is \emph{not} the default behaviour, because at
the time of rendering the image, the code
does not know how much (or whether) sky noise will be added.

An additional complication with the above prescription is that the
code does not know the value of $I_\mathrm{max}$ \emph{a priori}.  So
the actual algorithm starts by shooting enough 
photons to get a decent estimate for $I_\mathrm{max}$.  Then it
continues shooting until it reaches the value given by
\eqn{eq:nphotwithnu}, while periodically updating the estimated 
$I_\mathrm{max}$.  At the end of this process, it rescales 
the flux of all the photons according to \eqn{eq:gwithnu} using the
final value of $N_\mathrm{photons}$.

\subsection{Setting tolerances on rendering accuracy}\label{sect:tolerances}

Some of the algorithms involved in rendering images involve tradeoffs
between speed and accuracy.  In general, we have set accuracy
targets appropriate for the weak lensing requirements of Stage IV surveys
such as LSST, \emph{Euclid} \citep{2011arXiv1110.3193L}, and \emph{WFIRST}
(cf.\ \S\ref{sect:validation}). 
However, users may prefer faster code that is slightly less accurate for some
purposes, or more accurate but slower for others.
This is possible in \galsim\ using a structure called \code{GSParams},
which includes many parameters pertaining to the speed versus accuracy
tradeoff,
{\edit some of which are described below:}
\begin{itemize}
\item \code{folding\_threshold}\footnote{In earlier versions of
    \galsim\ this parameter was named \code{alias\_threshold}. It was
    renamed for clarity and to avoid confusion with the phenomenon of
    aliasing in undersampled data.}  sets
  a maximum permitted amount of real space image folding due to the
  periodic nature of DFTs, as described in
  \S\ref{sect:bandlimitingfolding}.  It is the critical parameter for
  defining the appropriate step size $k_{\rm step}$.  Additionally, it
  is relevant when letting \galsim\ choose output image sizes: the
  image will be large enough that at most a fraction
  \code{folding\_threshold} of the total flux falls off the edge of
  the image.  The default is \code{5.e-3}.

\item \code{maxk\_threshold} sets the maximum amplitude of the high
  frequency modes in Fourier space that are excluded by truncating the
  DFT at some maximum value $k_{\rm max}$.  As described in
  \S\ref{sect:bandlimitingfolding}, this truncation is required for
  profiles that are not formally band-limited.  Reducing \code{maxk\_threshold} can
  help minimize the effect of ``ringing'' in images
  for which this consideration applies.  The default is \code{1.e-3}.

\item
\code{xvalue\_accuracy} and \code{kvalue\_accuracy} set the accuracies of values
in real and Fourier space respectively.  When an approximation
must be made for some calculation, such as when to transition to Taylor
approximation as small $r$ or $k$, the error in the approximation is constrained to 
be no more than one of these values times the total flux.
The default for each is \code{1.e-5}.

\item
\code{realspace\_relerr} and \code{realspace\_abserr} set the relative and absolute
error tolerances for real-space convolution.  The estimated integration error for the 
flux value in each pixel is constrained to be less than either \code{realspace\_relerr}
times its own flux or \code{realspace\_abserr} times the object's total flux.
The default values are \code{1.e-4} and \code{1.e-6} respectively.

\item
\code{shoot\_accuracy} sets the relative accuracy on the total flux when photon shooting.
When the photon-shooting algorithm needs to make approximations, such as how high
in radius it must sample the radial profile, 
the resulting fractional error in the flux is limited to \code{shoot\_accuracy}.
The default value is \code{1.e-5}.
\end{itemize}

There are several less important parameters that can also be
set similarly, but the above parameters are
the most relevant for most use cases. 
{\edit All the \galsim\ objects described in \S\ref{sect:objects} have an
optional parameter \code{gsparams} that can set any number of these
parameters to a non-default value when initializing the object.}

{\edit
\subsection{Representing realistic galaxies}\label{sect:shera}
}

{\edit
\galsim\ can carry out a process (`reconvolution') to render an
image of a real galaxy observed at high resolution (with the \hst),
{\em removing} the \hst\ PSF, with an applied lensing shear and/or
magnification, as it would be observed with a lower-resolution
telescope.  Reconvolution was mentioned in \cite{2000ApJ...537..555K}
and implemented in \cite{2012MNRAS.420.1518M} in the \shera\ (SHEar Reconvolution Analysis)
software\footnote{\url{http://www.astro.princeton.edu/~rmandelb/shera/shera.html}}.
}

{\edit
Reconvolution is possible when the target band limit
}
$k_\text{max,targ}$ relates to the original \hst\ band limit
$k_\text{max,HST}$ via
\begin{equation}
k_\text{max,targ} < \left(1-\sqrt{\kappa^2+\gamma^2}\right) k_\text{max,HST}.
\end{equation}
For weak shears and convergences, this condition is easily
satisfied by all upcoming lensing surveys, even those from space.

{\edit
In \galsim, the \code{RealGalaxy} class represents the \hst\ galaxy
deconvolved by its original PSF.  This can then be transformed
(sheared, etc.) and convolved by whatever final PSF is desired.
Observations of galaxies from the \hst\ COSMOS survey \citep[described
in][]{koekemoeretal07}, which form the basis of
the \code{RealGalaxy} class, are available for download from the
\galsim\
repository\footnote{\url{https://github.com/GalSim-developers/GalSim/wiki/RealGalaxy\%20Data}}. 
Padding the input postage stamp images is necessary, for the
reasons described in \S\ref{sect:fourierinterpolatedimage}.  We carry
out noise padding on the fly, with a noise field corresponding to 
that in the rest of the postage stamp (unlike \shera, which required
padded images as inputs). 
The amount of padding was extensively tested during the numerical
validation of the \galsim\ software, and results for the default
(strongly recommended) choice are described in 
\S\ref{sect:validate-ii}.
}

The implementation is updated compared to that
{\edit
in \shera\ in several ways. 
First, \shera\ only implemented a cubic
interpolant, whereas \galsim\ includes
many interpolants that can be
tuned to users' needs (c.f.\ \S\ref{sect:interpolatedimage}).  
Second, \galsim\ fully deconvolves the original \hst\
}
PSF.  \shera\ uses a
technique called pseudo-deconvolution, only deconvolving the
\hst\ PSF on scales accessible in the final low-resolution image.
In practice, the difference between pseudo-deconvolution
and deconvolution is irrelevant, and what matters is the ability to
render sheared galaxies very accurately 
{\edit
(see \S\ref{sect:validation}).
Finally, 
}
the original \hst\ images typically contain correlated
noise, and further correlations are introduced by the shearing and subsequent
convolution. 
In \S\ref{sect:whitening} we describe how \galsim\ 
{\edit
(unlike \shera)
}
can model this correlated noise throughout the reconvolution process
and then whiten the final rendered image so that it contains
only uncorrelated Gaussian noise.

%
% Section 7
% noise
%

\section{Noise models}\label{sect:noise}
Noise is a feature of all real imaging data, due to finite numbers of
photons arriving at each detector, thermal noise and electronic noise
within the detector and readout, and other lossy processes. 
\galsim\ therefore provides a number of
stochastic noise models that can be applied to simulated images.

It is worth noting that images rendered by photon shooting (see
\S\ref{sect:photon}) already contain noise: photon counts are drawn
from a Poisson distribution of mean equal to the expected flux at each
pixel.  
But there may still be occasions when it is desirable to add
further noise, such as to simulate a shallower image or to add
detector read noise.  Images rendered by DFT 
(\S\ref{sect:dft}) need noise to be added to be made realistic.

\subsection{Random deviates}\label{sect:deviates}
Underlying all the noise models in \galsim\ are implementations of random
deviates that sample from a range of probability distributions.  These
include the uniform distribution $\mathcal{U}(0, 1)$ (implemented by
the \code{UniformDeviate} class), the
Gaussian/normal distribution $\mathcal{N}(\mu, \sigma^2)$ (\code{GaussianDeviate}),
and the Poisson distribution ${\rm Pois}(\lambda )$ (\code{PoissonDeviate}).
%implemented as the
%code{UniformDeviate}, \code{GaussianDeviate} and
%code{PoissonDeviate} classes, respectively.

These fundamentally important distributions form the basis of noise
models in \galsim, but they are also quite useful for generating
random variates for various galaxy or PSF parameters in large simulations.
Additionally, we include some other distributions that are more useful
in this latter context rather than for noise models:
the binomial distribution, the Gamma
distribution, the Weibull distribution (a generalization of the
exponential and Rayleigh distributions) and the $\chi^2$ distribution
\citep[see, e.g.,][]{johnsonetal05}.

Finally, \galsim\ also supports sampling from an arbitrary probability
distribution function, supplied by the user in the form of either a
lookup table and interpolating function, or a callable Python function
with a specified min and max value range.  This is the implemented by
the \code{DistDeviate} class.

\subsection{Uncorrelated noise models}
The majority of noise models that \galsim\ implements generate
uncorrelated noise, i.e.\ noise that takes a statistically independent
value from pixel to pixel.  
The simplest is the \code{GaussianNoise} class,
which adds values drawn from $\mathcal{N}(0,
\sigma^2)$ to an image.  The noise is thus spatially stationary,
{\edit i.e.\ having a constant variance throughout the image,}
as well as being uncorrelated.

The \code{PoissonNoise} class adds Poisson noise ${\rm
  Pois}(\lambda_i)$ to the $i$-th pixel of the image, where
$\lambda_i$ corresponds to the mean number of counts in that
pixel (an optional sky level can be supplied when simulating
sky-subtracted images).  The
\code{PoissonNoise} model is not stationary, since the variance varies 
across the image according to the different pixel fluxes.

The \code{CCDNoise} class provides a good approximation to the
noise found in normal CCD images.  The noise model has two components: Poisson noise
corresponding to the expected photon counts $\lambda_i$ in each pixel
(again with optional extra sky level) and stationary Gaussian
read noise $\mathcal{N}(0, \sigma^2)$.  It is also possible to
specify a gain value in photons/ADU (analog-to-digital units)\footnote{
  Normally the gain is given in e-/ADU, but in \galsim\ we
  combine the effect of the gain with the quantum efficiency,
  given in e-/photon
}, in which case the Poisson noise applies to the photon counts, but
the image gets values in ADU.

The \code{VariableGaussianNoise} class implements a non-stationary
Gaussian noise model, adding noise drawn from $\mathcal{N}(0,
\sigma_i^2)$, using a different variance for each pixel $i$ in the image.

Finally, any random deviate from \S\ref{sect:deviates}
can be used as the basis for a simple \galsim\ noise model, using
the \code{DeviateNoise} class, which draws from the supplied random
deviate independently for each pixel to add noise to the image.

\subsection{Correlated noise}
Astronomical images may contain noise that is spatially correlated.
This can occur at low levels due to detector crosstalk
\citep[e.g.][]{mooreetal04,antilogusetal14} or the use of pixel-based
corrections for charge transfer inefficiency (CTI,
\citealt{2010MNRAS.401..371M}), and it can be strikingly 
seen in images that have been created through the coaddition of two or
more dithered input images. 

Drizzled \hst\ images \citep[][]{2002PASP..114..144F,koekemoeretal02}
often have correlated noise due to the assignment of flux from single input
pixels to more than one output pixel.  The drizzled \hst\ COSMOS
survey images 
\citep[see][]{koekemoeretal07} used as the basis
for empirical galaxy models described in \S\ref{sect:shera}
contain significant inter-pixel noise
correlations. For faint objects, such as the galaxies typically of
interest for weak lensing, such noise is well approximated as a set
of correlated Gaussian random variables.

\subsubsection{Statistics of correlated Gaussian noise}\label{sect:corrnoise}
The statistical properties of correlated Gaussian noise
are fully described by a covariance matrix.  If we denote a noise field
as $\eta$, we may define a discrete pixel-noise autocorrelation
function
\begin{equation}\label{eq:corrfunc}
\xi_{\rm noise}[n, m] = \left\langle \eta[i,j]
  \eta[i',j']\right\rangle_{i'-i=n, \: j'-j=m},
\end{equation} 
where the angle brackets indicate the average over all pairs of pixel
locations $[i,j]$ and $[i',j']$ for which $i'-i=n$ and $j'-j=m$.
(Here we use square brackets for
functions with discrete, integer input variables.)  Therefore $n$, $m$
denote the integer separation between pixels in the positive $x$ and
$y$ directions, respectively.  All physical $\xi_{\rm noise}[n,m]$ functions
have two-fold rotational symmetry, so that $\xi_{\rm noise}[-n, -m] =
\xi_{\rm noise}[n, m]$, and are peaked at $n=m=0$.  If the noise is
stationary, i.e.\ its variance and covariances do not depend upon
location in the image, then $\xi_{\rm noise}[n, m]$ fully determines
the covariance matrix for all pairs of pixels in an image.

For any physical, discrete pixel-noise correlation function $\xi_{\rm
  noise}[n,m]$ there is a positive, real-valued noise power spectrum
$P_{\rm noise}[p, q]$ that is its Fourier counterpart (i.e.\ the two
are related by a DFT).  
We use $p$, $q$ to label array locations in Fourier space, and $n$, $m$ in
real space.

The function $P_{\rm noise}[p, q]$ can be used to generate random
noise as a Gaussian field.  We may construct an array
$\fourier{\eta}[p,q]$ as
\begin{equation}\label{eq:corrnoisegen}
\fourier{\eta}[p, q] = \sqrt{\frac{P_{\rm noise}[p, q] N}{2}} \left\{
  \mathcal{N}(0, 1) + {\rm i}\mathcal{N}(0, 1)\right\}
\end{equation}
where samples from the standard Gaussian distribution $\mathcal{N}(0,
1)$ are drawn independently at each location $p$, $q$ (subject to the
condition that $\fourier{\eta}[p,q]$ is Hermitian, see below), and $N$
is the total array dimension in either direction (we assume square
arrays for simplicity).  The inverse DFT, as defined in
\eqn{eq:dftnumpy}, is then applied to generate a real-valued noise
field realization in real space $\eta[n, m]$.  The enforced Hermitian
symmetry of $\fourier{\eta}[p, q]$ is exploited to enable the use of
efficient inverse-real DFT algorithms.  It can be seen that the
resultant $\eta[n, m]$ will have the required discrete correlation function
$\xi_{\rm noise}[n,m]$.

\subsubsection{Representing correlated Gaussian noise}\label{sect:repcorrnoise}

Stationary correlated noise is modelled and generated in \galsim\ by
the \code{CorrelatedNoise} class.  Internally the noise correlation
function $\xi_{\rm noise}[n, m]$ is represented as a 2D distribution
using the same routines used to describe astronomical objects,
described in \S\ref{sect:objects}.  Using these to calculate the noise
power spectrum, and then applying the expression in
\eqn{eq:corrnoisegen} followed by an inverse DFT, the
\code{CorrelatedNoise} class can be used to generate a Gaussian random
field realization with any physical power spectrum or correlation
function.

%\subsubsection{Operations on Correlated Gaussian Noise}\label{sect:corrnoiseopps}

Because the \code{CorrelatedNoise} class internally uses a regular
\galsim\ surface brightness profile to describe the correlation
function, it is easy to effect the transformation, as in
\S\ref{sect:transformations}, since the noise is transformed in the
same manner.  Similarly, multiple noise fields may be summed (as
  will be required in \S\ref{sect:whitening}), since the resultant
$\xi_{\rm noise}$ is given by the sum of the individual noise
correlation functions.

The effect of convolution of a noise model by some kernel is not quite the
same as what happens to surface brightness.  Instead,
$\xi_{\rm noise}$ should be convolved by the autocorrelation of the kernel.
The ability to describe transformations,
combinations and convolutions of noise is valuable, as will be
discussed in \S\ref{sect:whitening}.

\subsubsection{Describing discrete correlation functions in $\mathbb{R}^2$}\label{sect:describexi}
The description of the discrete correlation function $\xi_{\rm noise}
[n,m]$ as a distribution in $\mathbb{R}^2$ is worthy of some
discussion. As will be seen in \S\ref{sect:whitening} it is important
to be able to take into account the finite size of pixels in images
that may contain correlated noise, particularly when these images are
being used to form galaxy models that will ultimately be rendered onto
a pixel grid of a different scale.  We therefore wish to represent
$\xi_{\rm noise} [n,m]$ as a function of continuous spatial
variables, $\xi_{\rm noise}(x, y)$ for which we require
\begin{equation}\label{eq:corrnoisecontreq}
\xi_{\rm
  noise}(n s, m s) = \xi_{\rm noise} [n,m],
\end{equation}
where $s$ is the pixel scale.

Let us consider an image containing noise.  In the absence of any
smoothing of the image, it is described mathematically
as an array of delta function samples (located at the pixel centres)
convolved by the pixel response.
This could also be considered the formally correct
model to use for $\xi_{\rm noise}(x, y)$: its values, like those in an
image, should change discontinuously across the boundaries between
pixels as separation increases continuously.

However, this description of $\xi_{\rm noise}(x, y)$ presents
difficulties when performing certain of the operations described in
\S\ref{sect:repcorrnoise}, e.g.\ convolutions, or transformations
followed by a convolution.  The presence of sharp discontinuities at
every pixel-associated boundary makes the necessary Fourier space
representation of $\xi_{\rm noise}(x, y)$ extremely non-compact, and
prohibitive in terms of memory and computing resources.

Therefore, to reduce the computational costs of handling Gaussian
correlated noise to tolerable levels, \galsim\ adopts bilinear
interpolation between the sampled values $\xi_{\rm noise}(n
s, m s)$ to define the continuous-valued function
$\xi_{\rm noise}(x, y)$ within the \code{CorrelatedNoise} class.

In practice, the impact of this efficiency-driven approximation may
often be small.  The primary use case for the representation of
correlated noise within \galsim\ is in handling noise 
in models derived from direct observations of galaxies (see
\S\ref{sect:shera} and \S\ref{sect:whitening}).  In these
applications, both the applied PSF and the pixel scale on which the
final image will be rendered are typically significantly larger than
the pixel scale of the input images.  The use of bilinear
interpolation represents a small additional convolution kernel in
these cases.

\subsection{COSMOS noise}\label{sect:cosmosnoise}
The \hst\ COSMOS galaxy images made available for download with \galsim\ for
use in the \code{RealGalaxy} class
\citep[see][]{koekemoeretal07,leauthaudetal07,2012MNRAS.420.1518M}
contain noise that is spatially correlated due to drizzling.  
To construct a model for this correlated noise, we estimated the
discrete pixel noise correlation function in $\sim$100 
rectangular regions selected from empty sky in the public, unrotated,
ACS-WFC F814W science images (0.03 arcsec/pixel) from COSMOS
\citep[described in detail by][]{koekemoeretal07,leauthaudetal07}.

\begin{figure}
%\epsscale{1.0}
\includegraphics[width=\columnwidth]{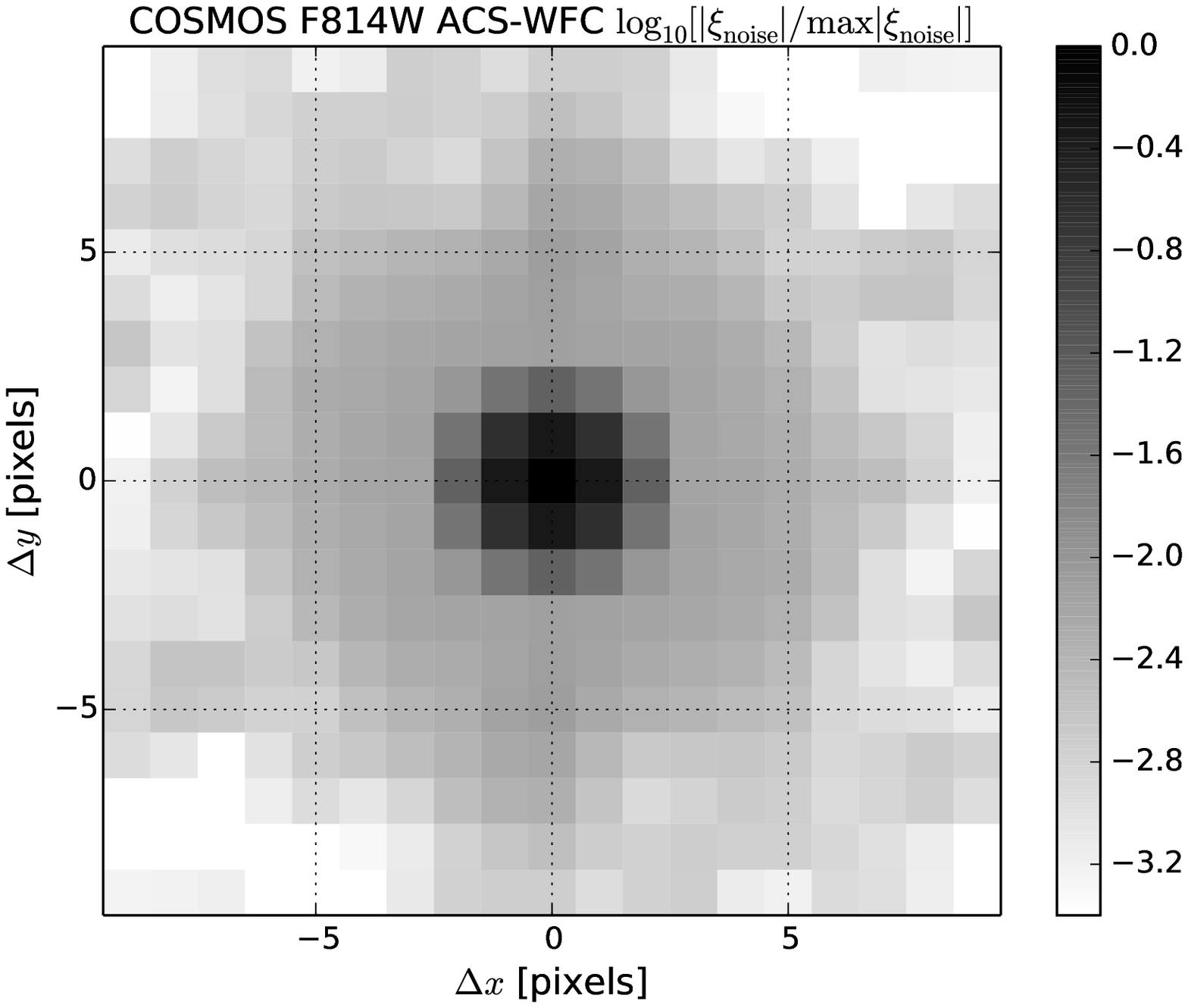}
\caption{Illustration of the \galsim\ estimate of the discrete pixel
  noise correlation function $\xi_{\rm noise}$ in the 0.03 arcsec/pixel,
  unrotated COSMOS F814W science images of
  \citet{leauthaudetal07}, made by averaging estimates from blank sky
  fields as described in \S\ref{sect:cosmosnoise}. The plot depicts
  the log of the correlation function normalized so that $\xi_{\rm noise}[0,
  0]=1$ for clarity, and shows only the central region corresponding
  to small separations $\Delta x$, $\Delta y$, which dominate
  covariances.\label{fig:cosmoscf}}
\end{figure}

Each rectangular region varied in size and dimensions, being chosen so
as not to include any detected objects, but were typically square
regions $\sim$50-200 pixels along a side.  The mean noise correlation
function in these images (calculated by summing over each individual
estimate, see \S\ref{sect:repcorrnoise}) provides an estimate of the
mean noise correlation function in COSMOS F814W galaxy images used by
\galsim\ as the basis for galaxy models.  The resulting estimate of
$\xi_{\rm noise}$ is depicted in Fig.~\ref{fig:cosmoscf}.

This model is included with the \galsim\ software for general use in
handling images that contain noise like those in the COSMOS F814W
images, and is accessible via the \code{getCOSMOSNoise()} convenience
function.

\subsection{Noise whitening}\label{sect:whitening}
All images of galaxies from telescope observations will contain noise.
As pointed out by \citet{2012MNRAS.420.1518M}, this noise will become
sheared and correlated (in an anisotropic way) by the action of the
reconvolution algorithm described in \S\ref{sect:shera}, which can
result in a significant systematic offset in the results from galaxy
shear simulations.  \citet{2012MNRAS.420.1518M} found that the
presence of correlated noise caused $\sim$1\% level errors in the
determination of calibration biases for weak lensing shear for
high-$S/N$ galaxies; the effect is worse at lower $S/N$, and cannot be
ignored for high-precision shear simulations. 

\galsim\ addresses the presence of correlated noise in simulated
images, such as the sheared, convolved noise fields created by
the reconvolution algorithm, through a process referred to as ``noise
whitening''.  This technique adds further noise to images, with a
correlation function chosen to make the combination of the two noise
fields approximately uncorrelated and stationary (aka ``white'').

We consider an image containing correlated Gaussian noise that we
label as $\eta$.  We assume $\langle \eta \rangle = 0$ and we model
$\eta$ as stationary across the image so that its covariance matrix is
fully described by its correlation function $\xi_{\rm noise}[n, m]$ or
power spectrum $P_{\rm noise}[p, q]$.  We add additional
whitening noise which we
label $\eta_{\rm whitening}$, giving a combined noise field
\begin{equation}\label{eq:etatotal}
\eta_{\rm total} = \eta + \eta_{\rm whitening}.
\end{equation}
The statistics of $\eta_{\rm whitening}$ are also
determined by its power spectrum, $P_{\rm whitening}[p, q]$, and $\eta_{\rm
  whitening}$ will be physically realizable provided
that
\begin{equation}
P_{\rm whitening}[p, q] \ge 0
\end{equation}
for all $p$, $q$.  The power spectrum of the combined noise field is simply
\begin{equation}
P_{\rm total} = P_{\rm noise} + P_{\rm whitening}.
\end{equation}

We may choose $P_{\rm whitening}$ so that $\eta_{\rm total}$ is
uncorrelated by requiring $P_{\rm total}$ to be a constant (a flat power
spectrum corresponds to a delta function correlation function, which
is white noise).  This,
and the requirement $P_{\rm whitening} \ge 0$, is satisfied by
choosing
\begin{equation}
P_{\rm whitening}[p, q] = \max_{p, q}\left\{ P_{\rm noise}[p, q]
\right\} - P_{\rm noise}[p, q].
\end{equation}
In practice, \galsim\ adds a small amount of
additional variance so that $P_{\rm whitening} > 0$, defining
\begin{align}\label{eq:pwhitening}
  P_{\rm whitening}[p, q] &= (1 + \epsilon) \times \max_{p, q}\left\{ P_{\rm noise}[p, q]
  \right\} \nonumber \\
&\qquad - P_{\rm noise}[p, q]
\end{align}
where $\epsilon = 0.05$ by default.  The whitening noise $\eta_{\rm
  whitening}$ can then be added following the prescription described
in \S\ref{sect:corrnoise}, using the expression for $P_{\rm
  whitening}$ above.  At a relatively small fractional cost in
  additional variance, the `headroom' parameter $\epsilon$ effectively
  guarantees $P_{\rm whitening}[p, q] \ge 0$.  Without $\epsilon$ this
  important condition might not be met, due to numerical rounding and
  approximations in the description of $\xi[n, m]$ (see
  \S\ref{sect:describexi}).

\begin{figure}
%\epsscale{1.0}
\includegraphics[width=\columnwidth]{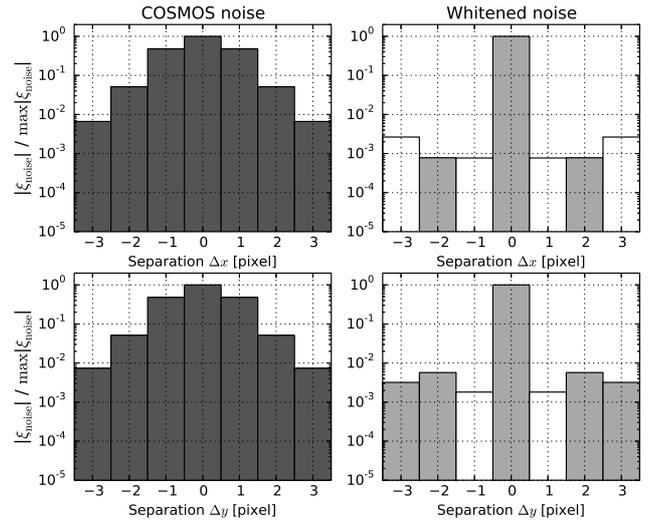}
\caption{Left panels: cross section through the discrete pixel noise
  correlation function $\xi_{\rm noise}$ for the COSMOS F814W blank
  sky fields described in \S\ref{sect:cosmosnoise} and shown in
  Fig.~\ref{fig:cosmoscf}, normalized to unit zero-lag variance; the
  upper left panel shows COSMOS noise correlations in the direction
  $\Delta y=0$, and the lower left panel shows COSMOS noise
  correlations in the direction $\Delta x=0$.  Right panels: cross
  section through the normalized discrete pixel noise correlation
  functions for a single $\sim$400~pixel~$\times$~300~pixel patch of
  blank sky, taken from COSMOS, to which further whitening noise has
  been added using the \S\ref{sect:cosmosnoise} model for $\xi_{\rm
    noise}$, using the procedure described in \S\ref{sect:whitening}.
  As for the left panels, the upper right and lower right panels show
  correlations along the directions $\Delta y = 0$ and $\Delta x = 0$,
  respectively. Unfilled bars represent negative correlation function
  values.\label{fig:cosmoswhitening}}
\end{figure}

Fig.~\ref{fig:cosmoswhitening} shows cross sections through $\xi_{\rm
  total}[n, m]$ from a single
$\sim 400\times 300$~pixel patch of blank sky taken from
COSMOS F814W ACS-WFC images, to which further whitening noise
has been added.  $P_{\rm whitening}$ was
calculated using the model for $\xi_{\rm noise}$ (and thus $P_{\rm noise}$)
described in \S\ref{sect:cosmosnoise} along with
\eqn{eq:pwhitening}.  Here \galsim\ noise whitening has lowered
interpixel covariances to $\sim$10$^{-2}$--10$^{-3}$ of the zero-lag
variance in the whitened image.

We have tested that applying this whitening procedure to COSMOS
noise fields that have been sheared and convolved (like those produced
by the reconvolution algorithm of \S\ref{sect:shera}) similarly
reduce the noise correlations by 2--3 orders of
magnitude, making them statistically undetectable in all but the
largest simulated images.

 The price paid for noise whitening is additional
  variance in the output image.  The noise added in \eqn{eq:etatotal}
  is an additional stochastic degradation of an image.  For the COSMOS
  F814W noise field whitened in Fig.~\ref{fig:cosmoswhitening}, the
  final noise field therefore has a variance that is greater than the
  variance in the original images.  However, the correlated noise in a
  COSMOS image itself results in a lower effective $S/N$ for estimates
  of object properties such as flux, or shape, relative to the case of
  an image with the same variance but with pure white noise \citep[for
  a discussion of these effects, which depend on object size and the
  properties of the correlated noise, see,
  e.g.,][]{casertanoetal00,leauthaudetal07}.  The effect of additional
  whitening of the noise in such an image, therefore, reduces the
  $S/N$ of estimates of object properties by less than implied by the
  ratio of pre- and post-whitening image variances.  The amount of
  information lost in whitening will depend on the property in
  question, the object profile, and the noise correlations in the
  original image.

In the case of noise fields that have been sheared, convolved by a
PSF, and then rendered on an image with a larger pixel scale (as
described in \S\ref{sect:shera}), the noise added in whitening depends
not only on the original noise correlation function, but also on the
applied shear, PSF, and final pixel scale.  Interestingly, it is found to be
typically substantially lower in overall variance than in the case of
whitening raw COSMOS noise at native resolution (the case described
above and shown in Fig.~\ref{fig:cosmoswhitening}).

On investigation, this was found to be due to a combination of
effects.  The shear and convolution both increase the noise
correlations at large scales.  However, the final image rendering
typically has a pixel scale $s$ significantly larger than the 0.03
arcsec/pixel of the COSMOS images, which means that the values of the
correlation function at integer multiples of $s$ in separation are
often smaller than in the original COSMOS image.  Also, convolution
with a broad PSF reduces the overall noise variance considerably,
being essentially a smoothing kernel.  These effects combine
fortuitously in practice: \citet{handbook} found that only a
relatively modest amount of whitening noise is required to treat the
sheared, correlated noise in images generated by the reconvolution
algorithm of \S\ref{sect:shera} with a ground-based target PSF and
pixel scale.

In \galsim\, the \code{Image} class has a method
\code{whitenNoise} that applies the above whitening algorithm
to a given image.
This method also returns the theoretical
estimate of the noise variance in the final whitened image (namely
$(1 + \epsilon) \max_{p, q}\left\{ P_{\rm noise}[p, q] \right\}$).
In addition, there is a method \code{symmetrizeNoise} that imposes
$N$-fold symmetry (for even $N\ge 4$) on the noise field rather than
making it entirely white.  Typically far less noise must be added by
\code{symmetrizeNoise} compared to \code{whitenNoise}, which makes it
a useful option for those applications that do not require fully white
noise.

The \code{RealGalaxy} class has an attribute \code{noise} that
keeps track of the estimated correlated noise in the profile.
When this object is transformed or convolved, the new object
will also have a \code{noise} attribute that has had the
appropriate transformation done to propagate the noise in the new profile.
This allows \galsim\ to automatically track the effect of
transformations on noise fields without much user input required.
The final profile's \code{noise} attribute can then be used
to whiten the rendered image.
Furthermore, users can add a \code{noise} attribute themselves to any
object they create and get the same behaviour.

In a larger image with many scattered, possibly overlapping galaxies, the correlated noise
in the original images may differ, further complicating the task of
getting uniform white noise in the final image.
\galsim\ handles this by drawing each galaxy on its own
postage stamp image and whitening that individually.  Still, the final
image has a complicated patchwork of noise with different regions having different
variances, but it is relatively straightforward in \galsim\ to keep track of 
these noise levels: {\edit the configuration file interface described
  in \S\ref{sect:gsstructure} allows this to be handled automatically.}
Then the \code{VariableGaussianNoise} class
can add a variable amount of noise to each pixel to bring the noise up to
a uniform value across the entire image.

%
% Section 8
% hsm
%

\section{Shear estimation}\label{sect:hsm}

\galsim\ includes routines
for estimation of weighted moments of galaxy shapes and 
for estimation of PSF-corrected shears.
The code for these
routines was originally introduced by \cite{2003MNRAS.343..459H} (hereafter HS03),
including an entirely new PSF correction method known as
re-Gaussianization.  These routines were later tested, improved, and
optimized in subsequent work
\citep{2005MNRAS.361.1287M,2012MNRAS.420.1518M,2012MNRAS.425.2610R}.
{\edit The code was developed as a free-standing C package named \textsc{hsm}, and
was released publicly under an open source license
simultaneously with its incorporation into \galsim.}

The function \code{FindAdaptiveMoments} implements the ``adaptive moments''
method described in \S2.1 of HS03,
based on \cite{BJ02}.  It iteratively computes the moments of the
best-fitting elliptical Gaussian, using the fitted elliptical
Gaussian as a weight function.

The function \code{EstimateShear} implements
multiple methods that were tested in HS03:
\begin{enumerate}
\item The PSF correction method described in Appendix C of
  \cite{BJ02} tries to analytically account for the effect of
  the kurtosis of the PSF
  (in addition to the second order moments) on the galaxy shapes.
\item The ``linear'' method (described in Appendix B of HS03) is a
  variant of the previous method that 
  accounts for the first-order departure of both the PSF and galaxy from
  Gaussianity.
\item The ``re-Gaussianization'' PSF correction method described in
\S2.4 of HS03 models the PSF as a Gaussian plus a residual, and
subtracts from the observed image an estimate of the residual
convolved with the pre-seeing galaxy.  The resulting image has roughly
a Gaussian PSF, for which the methods described above can be used for
the final PSF correction.
\item A specific implementation of the KSB method
  \citep{1995ApJ...449..460K,1997ApJ...475...20L}, as described in
  Appendix C of HS03, is also provided.
\end{enumerate}

These routines work directly on \galsim\ images.
{\edit The first three PSF correction methods output a measure of 
per-galaxy shape called a distortion, which for an ensemble of galaxies can be 
converted to shear estimates via a responsivity factor (cf.\ \citealt{BJ02}). The
outputs are clearly labeled as being distinct from per-galaxy shear
estimates, such as those output by the KSB routine.}

As is typical for all shape measurement algorithms, there are several 
tunable parameters for the above methods and for moment estimation
overall (with adaptive moment estimation playing a role in all but the
KSB method of PSF estimation).  \galsim\ therefore includes two ways
of tuning these parameters.
There is an \code{HSMParams} structure that allows users 
to change parameters that affect how the submodule works overall.
Other parameters are given as arguments to the specific functions.

Aside from the obvious utility of being able to operate directly on
images in memory, rather than files,
the versions of these routines in \galsim\
come with some improvements over the free-standing C code.
These include an intuitive and more flexible user interface;
optimization of convolutions, the \code{exp} function (via reduction of the
number of function calls), and Fourier transforms (using the
\fftw\ package, \citealp{FFTW05}); a clean way of
including masks within images that is easily extensible to variable
weight maps; and a fix for a bug that caused the original C code to
work incorrectly if the input PSF images were not square.

%
% Section 9
% validation
%

\section{Numerical validation}\label{sect:validation}

In this Section we describe the investigations that were
undertaken to validate the accuracy of \galsim\ image simulations.
Although an exhaustive validation of the rendering of every
combination of galaxy/PSF profiles and observing conditions is 
impractical, certain key aspects of \galsim\ performance are shown
here.  Emphasis is placed on confirming that \galsim\ meets the
stringent requirements on image transformations for lensing 
shear and magnification simulation.

In particular, our metric for validating that the rendered
images are sufficiently accurate is based on
measurements of the size and ellipticity of the rendered profiles,
calculated using the adaptive moment routines described in
\S\ref{sect:hsm}.

We define the following ``STEP-like'' \citep[see][]{heymansetal06} models for
the errors in the estimates of
object ellipticity $g_1$ and $g_2$ and size $\sigma$:
\begin{align}
\label{eq:mc1}
  \Delta g_i &= m_i g_i + c_i, \\
\label{eq:mc2}
  \Delta \sigma &= m_\sigma \sigma + c_\sigma,
\end{align}
where $i = 1,2$.  The method of estimating the errors $\Delta g_i$ and
$\Delta \sigma$ varies for each of the validation tests described
below, but a common component is adaptive moments estimates of
rendered object shapes from images (see \S\ref{sect:hsm}).  We will
use the formulae above when describing the nature of the errors in
each test.

As discussed in \citet{handbook}, a well-motivated target for
simulations capable of testing weak lensing measurement is to
demonstrate consistency at a level well within the overall
requirements for shear estimation systematics set by
\emph{Euclid} \citep[e.g.][]{2013MNRAS.431.3103C,masseyetal13}: $m_i \simeq 2\times
10^{-3}$ and $c_i \simeq 2\times 10^{-4}$.  Such values also place
conservative requirements on galaxy size estimation, as the
signal-to-noise expected for cosmological magnification measurements
has been estimated as $\lesssim 50 \%$ relative to shear 
\citep[e.g.][]{waerbeke10,2012ApJ...744L..22S,duncanetal14}.

Only if these stringent \emph{Euclid} conditions are met comfortably
will simulations be widely usable for testing weak lensing shear
estimation, and other precision cosmological applications, in the
mid-term future.  For each validation test we therefore 
require that \galsim\ {\edit can produce images showing} discrepancies that
are a factor of 10 or more below the \emph{Euclid} requirements,
i.e.\ $m_x < 2
\times 10^{-4}$, $c_x < 2 \times 10^{-5}$, where $x = 1,
2, \sigma$ corresponding to $g_1$, $g_2$ and $\sigma$, respectively.
{\edit Our intention is to show that rendering accuracy improves with
  more stringent rendering parameter settings, and can be raised to meet the
  requirement above.  \galsim\ default settings for these parameters,
  however, are chosen to reflect a balance between speed and accuracy
  for general use.
  Although these defaults typically take values that guarantee meeting 1/10th
  \emph{Euclid} requirements, this is not exclusively the case (see
  \S\ref{sect:validate-reconv}).}

{\edit For our tests, we use comparisons of rendering methods using
  adaptive moments measurements. Although there are inaccuracies
  inherent in such measurements in an individual sense, we take care
  to construct our comparisons to be sensitive only to differences
  in the image rendering. For example, the tests always compare
  noise-free or \emph{extremely} low noise images, with the same
  underlying galaxy model.  Noise bias and model bias therefore affect
  each test subject in the same way, and differences are
  due solely to how the test images were rendered.}

The tests in this Section were conducted over a period of extended
validation of the \galsim\ software between July 2013 and the time of
writing this paper.  During this time period, corresponding approximately to
versions 1.0 and 1.1 of \galsim, the routines for rendering objects
did not change significantly (except where modifications were found
necessary to meet the validation criteria on $m_x$ and $c_x$ defined above).

\subsection{Equivalence of DFT rendering and photon shooting}
\label{sect:validate-dft}

One of the principal advantages of the photon shooting method
(see \S\ref{sect:photon}) is that the implementations of the
various transformations described in \S\ref{sect:transformations}
are very simple.  Photons are just moved from their original position
to a new position.  Convolutions are similarly straightforward.
On the other hand, DFT rendering (see \S\ref{sect:dft}) needs
to deal with issues such as band limiting and aliasing due to
folding (cf.\ \S\ref{sect:bandlimitingfolding}).

Thus a powerful test of the accuracy of our DFT implementation
is that the the two rendering methods give equivalent results
in terms of measured sizes and shapes of the rendered objects.
An unlikely conspiracy of complementing errors on
both sides would be required for this test to yield false positive
results.

Of all the objects in Table~\ref{tab:analytic-objects},
\sersic\ profiles are the most numerically challenging to render using
Fourier methods.  Especially for $n\gtrsim 3$,
the profiles are extremely cuspy in the centre and have very broad wings,
which means that they require a large dynamic range of $k$ values
when performing the DFT.  They therefore provide a good test of our
choices for parameters such as \code{folding\_threshold} and
\code{maxk\_threshold} (see \S\ref{sect:tolerances}) as well as general
validation of the DFT implementation strategies.

For our test, we built \code{Sersic} objects with \sersic\ indices
in the range $1.5 \le n \le 6.2$.
The half-light
radii and intrinsic ellipticities $|g^{(s)}|$ were drawn from a
distribution that matches observed COSMOS galaxies, as described in
\citet{handbook}.  The galaxies were then rotated to a random 
orientation, convolved with a COSMOS-like PSF (a circular \code{Airy} profile),
and then rendered onto an image via both
DFT and photon shooting.

The error estimates were taken to be the difference between the
adaptive moments shape and size estimates from the two images:
\begin{align}
\Delta g_i &= g_{i, \text{DFT}} - g_{i, \text{phot}} \\
\Delta \sigma &= \sigma_{\text{DFT}} - \sigma_{\text{phot}}
\end{align}
For each
galaxy model, {\edit we made} multiple trial photon shooting images,
each with very high $S/N$ to avoid noise biases
($10^7$ photons shot per trial image).  The mean and standard error of
$\Delta g_i$ and $\Delta \sigma$ from these trials were used to
estimate values and uncertainties for $m_{x, \text{DFT}}$ and $c_{x,
\text{DFT}}$ using standard linear regression.

\begin{figure}
\begin{center}
%\epsscale{1.0}
\includegraphics[width=\columnwidth]{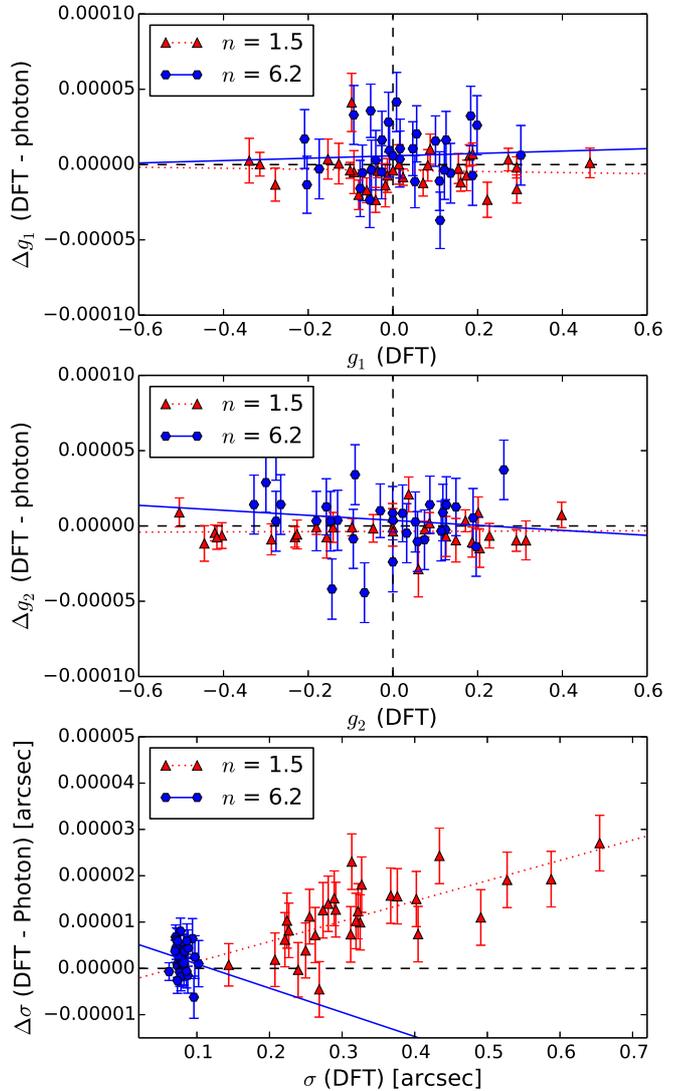}
%\includegraphics[width=\columnwidth][width=0.48\columnwidth]{figs/sersic_highn_basic_zoomin_sigma.png}
% This figure came from GalSim repo, master branch
% devel/external/test_sersic_highn/plots/.  Later converted from PNG
% to PDF using Preview.
\caption{\label{fig:sersic-validation} Difference between measured
  shears (upper panel: $g_1$; central panel: $g_2$; lower panel:
  $\sigma$) for \sersic\ profiles simulated using the photon-shooting
  and DFT rendering methods, plotted against the (shot noise free) shear and size
  measured from the DFT image. Results are shown for 30 galaxies with
  realistic size and shape distribution, and \sersic\ index values
  $n=1.5,6.2$.  (Note that the `peakiness' of the high-$n$ profiles results in their
  low $\sigma$ estimates. {\edit The two samples share the same
    distributions of half light radii.})  The best-fitting lines are shown, and estimates of
  the slopes $m_{x, \text{DFT}}$ for these and other values of $n$ are
  plotted in Fig.~\ref{fig:mvsn}.}
\end{center}
\end{figure}

\begin{figure}
\begin{center}
%\epsscale{1.0}
\includegraphics[width=\columnwidth]{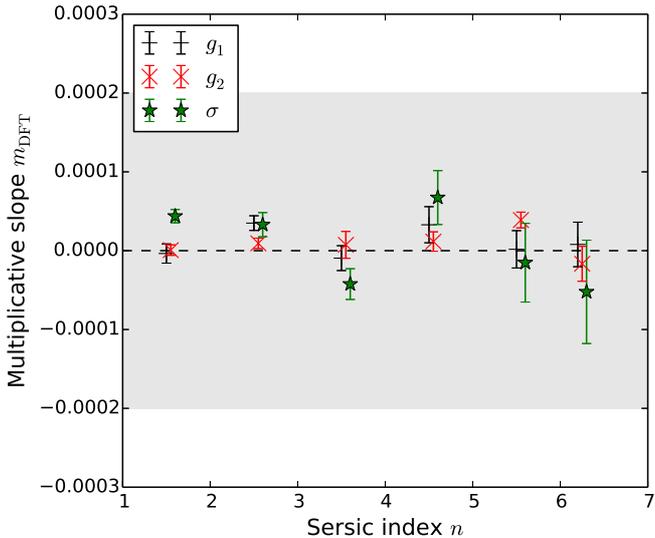}
\caption{Estimates of $m_{1, \text{DFT}}$, $m_{2, \text{DFT}}$ and
  $m_{\sigma, \text{DFT}}$, corresponding to rendering-induced discrepancies
  in ellipticity $g_1$, $g_2$ and size $\sigma$, respectively, as a
  function of \sersic\ index $n$.  These slope parameters are defined
  by \eqnb{eq:mc1}{eq:mc2} for the differences between
  measurements from DFT and photon-shooting-rendered images of
  \sersic\ profiles.  The shaded region shows the
 target for \galsim\ based on not exceeding one tenth of 
 weak lensing accuracy requirements for Stage IV
  surveys such as \emph{Euclid} (see
  \S\ref{sect:validation}).\label{fig:mvsn}}
\end{center}
\end{figure}

Differences between shape and size estimates are illustrated in
Fig.~\ref{fig:sersic-validation}, for $n = 1.5$ and $n=6.2$.
Fig.~\ref{fig:mvsn} shows derived estimates of $m_{x, \text{DFT}}$ for
these and other \sersic\ indices tested.   Tolerances
are met on $m$-type biases.  It was found that
$c$-type additive biases were consistent with zero for all $n$ indices.

\begin{figure}
\begin{center}
%\epsscale{1.0}
\includegraphics[width=\columnwidth]{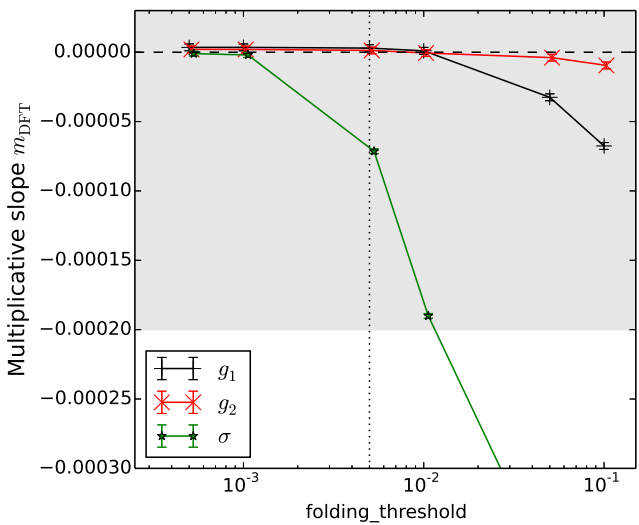}
\caption{Estimates of $m_{1, \text{DFT}}$, $m_{2, \text{DFT}}$ and
  $m_{\sigma, \text{DFT}}$, corresponding to rendering-induced
  discrepancies in ellipticity $g_1$, $g_2$ and size $\sigma$,
  respectively, as a function of the \code{GSParams} parameter
  \code{folding\_threshold}. Each point shows the average from the
  randomly-selected sample of 270 unmodified COSMOS galaxy models
  described in \S\ref{sect:validate-dft}.  The rendering parameter
  \code{folding\_threshold} is described in \S\ref{sect:tolerances}
  and takes a default value of $5\times10^{-3}$, indicated by the
  dotted line.  As for Fig.~\ref{fig:mvsn} these parameters are
  defined by the model of \eqnb{eq:mc1}{eq:mc2}.  The shaded region
  shows the target for \galsim\ based on not exceeding one tenth
    of weak lensing accuracy requirements for Stage IV surveys (see
    \S\ref{sect:validation}).\label{fig:folding_threshold}}
\end{center}
\end{figure}

Fig.~\ref{fig:folding_threshold} shows results from a high-precision
investigation of $m_{x, {\rm DFT}}$ as a function of \code{GSParams}
parameters (see \S\ref{sect:tolerances}), using a randomly selected
sample of 270 galaxies from COSMOS at each parameter value and large
numbers of photons.  Each galaxy was generated in an 8-fold ring test
configuration \citep{nakajimabernstein07} to further reduce
statistical uncertainty.  The plot in Fig.~\ref{fig:folding_threshold}
shows the impact of increasing the \code{folding\_threshold} parameter:
as expected, the rendering agreement decreases as
\code{folding\_threshold} increases, and the representation of object
size is most affected. Analogous results were achieved for many of the
parameters discussed in \S\ref{sect:tolerances}, and the default
\code{GSParams} parameters were found to give conservatively good
performance in all tests.

\subsection{Accuracy of transformed interpolated images}\label{sect:validate-ii}

Interpolated images pose unique challenges for DFT rendering.  As
we discussed in \S\ref{sect:fourierinterpolatedimage}, details such as
the choice of the Fourier-space interpolant and how much padding to
use around the original image significantly affect the accuracy of the
rendering.  {\edit We thus need to validate that these choices can be varied 
to produce sufficiently accurate results.}

For this test, {\edit we randomly selected a sample 6000 \sersic\ models from
the catalogue COSMOS galaxy fits described in \citet{handbook}}.  For
each galaxy in the sample, we constructed a \code{Sersic} object using
the best-fitting parameters in the catalogue, which was then convolved
by a COSMOS-like PSF (a circular \code{Airy} profile).  This profile
was drawn onto an image with 0.03 arcsec resolution (matching the
COSMOS images).  That image was then used to construct an
\code{InterpolatedImage} profile.

{\edit For each object in the sample we therefore had
the same profile modelled both as an analytic convolution
and as an interpolated image.}  Small shears were then applied to both
models and the resulting profiles were drawn onto images using the DFT
rendering algorithm, although without including the integration by a
pixel.  Each profile was convolved by a tiny Gaussian to force
\galsim\ to use the DFT rendering method (as direct rendering would
otherwise be used when omitting the pixel convolution).

Adaptive moments estimates of the change in ellipticity were
seen to be consistent between the parametric model and
interpolated image at better than 1/10th \emph{Euclid} requirements,
for all real and Fourier space interpolants of higher order than
bilinear. Crucially, both were consistent with the known applied
shear. Results for changes in object size were similar. These
tests validated the \code{InterpolatedImage} handling of
transformations for simple input images derived from parametric
profiles (such as a \sersic\ galaxy convolved with a COSMOS PSF).

The test was then extended to the more complicated case of 
\code{RealGalaxy} profiles.  Once again, adaptive moments estimates
of the object ellipticity were made before and after applying a
shear. This was compared to the expected shear, calculable given an adaptive
moments estimate of the intrinsic ellipticity prior to shearing, and the
known applied shear. The \code{RealGalaxy} profile was again convolved
with a tiny Gaussian PSF prior to rendering (see above), but in this
case additional whitening noise was applied to verify that the
presence of sheared, correlated noise in the output was being
appropriately handled.

\begin{figure}
\begin{center}
%\epsscale{1.0}
\includegraphics[width=\columnwidth]{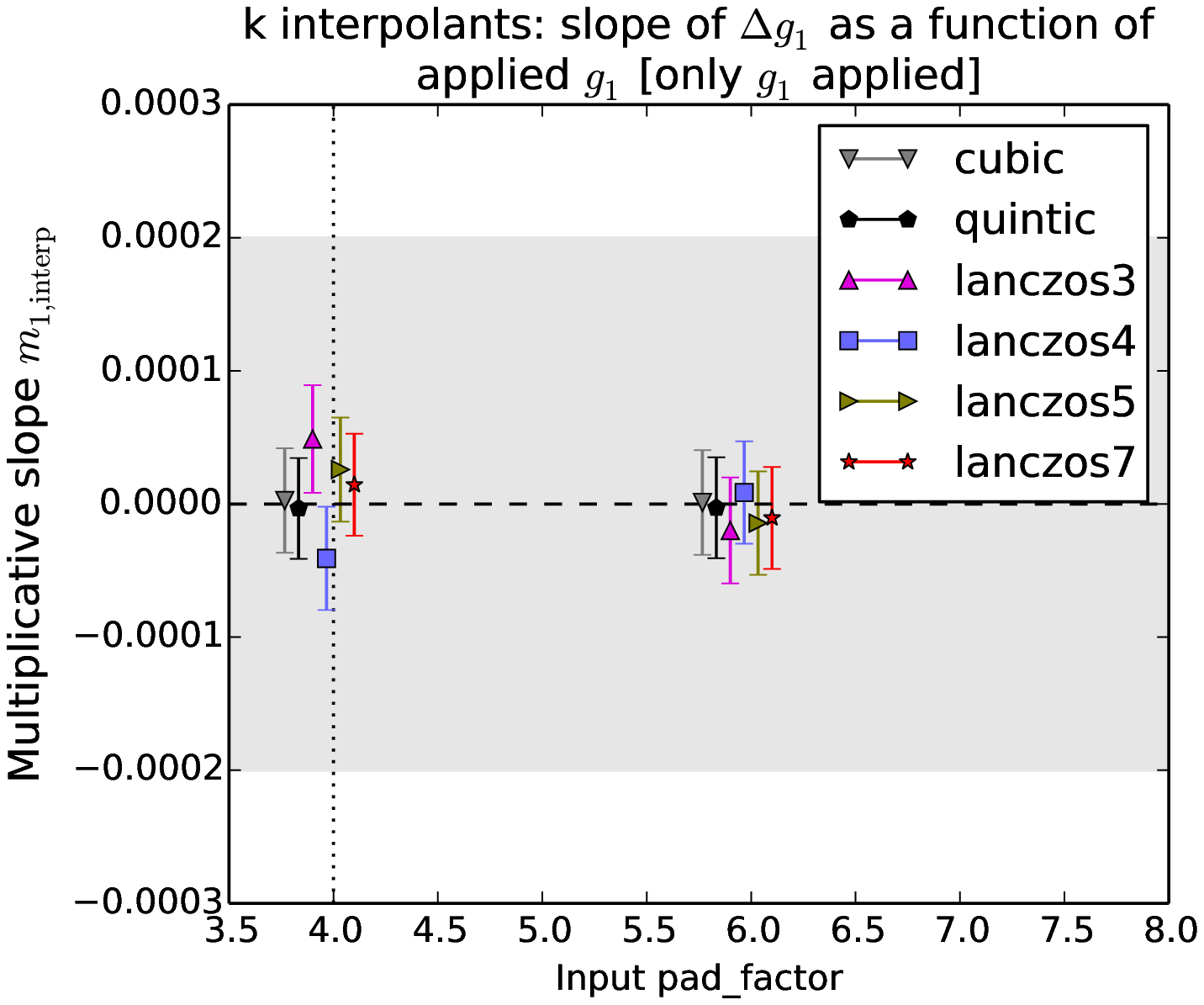}
\includegraphics[width=\columnwidth]{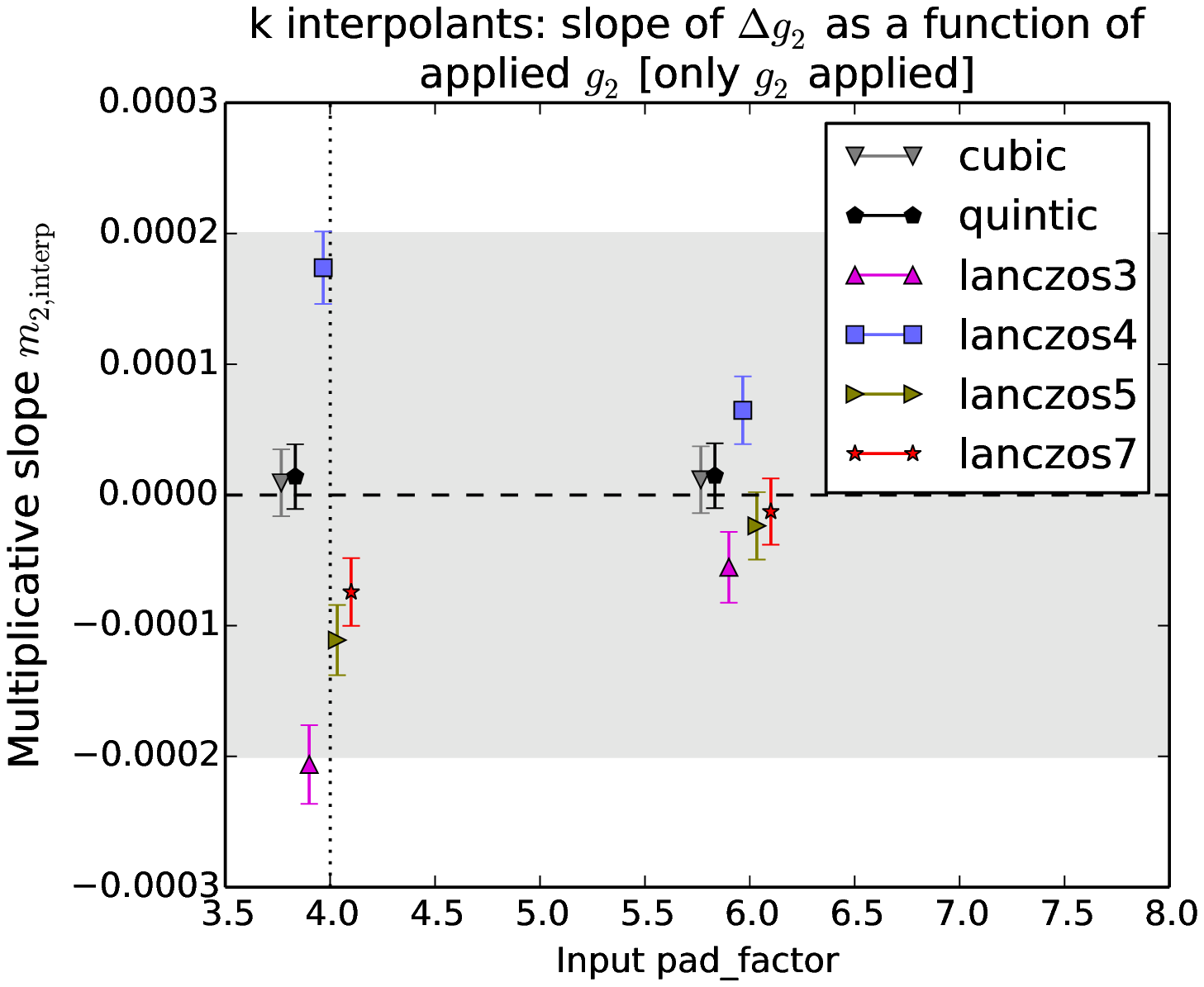}
\caption{Multiplicative bias $m_{i, \text{interp}}$ in the
  relationship between an applied shear and an adaptive moments
  estimate of the observed shear for \code{RealGalaxy} objects, for
  various Fourier space interpolants, e.g.\ cubic, quintic etc., and
  the padding factor (default = 4, indicated by the dotted line).
  These results include the handling of correlated noise via the noise
  whitening procedure described in \S\ref{sect:whitening}. The shaded region
  shows the target for \galsim\ based on not exceeding one tenth
    of weak lensing accuracy requirements for Stage IV surveys (see
    \S\ref{sect:validation}).\label{fig:minterp}}
\end{center}
\end{figure}

This test was repeated for the same sample of 6000 COSMOS galaxy
images. Fig.~\ref{fig:minterp} shows estimates of $m_{i,
  \textrm{interp}}$ for a range of Fourier space interpolants, and for
two values of the \code{pad\_factor} parameter, which specifies how
large a region around the original image to pad with zeroes.  A larger
value for \code{pad\_factor} produces more accurate results, as shown
in Fig.~\ref{fig:minterp}, but is accompanied by large increases in
computation time.

We performed many similar tests to this for a range of transformations
(in various combinations) and input parameters.  In all tests
performed, setting \code{pad\_factor} = 4 and using the two-dimensional quintic
interpolant gave results that satisfied $m_{i, \text{interp}}< 2
\times 10^{-4}$, $c_{i, \text{interp}}< 2 \times 10^{-5}$.
These results justify the choice of these as the defaults in \galsim. 
Values of $c_{i, \text{interp}}$ were found to be extremely
small in general. 

In addition to verifying direct shear response, we also checked for
leakage between shear components, i.e., that applying shear in one
component does not result in an incorrect level of shear in the other
component.  The leakage was found to be consistent with zero and
comfortably within requirements for \code{pad\_factor} values of 4 and
6, for all interpolants tested.

The only {\edit \code{RealGalaxy}} test that did not pass our requirement was
when we simultaneously simulated shears in both components and
a magnification.  In this case we found
$m_{\sigma, \textrm{interp}}\sim (4 \pm 0.4) \times 10^{-4}$,
largely irrespective of the choice of interpolant.  The
source of this bias, which is larger than the target
$2 \times 10^{-4}$, has not yet been
identified. However, as magnification is a cosmological probe that is
typically noisier than shear, this degree of discrepancy is likely to
be tolerable for simulations of magnification in Stage IV survey data.

\subsection{Accuracy of reconvolution}\label{sect:validate-reconv}

As a final demonstration of \galsim's high precision operation, we
tested that we can accurately apply the reconvolution algorithm
of \S\ref{sect:shera} \citep{2012MNRAS.420.1518M}.
The aim is to represent the appearance of a test object 
following an applied shear $g_\text{applied}$, when viewed at lower
resolution.

This test was carried out using \code{Sersic} profiles convolved by a
known COSMOS-like PSF (a circular \code{Airy} profile),
rendered at high resolution (0.03 arcsec/pixel).
These images, along with images of the PSF, were then used as inputs to
initialize \code{RealGalaxy} objects, mimicking the use of real COSMOS
galaxy images.  In the usual manner these objects were sheared and
reconvolved by a broader (ground-based or Stage IV space-based survey)
PSF, then rendered at lower resolution.

Because of the use of an underlying parametric \sersic\ profile, the
rendering of which has been validated in \S\ref{sect:validate-dft}, we
can also render the convolved, sheared object \emph{directly} at lower
resolution to provide a reference for comparison.  We quantify any
error in the effectively applied shear due to the reconvolution
process as $m_{i, \text{reconv}}$ and $c_{i, \text{reconv}}$,
defined according to \eqn{eq:mc1}.

The test was done for 200 profiles whose parameters were selected from
the real COSMOS galaxy catalogue described in \citet{handbook}, using
random galaxy rotations in an 8-fold ring test configuration
\citep{nakajimabernstein07}.

Since galaxies with different light profiles might be more or less
difficult to accurately render using reconvolution, we must consider
not only the mean values of $m$ and $c$, but also investigate their ranges, which
could identify galaxy types for which the method fails to work
sufficiently accurately even if it is successful for most galaxies.

\begin{figure}
\begin{center}
%\epsscale{1.0}
\includegraphics[width=\columnwidth]{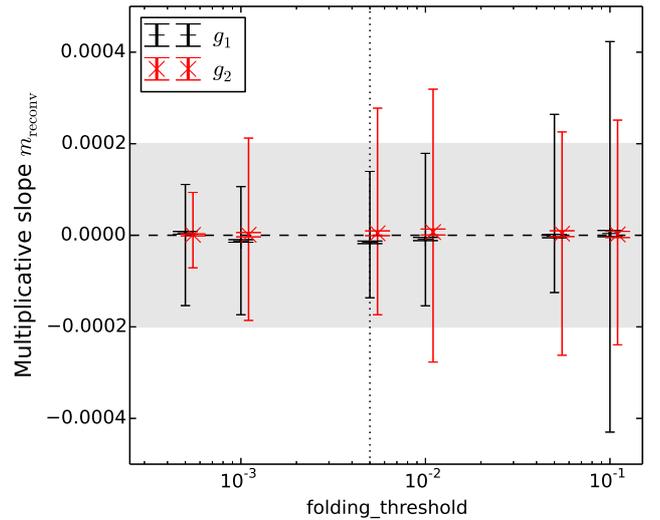}
\caption{Multiplicative slope $m_{i, \textrm{reconv}}$ for the
  reconvolution test of \S\ref{sect:validate-reconv}, for $g_1$ and
  $g_2$, as a function of the \code{folding\_threshold} parameter
  (cf.\ \S\ref{sect:tolerances}).
  The exterior error bars show the full range of values for the 200 models tested,
  and the points and interior error bars show the
  mean and standard error.  The shaded region
  shows the target for \galsim\ based on not exceeding one tenth
    of weak lensing accuracy requirements for Stage IV surveys (see
    \S\ref{sect:validation}).
  The default value of \code{folding\_threshold} is indicated by the
  dotted line.
  \label{fig:mreconv}}
\end{center}
\end{figure}

Fig.~\ref{fig:mreconv} shows $m_{i,
  \textrm{reconv}}$ as a function of the \code{folding\_threshold}
parameter described in \S\ref{sect:tolerances}.  Near the \galsim\
default value of $5 \times 10^{-3}$, our requirement $m_{i,
  \text{reconv}} < 2\times 10^{-4}$ is met comfortably in the ensemble
average.  Across the sample of 200 COSMOS galaxies a small fraction
(3/200) exceeded our requirement for the default
\code{folding\_threshold} value for $m_{2, \text{reconv}}$.  However, we
do not believe that this represents enough of a concern to change the
default \code{GSParams} settings. Provided that a representative
training set of galaxy models (such as the COSMOS sample), of
sufficient size, is used, the variation in $m_{i, \text{reconv}}$ seen
in Fig.~\ref{fig:mreconv} should not prevent simulations using the
reconvolution algorithm from being accurate to Stage IV
requirements for weak lensing.

If greater accuracy is required, users wishing to reduce the impact of
these effects can modify the values of the \code{GSParams} according
to their needs.  In this case, reducing \code{folding\_threshold} by a
factor of 10 brings $m_{2, \text{reconv}}$ within requirements for all
200 galaxies tested.  Additive biases $c_{i, \text{reconv}}$ were
found to be extremely small (and consistent with zero) in all cases.

These results shows that the approximations inherent in the
reconvolution process do not significantly interfere with \galsim's
ability to render accurate images suitable for weak lensing
simulations, for a realistic range of galaxy profiles drawn from
COSMOS.

\subsection{Limitations}

While results presented in this Section are encouraging, 
with the default settings providing accuracy that comfortably exceeds
our requirements by a factor of 5--10 in many cases, it must be 
remembered that no set of tests presented in this article could be
sufficient to positively validate \galsim's performance for all
possible future applications.  Users of \galsim\ are strongly advised
to conduct their own tests, tailored to their specific requirements.

One specific caveat worthy of mention is the adoption of a circular
PSF for all tests presented.  A circular \code{Airy} was chosen as a
simple approximation to the PSF found in COSMOS and other \hst\
images.  \galsim\ makes no distinction between those objects
describing PSFs and those describing galaxies when rendering
convolutions of multiple profiles. However, it is
possible that a subtle bug or {\edit genuine rendering} issue might only be
activated for cases where \emph{both} galaxy and PSF break circular
symmetry.  {\edit We stress again that no set of tests
  performed here can cover all
  possible uses of the code.  Where data properties such as the PSF
  are known we encourage \galsim\ users to perform their own
  validation and adjust the rendering parameters as required for their
  particular use case.}

Another caveat is that we only used a particular set of COSMOS
galaxies for the training sample.  It is plausible that
galaxy models drawn from a population with a different redshift
distribution to the COSMOS sample, or imaged in a filter other than
F814W, might have sufficiently different morphological characteristics
to fail the rendering requirements adopted in this work.  {\edit
  In future it will be of value to understand the marginal,
  $m_{\sigma,{\rm interp}} \simeq 4 \times 10^{-4}$ failure of galaxy
    size transformations in the \code{RealGalaxy} tests when both
    shear and magnification were applied (see \S\ref{sect:validate-ii})}.

In many cases, therefore, users may find it necessary to modify {\rm
  and refine} the
tests presented here, especially where the inputs and requirements of
their analyses differ significantly from the assumptions adopted. 
Some of the tests in this Section will hopefully serve as a
useful starting point for these investigations.

%
% Section 10
% performance
%

\section{Performance}\label{sect:performance}

While the paramount consideration in \galsim\ is the accuracy of the 
renderings, especially with respect to the shapes, we have also paid
considerable attention to making the code as fast as possible.
Code optimization is of course a wide ranging topic, and we will not
attempt to go into detail about all the optimizations included in
the \galsim\ code.  We just mention a few optimizations that
were particularly helpful and give some estimates of the timings for
typical use cases.

\subsection{Optimizations}

The most important performance decision was to use C++ for all time-critical
operations.  The Python layer provides a nice front-end user interface, but
it is notoriously difficult to do serious coding optimization in Python.
So for all significant calculations, \galsim\ calls C++ functions from 
within Python, where we have used the normal kinds of optimization
techniques that are standard with C and C++ code:
precomputing values that will be used multiple times,
having tight inner loops, using iterators to be more amenable to vectorization
by the compiler, using lookup tables and LRU caches, etc.

The FFTs for computing convolutions in DFT rendering normally
constitute the bulk of the calculation time.  For these, we use the
excellent \fftw\ package \citep{FFTW05}.  The calculations are fastest
when the image size is either an exact power of 2, or 3 times a power
of 2, so \galsim\ pads the images to meet this criterion before
performing the FFT.  Aside from this consideration, we only use as
large an FFT image as is required to achieve the desired accuracy.
This balance is one of the principal speed/accuracy tradeoffs
discussed in \S\ref{sect:tolerances}.

For some DFT calculations, we can avoid the two-dimensional FFT
entirely by turning it into a one-dimensional Hankel transform.  This
is possible when the profile being transformed is radially symmetric,
as are many of the analytic profiles.  If $f(r,\theta) = f(r)$, then
the Fourier transform is similarly radially symmetric:
\begin{align}
\fourier{f}(k) &= \int_0^\infty \!  \! r \, \rmd r\int_0^{2\pi} \! 
f(r,\theta)\, e^{{\rm i} k r \cos(\theta)} \, \rmd \theta\\
&= 2\pi \int_0^\infty  \! f(r)\, J_0(k r) \, r \, \rmd r
\end{align}
This optimization is particularly effective for \sersic\ profiles.
The large dynamic range of these profiles, especially for $n>3$, means
that a two-dimensional FFT would require a very large image to achieve
the appropriate accuracy.  It is significantly faster to precompute
the one-dimensional Hankel transform $\fourier{f}(k)$, and use that to
fill in the $k$-space values.

The integration package in \galsim\ uses an efficient, adaptive
Gauss-Kronrod-Patterson algorithm \citep{Patterson68}.  It starts with the 10-point
Gaussian quadrature abscissae and continues adding points at optimally
spaced additional locations until it either converges or reaches 175
points.  If the integral has not converged at that point, it splits
the original region in half and tries again on each sub-region,
continuing in this manner until all regions have converged.  This
algorithm is used in \galsim\ for Hankel transforms, real-space
integration, cumulative probability integrals for photon shooting,
calculating the half-light-radius of \sersic\ profiles, and a few
other places where integrations are required.

Matrix calculations do not generally account for a large share of the
running time, but to make these as efficient as possible, we use the
TMV library\footnote{\url{https://code.google.com/p/tmv-cpp/}}, which calls
the system BLAS library when appropriate if such a system library is
available (reverting to native code otherwise).  The matrix
calculations are thus about as efficient as can be expected on a given
platform.

One calculation required us to devise our own efficient function evaluation. 
For interpolated images using the Lanczos interpolant, Fourier-space calculations
require evaluation of the Sine integral:
\begin{equation}
\Si(x) = \int_0^x \frac{\sin t}{t} \rmd t
\end{equation}
The standard formulae from \citet{AS} \S5.2 for computing this
efficiently were only accurate to about $10^{-6}$.  This was not
accurate enough for our needs, and we could not find any other source
for a more accurate calculation.  We therefore developed our own
formulae for efficiently evaluating $\Si(x)$, accurate to $10^{-16}$.
The formulae are given in \app{sect:si}.

{\edit
\subsection{Parallelization}
}

Another important optimization happens in the Python layer.  When
running \galsim\ using a configuration file, rather than writing
Python code, it is very easy to get the code to run using multiple
{\edit
processes using the configuration parameter
\code{nproc}.
Of course, experienced Python programmers can also do this manually,
but multiprocessing in Python
}
is not extremely user friendly, so it can be convenient to take
advantage of this parallelism already being implemented in \galsim's
configuration file parser.

\subsection{Timings}

The time it takes for \galsim\ to render an image depends on many details of the 
profiles and the image on which they are being rendered.  Some profiles tend to take
longer than others, and the size of the object relative to the pixel scale of the final
image also matters.

For most analytic profiles, if the image scale is not much smaller than the Nyquist scale
of the object, rendering using the DFT method will generally take of order
0.01 seconds or less per object.  For very many purposes, this kind of profile is perfectly
appropriate, and it can include a reasonable amount of complication including multiple
components to the galaxy such a bulge plus \disk\ model, offsets of those components, 
multi-component PSFs, etc.

Models that include interpolated images take longer. This includes
models with an \code{OpticalPSF} component or a \code{RealGalaxy}.  Such profiles typically
take of order 0.1 seconds per object for typical sizes of the interpolated images, but
the time is highly dependent on the size of the image being used to define the profile.

The running time for photon shooting of course scales linearly with
the number of photons.  The crossover point where photon shooting is
faster than DFT rendering depends on the profile being rendered, but
it is typically in the range of $10^3$--$10^4$ photons.  So for faint
objects, it can be significantly faster to use the photon-shooting
method.

{\edit
\subsection{Potential speed pitfalls}

There are a number of potential pitfalls that can lead to slower than normal
rendering, which are worth highlighting:
}

\begin{itemize}
{\edit
\item \emph{Whitening.}
}
Profiles that use a \code{RealGalaxy} can be particularly
slow if the resulting image is whitened (cf.\ \S\ref{sect:whitening})
to remove the correlated noise that is
present in the original \hst\ images (and gets exacerbated by the PSF convolution).
The whitening step can increase the time per object to about 1 second,
although it varies depending on details such as the final pixel scale, what kind of 
PSF is used, etc.

{\edit
\item \emph{Pixel size.}
}
{\edit
Using a DFT to render images with very small pixels relative to the size of the galaxy
can be very slow.
The time required to perform the FFTs scales as $O(N^2 \log N)$ where $N$
is the number of pixels across for the image, so a
}
moderate increase in the size of the galaxy relative to the pixel
scale can make a big difference in the running time.

{\edit
\item \emph{\sersic\ profiles with varying index.}
}
\sersic\ galaxies require pre-computation of some
integrals for each \sersic\ index used.  When simulating many $n=3.3$
galaxies, the setup time will be amortized over many galaxies and is
thus irrelevant.  When using a different index $n$ for each galaxy
though, the setup must be done for each galaxy, potentially dominating
the running time.
{\edit
Thus, it is better to select $n$
from a list of a finite number of values
rather than letting it varying continuously within some range.
}

{\edit
\item \emph{Variable optical PSF parameters.}
}
The \code{OpticalPSF} class similarly has a large setup cost to
compute the Fourier transform of the wavefront in the pupil plane.  If
there is a single PSF for the image, this setup cost is amortized over
all the objects, but for a spatially variable PSF (which is more
realistic), each object will require a new setup calculation.  The
variable PSF branches of the GREAT3 challenge \citep{handbook} were
the most time-consuming to generate for this reason.

{\edit
\item \emph{Complicated wavelength dependence.} 
Drawing chromatic objects with complex SEDs through non-trivial 
bandpasses can take quite a long time.  The surface brightness profile
needs to be integrated over the relevant range of wavelengths.
The more complicated the SED and bandpass are, the more samples are
required to perform the integration.  Thus, depending on the purpose
of the images being rendered, it may be advisable to use simple SEDs and/or bandpasses,
instead of more realistic ones. 
  Optimizations to this process for non-trivial chromatic objects are being incorporated into
future versions of \galsim.}

\end{itemize}

\subsection{Example output}

\begin{figure*}
\begin{center}
\includegraphics[width=\textwidth]{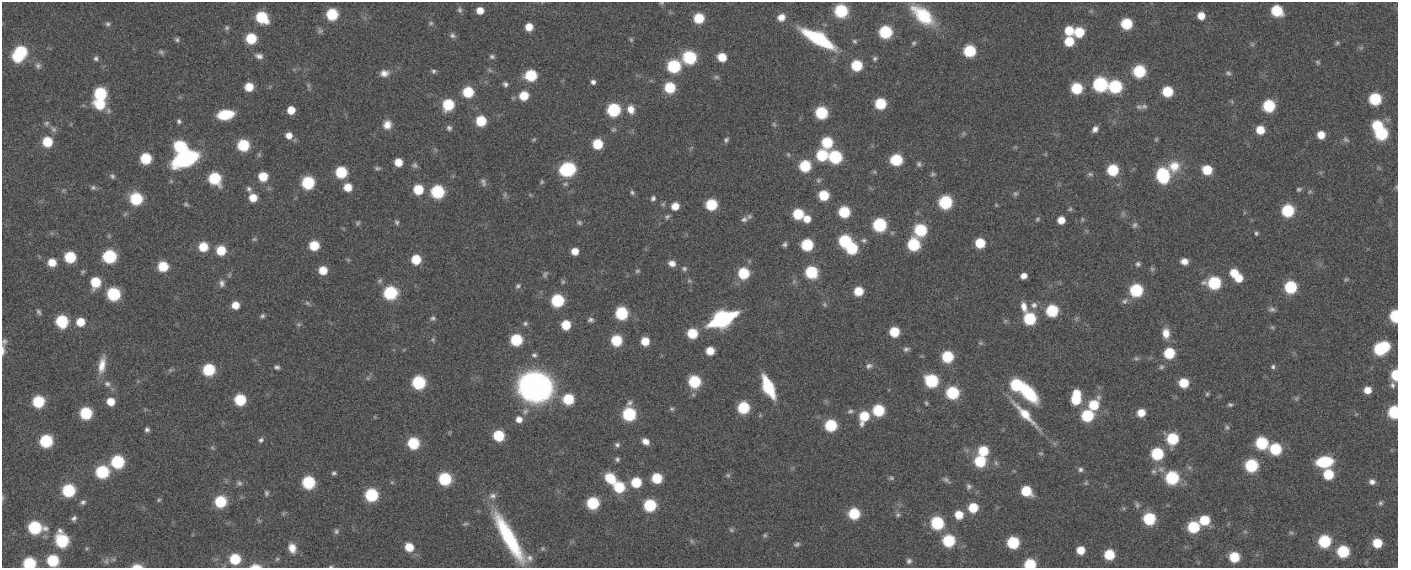}
\caption{\label{fig:lsst} A portion of a \galsim\ simulation of an LSST focal plane.
  This section is about 5 arcminutes across with a number
  density of 70 objects per square arcmin, of which 80\% are galaxies and 20\%
  are stars.
}
\end{center}
\end{figure*}

Figure~\ref{fig:lsst} shows a portion of an image that can be produced by \galsim.
This particular image is from a simulation that is intended to approximate an LSST focal plane.

The simulation has an object density of 70 objects per square arcminute.
The galaxies are bulge plus
\disk\ models, and the PSF includes both atmospheric seeing and an optical component with
the appropriate obscuration for the LSST telescope, struts, and plausible aberrations.
20\% of the objects are
stars.  The fluxes of both the galaxies and the stars follow a power law distribution,
as does the size distribution of the galaxies, and
there is an applied cosmological lensing field including both shear and magnification.
The noise model includes both Gaussian read
noise and Poisson noise on the counts of the detected electrons.

{\edit
The configuration file to build this image\footnote{
Available publicly at \url{http://ls.st/xj0}.}
is remarkably short (just 122 lines),
especially considering how many realistic data features are included.
This example showcases why we consider the configuration mode to be one
of the strengths of \galsim.
Even for users who are already comfortable writing Python code, the
configuration files can be easier to set up and get running in many
cases.
}

Each 4K$\times$4K chip in this simulation takes about
3 minutes to render on a modern laptop using a single core.
The simulation is set to use as many {\edit processors} as
are available, and with 8 cores running at once, the whole field of
view (189 chips in all) can be rendered in a bit more than an hour.

Currently \galsim\ is unable to simulate many of the features of real images including
saturation, bleeding, vignetting, cosmic rays, and the like (cf.\ \S\ref{sect:notingalsim}),
so those features are not shown here.  For this image, we also chose to
cut off the size distribution at 20 arcsec to avoid the rare objects
that would require an extremely large FFT.

Despite these apparent shortcomings, this kind of image is still very useful for
studying image processing algorithms.  Indeed, it is often helpful to
intentionally make these and other simplifications to
study the behavior of a particular algorithm.

Images similar to the one shown in 
Figure~\ref{fig:lsst} have been used to study the effects of blending,
star-galaxy separation algorithms, and shear estimation.  For these and many other 
investigations, complete realism of the simulated images is not required.

%
% Section 11
% notingalsim
%

\section{Effects not in \galsim}\label{sect:notingalsim}
It is worth mentioning some of the many physical effects in optical and
near-infrared astronomical observations that \galsim\ cannot yet
model.  Many of these effects will be important to model and explore 
to help ensure that accurate measurements will be possible from
future extragalactic survey data:
\begin{itemize}
\item \emph{Physical atmospheric PSF models.}  Examples of public
  software that can be used to generate physically motivated, multi-layer
  phase screen models of atmospheric PSFs include
  \textsc{Arroyo}\footnote{\url{http://cfao.ucolick.org/software/arroyo.php}},
  and the LSST
  \textsc{ImSim}\footnote{\url{http://lsst.astro.washington.edu/}} software
  (Peterson et~al. in prep.).
  While the possibility of adding such an atmospheric PSF module to
  \galsim\ was investigated as part of preparatory work for the GREAT3
  project \citep{handbook}, this functionality is still lacking.
\item \emph{Non-linear detector effects}.  \galsim\ cannot
  yet model pixel saturation or bleeding \citep[e.g.][]{bertin09},
  interpixel capacitance \citep[e.g.][]{mooreetal04} and other forms
  of cross-talk, all of which affect image flux and shape determination.
\item \emph{Image artifacts such as ghosts, detector persistence}.
  These effects (\citealp[e.g.][]{wynneetal84,mengetal13,longetal10};
  \citealp*{barricketal12}; \citealp{andersonetal14}) can be difficult
  to model, or even classify and remove, when measuring the properties
  of astronomical objects in an object-by-object fashion (the current
  standard procedure\footnote{Although see \url{http://thetractor.org/}}).
  Simulating observations with these effects will therefore be
  important to be able to demonstrate reliable measurements in their
  presence.
\item \emph{Cosmic rays}. Simulations of these effects
  \citep[such as, e.g.,][]{rollandetal08} are likely to be an important
  consideration for imaging in upcoming Stage IV surveys, particularly
  for space missions at the second Lagrange point
  \citep*[e.g.][]{barthetal00}.
\item \emph{Near field contaminants such as satellite trails and
    meteors.}  These effects, of relevance to ground-based imaging,
  are not currently supported by \galsim.  Like cosmic rays and
  instrumental artifacts such as ghosts and persistence, these effects
  will place stringent demands on object classification procedures for
  upcoming survey experiments.
\item \emph{Flexion.} Although trivially described using a
  photon-shooting approach, simulations of non-affine image
  transformations such as weak gravitational flexion are incompatible
  with the Fourier space rendering of individual profiles described in
  \S\ref{sect:rendering}.  Flexion is not currently implemented in
  \galsim. However, the
  \textsc{GalFlex}\footnote{\url{http://physics.drexel.edu/~jbird/galflex/}}
  package is a standalone, open-source Python module that can be
  integrated with \galsim\ via a \numpy\ array interface.
\item \emph{Vignetting, variable quantum efficiency, and flat fielding
    effects.}  No variation in the efficiency of photon detection as a
  function of position in the output image is currently included in
  \galsim, including variation within pixels.  It will be
    possible to address large scale effects in a relatively
    straightforward post-processing step.  However, significant
    intra-pixel quantum efficiency variation may require careful
    design to render efficiently.
\item \emph{Complicated WCS functions.} Recent studies of
  the DECam and LSST detectors \citep{plazas14,rasmussen14}
  have revealed complicated astrometric shifts, including edge
  distortions and tree-rings.  These are not currently modeled by
  any existing WCS package, so it would require a custom implementation
  to include these effects in \galsim. 
\end{itemize}

Many further possibilities for expanding and improving \galsim\
capabilities can be found in already existing functionality.  The
\code{OpticalPSF} class described in \S\ref{sect:opticalpsf} could be
extended to a more general treatment of optical wavefronts by allowing
an arbitrary number of Zernike coefficients to be specified on input.
Support for photon shooting of \code{Shapelet} objects (see
\S\ref{sect:shapelets}), although a considerable task, would also be a
valuable addition to \galsim.  It is hoped that, through the open
and collaborative development pattern already established within the
\galsim\ project, many of these capabilities will be added to \galsim\
in the future.

%
% Section 12
% conclusions
%

\section{Conclusions}\label{sect:conclusions}
We have described \galsim: the modular galaxy image simulation
toolkit.  A primary motivation for the project was the desire for an
open, transparent, and community-maintained standard toolkit for high
precision simulations of the deep extragalactic imaging expected in
upcoming Stage III and Stage IV surveys.  We have described the range
of physical models chosen as the basis for describing astronomical
objects in \galsim, and the support for their description in a range
of world coordinate systems.  We have also described the mathematical
principles used in the multiple strategies supported by \galsim\ for
rendering these models onto images, and for adding noise.

A number of the techniques employed by \galsim\ are novel, including
the photon shooting of a range of profiles including
\code{InterpolatedImage} objects (see \S\ref{sect:photon-interp}) and
noise whitening (\S\ref{sect:whitening}).  The consistent handling of
astronomical object transformation and rendering under a variety of
WCS transformations, described in \S\ref{sect:wcs}, is also new in the
field of astronomical image simulation.  \galsim\ is unique in
bringing all these features together under a simple, highly
flexible, class library interface.

An important part of any software project is testing and validation.
The provision of multiple parallel methods for rendering images (see
\S\ref{sect:rendering}), a capability unique to \galsim\ among
comparable simulation packages, played a vitally important role in
this process.  

In \S\ref{sect:validation} we showed results from a number of tests
demonstrating that \galsim\ output can describe weak lensing
transformations such as shear and magnification at the level of
accuracy required for Stage IV surveys such as \emph{Euclid},
\emph{LSST} and \emph{WFIRST-AFTA}.  This was demonstrated for highly
challenging analytic profiles (the \code{Sersic} class, see
\S\ref{sect:validate-dft}), for surface brightness profiles derived
from images such as the \code{InterpolatedImage} and \code{RealGalaxy}
classes (see \S\ref{sect:validate-ii}), and for the reconvolution
algorithm (see \S\ref{sect:validate-reconv}).

The accuracy of rendering results can be tuned using input parameters,
and using the \code{GSParams} structure described in
\S\ref{sect:tolerances}.  Less stringent requirements will yield
faster execution, but at the cost of lowering rendering accuracy in
output images. As no set of validation tests can completely cover all
usage scenarios, the tests of \S\ref{sect:validation} should be
modified by users of the \galsim\ software to tailor them to each
application's specific requirements.

As discussed in \S\ref{sect:notingalsim}, there are a number of
important effects that \galsim\ cannot yet simulate,
but that may significantly impact data analysis for upcoming
surveys.  Work is expected to continue in these interesting areas:
while the development and support of \galsim\ began in early 2012, it
is an ongoing project that continues to build on increasing numbers of
users (and developers).

As well as being used to generate the images for the GREAT3 challenge
\citep{handbook}, \galsim\ now provides a considerable open-source
toolkit for the simulation of extragalactic survey images.  Having
been shown to meet demanding tolerances on rendering accuracy in a
number of tests, \galsim\ simulations can be used as a benchmark for
development or common reference point for code comparison.  It also
provides an accessible, well-documented reference codebase.  \galsim\
is easily extended, or can be used simply as a class library in the
development of further astronomical image analysis applications.  In
these respects \galsim\ achieves its primary aims, and will become an
increasingly powerful toolkit for astronomical image simulation 
following development that is planned for the near future.

\section*{Acknowledgments}

{\edit We wish to thank the two anonymous referees whose insightful
comments significantly improved the paper.}
This project was supported in part by NASA via the Strategic
University Research Partnership (SURP) Program of the Jet Propulsion
Laboratory, California Institute of Technology. Part of BR's work was
done at the Jet Propulsion Laboratory, California Institute of
Technology, under contract with NASA.  BR, JZ and TK acknowledge
support from a European Research Council Starting Grant with number
240672.  
MJ acknowledges support from NSF award AST-1138729.
RM was supported in part by program HST-AR-12857.01-A,
provided by NASA through a grant from the Space Telescope Science
Institute, which is operated by the Association of Universities for
Research in Astronomy, Incorporated, under NASA contract NAS5-26555;
and in part by a Sloan Fellowship.  HM acknowledges support from Japan
Society for the Promotion of Science (JSPS) Postdoctoral Fellowships
for Research Abroad and JSPS Research Fellowships for Young
Scientists. PM is supported by the U.S.\ Department of Energy under
Contract No.\ DE- FG02-91ER40690.  The authors acknowledge the use of
the UCL Legion High Performance Computing Facility (Legion@UCL), and
associated support services, in the completion of this work.
% We discussed GalSim validation quite a bit at Aspen, so let's acknowledge them.
This work was supported in part by the National Science
Foundation under Grant No.\ PHYS-1066293 and the hospitality of the
Aspen Center for Physics.

\onecolumn

\appendix

\section{Conventions in the lensing engine}\label{sect:appendix}

\subsection{Grid parameters}

All calculations in the \galsim\ lensing engine (see
\S\ref{sect:lensingengine}) approximate real continuous functions
using a discrete, finite grid.  The real-space grid is defined by
\begin{align}
L &= \mbox{length of grid along one dimension (angular units)}\\
d &= \mbox{spacing between grid points (angular units)}\\
N &= \mbox{number of grid points along one dimension} = L/d.
\end{align}
There is a comparable grid in Fourier-space (i.e., same $N$).  Given
the parameters of the real-space grid, the \kmin\ along one dimension
is $2\pi/L$, so we can think of this quantity as the grid spacing in
Fourier space, i.e., $\kmin=\Delta k$.  This value of \kmin\
corresponds to a Fourier mode that exactly fits inside of our square
grid (in one dimension).  The $k$ range, again in one dimension, is
from $k_1=-\pi/d$ to $\pi/d$, i.e., $|k_1|<\kmax=\pi/d$.  This
corresponds to a mode that is sampled exactly twice (the minimum
possible) given our choice of grid spacing.  For the two-dimensional
grid, the maximum value of $|k|$ is then $\sqrt{2}\pi/d$.

\subsection{Fourier transform properties}

The convention for the Fourier transform defined by
\eqnb{eq:fwdft}{eq:invft} is adopted throughout, as this is the form
typically used for describing the statistics of cosmological shear and
matter fields.  Unlike the convention typically used in computing and
signals processing, and in the standard definition of the
Discrete Fourier Transform (DFT), this convention is non-unitary and
uses an angular (rather than normal) frequency in the
exponent\footnote{For a discussion of transform conventions see
  \url{http://en.wikipedia.org/wiki/Fourier_transform\#Other_conventions}}.
This introduces some additional factors of $2\pi$ into standard
results, and so here we re-express these results using the
non-unitary, angular frequency convention.

It can be readily shown that if we define $g(x L) \equiv
f(x)$, the following well-known identity holds:
\begin{equation} \mathcal{F}\{g(x L)\} = \frac{1}{L} \fourier{g}\left(\frac{k}{L}\right).
\label{eq:gmult} \end{equation} 
If we further define $s(h + x) \equiv g(x)$ then it is
straightforward to show that 
\begin{equation} \mathcal{F}\{ s(h + x) \} = \me^{\mi k h}
\fourier{s}(k).  \label{eq:sadd} 
\end{equation} 
There is an additional factor of $2\pi$ in the exponent when this
property of Fourier transforms is expressed using the normal (rather
than angular) frequency convention.

The most important result for approximating continuous functions using
DFTs is the Poisson summation formula, which under the conventions of
\eqnb{eq:fwdft}{eq:invft} may be stated as
\begin{equation}
\sum_{n=-\infty}^{\infty} f(n) = \sum_{q=-\infty}^{\infty} \fourier{f}(2
\pi q)
\end{equation}
for integers $n$ and $q$.
% This result can be derived by writing the expression for the inverse
% Fourier transform of $\fourier{f}(2 \pi q)$ and considering the
% Fourier series expansion of the periodic Dirac comb
%function
%\begin{equation}
%\Sha_L (x) = \sum_{n=-\infty}^{\infty} \delta(x - nL).
%\end{equation}
It should be noted that in the normal frequency, unitary transform
convention form of the Poisson summation formula the $2\pi$ within
$\fourier{f}(2 \pi q)$ is absent.  Using \eqnb{eq:gmult}{eq:sadd} we
arrive at the following useful expression of the Poisson summation
formula:
\begin{equation}
\sum_{n=-\infty}^{\infty} f(n L + x) = \frac{1}{L}
\sum_{q=-\infty}^{\infty} \fourier{f} \left(\frac{2 \pi  q}{L} \right)
\me^{\mi 2 \pi x q / L}.
\end{equation}
By substituting $\Delta k \equiv k_{\rm min} = 2 \pi / L$ we write
this as
\begin{equation}
\sum_{n=-\infty}^{\infty} f\left(\frac{2\pi n}{\Delta k} + x \right) = 
\left(\frac{\Delta k}{2 \pi} \right)
\sum_{q=-\infty}^{\infty} \fourier{f} \left(q \Delta k \right)
\me^{\mi x \Delta k q }. \label{eq:poisson}
\end{equation}

\subsection{Fourier transform of discrete samples of the power spectrum}
Let us define the following dimensionless function, which takes samples from a power spectrum:
\begin{equation}
P_{\Delta k} [q, p] \equiv (\Delta k)^2 P(q \Delta k, p \Delta k).
\end{equation}
We will use square brackets to denote functions with
discrete, integer input variables, as here.  We will also use $n, m$
for integer indices in
real space summations, and $q, p$ in Fourier space. % ($i,j$ are awkward
%due to the common notations for $\sqrt{-1}$).
It can then be shown that
\begin{equation}
  \mathcal{F}^{-1} \left\{\sum_{q,p = -\infty}^{\infty} \delta(k_1
    - q \Delta k) \, \delta(k_2
    - p \Delta k) \, P_{\Delta k} [q, p] \right\} = \sum_{n,m=-\infty}^{\infty} \xi_+\left(\theta_1
    - \frac{2 \pi n}{\Delta k},  \theta_2
    - \frac{2 \pi m}{\Delta k} \right), \label{eq:xisum}
\end{equation}
where we have used the Fourier series expression for the Dirac comb
function, and the convolution theorem.  We note that this expression
is still continuous, but describes an infinite, periodic summation (of
period $L = 2 \pi / \Delta k$) of copies of the correlation function
$\xi_+$.  For sufficiently small $\Delta k$, these copies may be
spaced sufficiently widely in the real domain to be able to learn much
about $\xi_+(\theta_1, \theta_2)$ in the non-overlapping regions.  We
therefore define this function as
\begin{equation}
  \xi_{\frac{2\pi}{\Delta k}}(\theta_1, \theta_2) \equiv \sum_{n = -\infty}^{\infty} \xi_+\left(\theta_1
    - \frac{2 \pi n}{\Delta k},  \theta_2
    - \frac{2 \pi m}{\Delta k} \right).
\end{equation}
Using the expression of the Poisson summation formula in \eqn{eq:poisson},
we can also write 
\begin{equation}
  \xi_{\frac{2\pi}{\Delta k}}(\theta_1, \theta_2) =
  \sum_{q,p=-\infty}^{\infty} \left( \frac{\Delta k}{2 \pi} \right)^2 P(q \Delta k, p \Delta k) \me^{\mi \Delta
    k (\theta_1 q + \theta_2 p)} = \frac{1}{( 2 \pi)^2}
  \sum_{q,p=-\infty}^{\infty} P_{\Delta k} [q, p] \me^{\mi \Delta
    k (\theta_1 q + \theta_2 p)}, \label{eq:dtft} 
\end{equation}
This is in fact an expression for the inverse of the Discrete Time
Fourier Transform (DTFT), although this result is normally derived
using unitary conventions for the transform pair.

The expression in \eqn{eq:dtft} is periodic with period $2 \pi /
\Delta k = L$.  All of the information it contains about $\xi_+
(\theta_1, \theta_2)$ is therefore also contained in one period of the
function only.  For approximating this information discretely, as
desired in numerical analysis, we can imagine taking $N$ equally
spaced samples of the function in \eqn{eq:dtft} along a single period
$L$ in each dimension.  These samples are therefore separated by
$\Delta \theta = 2 \pi / \Delta k N = d$ in real space, and we define
the sampled function itself as
\begin{equation}
\xi_{\Delta \theta}[n, m] \equiv \xi_{\frac{2 \pi}{\Delta
    k}} \left( n \Delta \theta, m \Delta \theta \right) = \frac{1}{( 2 \pi)^2}
\sum_{q,p=-\infty}^{\infty} P_{\Delta k} [q, p] \me^{\mi 2 \pi (qn +
  pm) / N},
\end{equation}
for the integer indices $n, m = 0, 1, \ldots, N-1$, by substitution
into \eqn{eq:dtft}.  Using the periodicity of the exponential term in
the expression above, this may be written as
\begin{equation}
\xi_{\Delta \theta}[n, m] = \frac{1}{( 2 \pi)^2}
\sum_{q,p=0}^{N-1} P_N [q, p] \me^{\mi 2 \pi (qn +
  pm) / N}
\end{equation}
where we have defined
\begin{equation}
P_N[q, p] \equiv \sum_{i,j=-\infty}^{\infty}  P_{\Delta k} [q - i N, p
- j N] .
\end{equation}
Needless to say, in order to be able to calculate the values of
$P_N[q, p]$ in practice we must also truncate the $P_{\Delta k}$
sequence to be finite in length.  A \emph{very} common choice is to
use the same number $N$ of samples in both real and Fourier space.
Choosing to use only $N$ samples from $P_{\Delta k}$ then gives
\begin{align}
\xi_{\Delta \theta}[n, m] &= \frac{1}{( 2 \pi)^2}
\sum_{q,p=0}^{N-1} P_{\Delta k} [q, p] \me^{\mi 2 \pi (qn +
  pm) / N} \nonumber \\
&= \frac{1}{ N^2 \left( \Delta \theta \right)^2} \sum_{q,p=0}^{N-1} P(q \Delta k, p \Delta k) \me^{\mi 2 \pi (qn +
  pm) / N}. \label{eq:dft}
\end{align}
For greater detail regarding this calculation, please see the lensing
engine document in the \galsim\
repository.\footnote{\url{https://github.com/GalSim-developers/GalSim/blob/master/devel/modules/lensing_engine.pdf}}

The relationship in \eqn{eq:dft} is in fact the inverse Discrete
Fourier Transform for our non-unitary convention. The more usual
definition of the inverse DFT, and that adopted by many DFT
implementations (including the \numpy\ package in Python), is
\begin{equation}
f[n, m] = \code{numpy.fft.ifft2}\left(\fourier{f}[p, q]\right) \equiv \frac{1}{N^2}
\sum_{q,p=0}^{N-1} \fourier{f}[q, p] \me^{\mi 2 \pi (qn +
  pm) / N} ~~~~ (\textrm{\numpy\ convention}). \label{eq:dftnumpy}
\end{equation}
(The factor of $1/N^2$ here is often found in conventions for the inverse DFT, and ensures
that the DFT followed by the inverse DFT yields the original, input
array.)  The Fast Fourier Transform algorithm allows the DFT to be
calculated very efficiently.

\subsection{Generating Gaussian fields}
Using \eqn{eq:dft} we see how to construct a 
realization of a Gaussian random lensing field on a grid, with an ensemble
average power spectrum that is approximately $P(k)$.  We create the following
complex, 2D array of dimension $N$ in Fourier space:
\begin{equation}
\fourier{\kappa}[p, q] = \fourier{\kappa}_E[p, q]+\textrm{i}\fourier{\kappa}_B[p, q],
\end{equation}
where $\fourier{\kappa}_E$ is defined in terms of the desired E-mode power
spectrum $P_E$ as
\begin{equation}
\fourier{\kappa}_E[p, q] = \frac{1}{N \Delta \theta}\sqrt{\frac{P_E(p\Delta
    k,q \Delta k)}{2}} \left\{
  \mathcal{N}(0, 1) + \textrm{i} \mathcal{N}(0, 1) \right\},
\end{equation}
and likewise for $\fourier{\kappa}_{B}$.  Here 
$\mathcal{N}(0, 1)$ denotes a standard Gaussian random deviate
drawn independently at each $[p, q]$ array location.  This
complex-valued Fourier-space convergence can be used to construct the
Fourier-space shear field via\footnote{See, for example,
  \cite{1993ApJ...404..441K}, but note that there is a sign error in
  equation 2.1.12 in that paper that is corrected in equation 2.1.15.}
\begin{equation}
\fourier{g} = \textrm{e}^{2\textrm{i}\psi} \fourier{\kappa},
\end{equation}
where $\textrm{e}^{2\textrm{i}\psi}$ is the phase of the $k$ vector
defined as $k=\Delta
k (p + \textrm{i} q)$.
 It can be seen
that $\langle \fourier{g}[p, q] \fourier{g}^*[p', q'] \rangle =
\delta_p^{p'} \delta_q^{q'} P(|k|) / (N \Delta \theta)^2$ where
$|k|=\Delta k \sqrt{p^2 + q^2} $.  Taking the inverse DFT of $\fourier{g}$
provides the realization of the Gaussian lensing field $g[n, m]$ on a grid in
real space, where the real (imaginary) parts of $g[n,m]$ correspond to
the first (second) shear components.  From
\eqn{eq:dft}, $g[n, m]$ will be correctly scaled to have discrete
correlation function $\xi_{\Delta \theta}[n, m]$ by construction.

\section{Efficient evaluation of the Sine and Cosine integrals}\label{sect:si}

The trigonometric integrals are defined as
\begin{align}
\Si(x) &= \int_0^x\frac{\sin t}{t}\, \rmd t = \frac{\pi}{2} -
\int_x^\infty \frac{\sin t}{t}\, \rmd t \\
\Ci(x) &= \gamma + \ln x + \int_0^x\frac{\cos t - 1}{t}\, \rmd t =
-\int_x^\infty \frac{\cos t}{t}\, \rmd t
\end{align}
where $\gamma$ is the Euler-Mascheroni constant.  In \galsim, calculations involving Lanczos
interpolants require an efficient calculation of $\Si(x)$.  As we were unable to find a source
for an efficient calculation that was suitably accurate, we developed our own,
which we present here.
We do not use $\Ci(x)$ anywhere in the code, but for completeness, we provide a similarly
efficient and accurate calculation of it as well.

For small arguments, we use Pad\'{e} approximants of the convergent Taylor series
\begin{align}
\Si(x) &= \sum_{n=0}^\infty \frac{(-1)^{n}x^{2n+1}}{(2n+1)(2n+1)!} \\
\Ci(x) &= \gamma+\ln x+\sum_{n=1}^{\infty}\frac{(-1)^{n}x^{2n}}{2n(2n)!}
\end{align}
Using the Maple software
package\footnote{\url{http://www.maplesoft.com/}}, we derived the
following formulae, which are accurate to better than $10^{-16}$ for $0 \le x \le 4$:
\begin{align}
\Si(x) &= x \cdot \left( 
\frac{
\begin{array}{l}
1 -4.54393409816329991\cdot 10^{-2} \cdot x^2 + 1.15457225751016682\cdot 10^{-3} \cdot x^4 \\
~~~ - 1.41018536821330254\cdot 10^{-5} \cdot x^6 + 9.43280809438713025 \cdot 10^{-8} \cdot x^8 \\
~~~ - 3.53201978997168357 \cdot 10^{-10} \cdot x^{10} + 7.08240282274875911 \cdot 10^{-13} \cdot x^{12} \\
~~~ - 6.05338212010422477 \cdot 10^{-16} \cdot x^{14}
\end{array}
}
{
\begin{array}{l}
1 + 1.01162145739225565 \cdot 10^{-2} \cdot x^2 + 4.99175116169755106 \cdot 10^{-5} \cdot x^4 \\
~~~ + 1.55654986308745614 \cdot 10^{-7} \cdot x^6 + 3.28067571055789734 \cdot 10^{-10} \cdot x^8 \\
~~~ + 4.5049097575386581 \cdot 10^{-13} \cdot x^{10} + 3.21107051193712168 \cdot 10^{-16} \cdot x^{12}
\end{array}
}
\right)\\
\Ci(x) &= \gamma + \ln(x) + \nonumber \\ 
& \quad x^2 \cdot \left(
\frac{
\begin{array}{l}
-0.25 + 7.51851524438898291 \cdot 10^{-3} \cdot x^2 - 1.27528342240267686 \cdot 10^{-4} \cdot x^4 \\
~~~ + 1.05297363846239184 \cdot 10^{-6} \cdot x^6 - 4.68889508144848019 \cdot 10^{-9} \cdot x^8 \\
~~~ + 1.06480802891189243 \cdot  10^{-11} \cdot x^{10} - 9.93728488857585407 \cdot 10^{-15} \cdot x^{12} \\
\end{array}
}
{
\begin{array}{l}
1 + 1.1592605689110735 \cdot 10^{-2} \cdot x^2 + 6.72126800814254432 \cdot 10^{-5} \cdot x^4 \\
~~~ + 2.55533277086129636 \cdot 10^{-7} \cdot x^6 + 6.97071295760958946 \cdot 10^{-10} \cdot x^8 \\
~~~ + 1.38536352772778619 \cdot 10^{-12} \cdot x^{10} + 1.89106054713059759 \cdot 10^{-15} \cdot x^{12} \\
~~~ + 1.39759616731376855 \cdot 10^{-18} \cdot x^{14} \\
\end{array}
}
\right)
\end{align}

For $x > 4$, one can use the helper functions:
\begin{align}
f(x) 
&\equiv \int_0^\infty \frac{\sin(t)}{t+x} \rmd t = \int_0^\infty
\frac{e^{-x t}}{t^2 + 1} \rmd t  = \Ci(x) \sin(x) + \left(\frac{\pi}{2} - \Si(x) \right) \cos(x) \\
g(x)
&\equiv \int_0^\infty \frac{\cos(t)}{t+x} \rmd t = \int_0^\infty \frac{t\,
  e^{-x t}}{t^2 + 1} \rmd t = -\Ci(x) \cos(x) + \left(\frac{\pi}{2} - \Si(x) \right) \sin(x)
\end{align}
using which, the trigonometric integrals may be expressed as
\begin{align}
\Si(x) &= \frac{\pi}{2} - f(x) \cos(x) - g(x) \sin(x) \\
\Ci(x) &= f(x) \sin(x) - g(x) \cos(x)
\end{align}

In the limit as $x \rightarrow \infty$, $f(x) \rightarrow 1/x$ and $g(x) \rightarrow 1/x^2$.  
Thus, we can use Chebyshev-Pad\'{e} expansions of $\frac{1}{\sqrt{y}} \; f\left(\frac{1}{\sqrt{y}} \right)$ and $\frac{1}{y} \; g\left(\frac{1}{\sqrt{y}} \right)$
in the interval $0..\frac{1}{4^2}$ to obtain the following approximants,
good to better than $10^{-16}$ for $x \ge 4$:
\begin{align}
f(x) &= \dfrac{1}{x} \cdot \left(\frac{
\begin{array}{l}
1 + 7.44437068161936700618 \cdot 10^2 \cdot x^{-2} + 1.96396372895146869801 \cdot 10^5 \cdot x^{-4} \\
~~~ + 2.37750310125431834034 \cdot 10^7 \cdot x^{-6} + 1.43073403821274636888 \cdot 10^9 \cdot x^{-8} \\
~~~ + 4.33736238870432522765 \cdot 10^{10} \cdot x^{-10} + 6.40533830574022022911 \cdot 10^{11} \cdot x^{-12} \\
~~~ + 4.20968180571076940208 \cdot 10^{12} \cdot x^{-14} + 1.00795182980368574617 \cdot 10^{13} \cdot x^{-16} \\
~~~ + 4.94816688199951963482 \cdot 10^{12} \cdot x^{-18} - 4.94701168645415959931 \cdot 10^{11} \cdot x^{-20}
\end{array}
}{
\begin{array}{l}
1 + 7.46437068161927678031 \cdot 10^2 \cdot x^{-2} + 1.97865247031583951450 \cdot 10^5 \cdot x^{-4} \\
~~~ + 2.41535670165126845144 \cdot 10^7 \cdot x^{-6} + 1.47478952192985464958 \cdot 10^9 \cdot x^{-8} \\
~~~ + 4.58595115847765779830 \cdot 10^{10} \cdot x^{-10} + 7.08501308149515401563 \cdot 10^{11} \cdot x^{-12} \\
~~~ + 5.06084464593475076774 \cdot 10^{12} \cdot x^{-14} + 1.43468549171581016479 \cdot 10^{13} \cdot x^{-16} \\
~~~ + 1.11535493509914254097 \cdot 10^{13} \cdot x^{-18}
\end{array}
}
\right) \\
g(x) &= \dfrac{1}{x^2} \cdot \left(\frac{
\begin{array}{l}
1 + 8.1359520115168615 \cdot 10^2 \cdot x^{-2} + 2.35239181626478200 \cdot 10^5 \cdot x^{-4} \\
~~~ + 3.12557570795778731 \cdot 10^7 \cdot x^{-6} + 2.06297595146763354 \cdot 10^9 \cdot x^{-8} \\
~~~ + 6.83052205423625007 \cdot 10^{10} \cdot x^{-10} + 1.09049528450362786 \cdot 10^{12} \cdot x^{-12} \\
~~~ + 7.57664583257834349 \cdot 10^{12} \cdot x^{-14} + 1.81004487464664575 \cdot 10^{13} \cdot x^{-16} \\
~~~ + 6.43291613143049485 \cdot 10^{12} \cdot x^{-18} - 1.36517137670871689 \cdot 10^{12} \cdot x^{-20}
\end{array}
}{
\begin{array}{l}
1 + 8.19595201151451564 \cdot 10^2 \cdot x^{-2} + 2.40036752835578777 \cdot 10^5 \cdot x^{-4} \\
~~~ + 3.26026661647090822 \cdot 10^7 \cdot x^{-6} + 2.23355543278099360 \cdot 10^9 \cdot x^{-8} \\
~~~ + 7.87465017341829930 \cdot 10^{10} \cdot x^{-10} + 1.39866710696414565 \cdot 10^{12} \cdot x^{-12} \\
~~~ + 1.17164723371736605 \cdot 10^{13} \cdot x^{-14} + 4.01839087307656620 \cdot 10^{13} \cdot x^{-16} \\
~~~ + 3.99653257887490811 \cdot 10^{13} \cdot x^{-18}
\end{array}
}
\right)
\end{align}
Thus we can calculate $\Si(x)$ to double precision accuracy for any value of $x$.
In \galsim, all of the above polynomials are implemented using Horner's rule, so they are very efficient 
to evaluate.

\bibliographystyle{elsarticle-harv}
\bibliography{ms}

\end{document}